\documentclass[twocolumn]{aastex62}
\usepackage[normalem]{ulem}
\useunder{\uline}{\ul}{}
\usepackage{hyperref}
\usepackage{lineno}

\usepackage{color}
\usepackage{amsmath,amssymb}

\hypersetup{linkcolor=cyan,citecolor=magenta,filecolor=yellow,urlcolor=blue}

\revised{\today}

\shorttitle{}
\shortauthors{}

\begin{document}

\title{Gravitomagnetic interaction of a Kerr black hole with a magnetic field as the source of the jetted GeV radiation of gamma-ray bursts}

\author{J.~A.~Rueda}
\affiliation{ICRA, Dip. di Fisica, Sapienza Universit\`a  di Roma, Piazzale Aldo Moro 5, I-00185 Roma, Italy}
\affiliation{ICRANet, Piazza della Repubblica 10, I-65122 Pescara, Italy}
\affiliation{ICRANet-Ferrara, Dip. di Fisica e Scienze della Terra, Universit\`a degli Studi di Ferrara, Via Saragat 1, I--44122 Ferrara, Italy}
\affiliation{Dip. di Fisica e Scienze della Terra, Universit\`a degli Studi di Ferrara, Via Saragat 1, I--44122 Ferrara, Italy}
\affiliation{INAF, Istituto di Astrofisica e Planetologia Spaziali, Via Fosso del Cavaliere 100, 00133 Rome, Italy}

\author{R.~Ruffini}
\affiliation{ICRANet, Piazza della Repubblica 10, I-65122 Pescara, Italy}
\affiliation{ICRA, Dip. di Fisica, Sapienza Universit\`a  di Roma, Piazzale Aldo Moro 5, I-00185 Roma, Italy}
\affiliation{Universit\'e de Nice Sophia-Antipolis, Grand Ch\^ateau Parc Valrose, Nice, CEDEX 2, France}
\affiliation{INAF,Viale del Parco Mellini 84, 00136 Rome, Italy}

\author{R.~P.~Kerr}
\affiliation{ICRANet, Piazza della Repubblica 10, I-65122 Pescara, Italy}
\affiliation{ICRA, Dip. di Fisica, Sapienza Universit\`a  di Roma, Piazzale Aldo Moro 5, I-00185 Roma, Italy}
\affiliation{University of Canterbury, Christchurch, New Zealand}

\email{jorge.rueda@icra.it, ruffini@icra.it, roy.kerr@canterbury.ac.nz}  

\begin{abstract}
We show that the gravitomagnetic interaction of a Kerr black hole (BH) with a surrounding magnetic field induces an electric field that accelerates charged particles to ultra-relativistic energies in the vicinity of the BH. Along the BH rotation axis, these electrons/protons can reach energies of even thousands of PeV, so stellar-mass BHs in long gamma-ray bursts (GRBs) and supermassive BHs in active galactic nuclei (AGN) can contribute to the ultrahigh-energy cosmic rays (UHECRs) thorough this mechanism. At off-axis latitudes, the particles accelerate to energies of hundreds of GeV and emit synchrotron radiation at GeV energies. This process occurs within $60^\circ$ around the BH rotation axis, and due to the equatorial-symmetry, it forms a double-cone emission. We outline the theoretical framework describing these acceleration and radiation processes, how they extract the rotational energy of the Kerr BH and the consequences for the astrophysics of GRBs.
\end{abstract}

\keywords{gamma-ray bursts: general -- black hole physics -- pulsars: general -- magnetic fields}

\section{Introduction}\label{sec:1}

In the absence of any observational evidence, the process of gravitational collapse was studied in the simplest possible mathematical solution of the Einstein equations, i.e., for a spherically symmetric and vacuum spacetime \citep{1939PhRv...56..455O, 1958PhRv..110..965F, 1960PhRv..119.1743K}. The renaissance of the physics of general relativity started with the new era of relativistic astrophysics heralded by three discoveries:
\begin{enumerate}
    \item
    On June $12$, $1962$, the first evidence of an X-ray source outside the solar system, Sco X1 \citep{PhysRevLett.9.439}, whose nature as a binary system was then identified by \citet{1968SvA....11..749S}. This was the first source of a long number of binary X-ray sources later discovered \citep[see, e.g.,][]{1978pans.proc.....G}. The identification of the detailed nature of Sco X1 stills today represents an open problem \citep[see, e.g.,][]{2020JHEAp..25....1J}.
    \item 
    On July $26$, $1962$, the discovery by \citet{1963PhRvL..11..237K} of an algebraically-special metric, among the solutions of the Einstein field equations in vacuum, that describes the gravitational field of a spinning mass. The \textit{Kerr solution} introduced a theoretical formalism of unprecedented complexity in general relativity to describe the effect of the rotation in the black hole (BH) geometry. From an observational point of view, the Kerr metric offered, in principle, since 1971 (see below) an unprecedented energy source for an astrophysical system, alternative, e.g., to nuclear energy sources.
    \item 
    On March $16$, $1963$, the epochal discovery of the quasar nature of the radio source 3C 273 by \citet{1963Natur.197.1040S}, at $z=0.158$, i.e., at a distance of about $760$~Mpc, implying energies of $10^{59}$~erg originating from the nuclear region of a galaxy. The explanation of the energy source in terms of the rotational energy of the BH is among the leading open astrophysical problems still today, and gaining additional crucial observations from closer AGN as M87 at $z=0.00436$ \citep{2010A&A...524A..71B} and IC 310 at $z=0.0189$ \citep{2002AJ....123.2990B}.
\end{enumerate}

The authentic shock of the pulsar discovery, the very quick identification of its nature as the first observed gravitationally collapsed object, and the long awaited explanation of the nature of the supernova (SN), marked the real birth of relativistic astrophysics and of the physics of gravitational collapse. The discovery of the Crab Nebula pulsar by \citet{1968Natur.217..709H} allowed:
\begin{enumerate}
    \item 
    The first, unequivocal observational identification of a pulsar as a rapidly rotating NS \citep{1968Natur.218..731G, 1968Natur.219..145P, 1969ApJ...155L.107F}.
    \item 
    The possibility to confirm the hypothesis of \citet{1934PhRv...46...76B} that NSs originate in SNe. In this case, the Crab pulsar had originated in the SN observed by Chinese astronomers in the year $1054$ \citep{1953DoSSR..90..983S,1969supe.book.....S}.
    \item
    To conceptually realize that, the process of gravitational collapse leading to NS formation is not an isolated event in time, but followed by a long-lasting emission of thousands of years. The conceptual role of this prolonged emission, traditionally neglected, is now acquiring a special role in the context of gamma-ray bursts (GRBs), where both the collapse to a NS and to a BH are observed daily with a long lasting X-ray and GeV emission (see \citealp{2021MNRAS.504.5301R}, references therein, and Eqs. \ref{eq:LX}--\ref{eq:Lgev} in this article).
\end{enumerate}

In Princeton, John A. Wheeler's group proceeded ahead in the study of BHs adopting as a background the geometry of the Kerr solution. This research program soon materialized in three steps:
\begin{enumerate}
    \item 
    The introduction in $1969$ of the effective potential technique to determine the properties of co-rotating and counter-rotating orbits, including the often quoted last stable circular orbits in the Kerr metric (see Ruffini \& Wheeler in \S 104 of \citealp{1975ctf..book.....L}).
    %
    \item 
    The identification, among these trajectories, of the ones corresponding to reversible and irreversible transformations presented in \citet{1970PhRvL..25.1596C}. 
    The derivation by Demetrios Christodoulou of the mass-energy formula of a Kerr BH was made possible by introducing the concept of \textit{irreducible} mass $M_{\rm irr}$ (from the Italian ``\textit{irriducibile}'') following the definition of \textit{ergosphere} introduced in \citet{1970PhRvL..25.1596C} by Ruffini and Wheeler. The further relation between the irreducible mass and the horizon area was soon shown in \citet{1971PhRvD...4.3552C} and \citet{1971PhRvL..26.1344H}. These results were obtained practicing on the \textit{gedanken} ``Penrose process'', see Fig. $2$ in \citet{1970PhRvL..25.1596C}, which we found to be  physically not implementable, as confirmed later by \citet{1971NPhS..229..177P}\footnote{The negative results on the ``Penrose process'' have in no way affected the derivation of the irreducible mass, uniquely based on the reversible and irreversible transformations, which never attracted Penrose's attention.}.
    \item 
    The publication of ``\textit{Introducing the black hole}'' \citep{1971PhT....24a..30R}, where the BH was introduced as a physical system characterized by three parameters, the total mass, $M$, the angular momentum, $J$, and the irreducible mass, $M_{\rm irr}$, with the geometry given by the Kerr metric.  
\end{enumerate}

A first step in this novel domain established the absolute upper limit to the value of the critical mass of a non-rotating NS, $3.2~M_\odot$, generalizing the first established value of $0.7~M_\odot$ by \citet{1939PhRv...55..374O}. The result was obtained, out of first principles, imposing the non-violation of causality, a fiducial nuclear density, and adopting the validity of general relativity \citep{1974PhRvL..32..324R}. A second step was the discovery of the first BH in our galaxy, Cygnus X1 \citep{1973ApJ...180L..15L, 1974asgr.proc..349R, 2003RvMP...75..995G}, using X-ray and optical data and imposing the above critical mass limit for BH formation.  Nevertheless, the discovery of such a Galactic object, although of great interest, did not offer any evidence for the new physics of the irreducible mass. It was clear that a binary X-ray source evolves on a timescale of $10^8$ yr, so we initially thought that we would have to wait for too long to see what occurs next in their evolution. But this was a big mistake. The gravitational collapse releases about $10^{52}$ erg, so such an event would be visible from any place in the Universe, and from the number of binary X-ray sources, that would lead to the observation of these events almost everyday. We had to wait a third gigantic step following the launch of Beppo-Sax and the discovery of the extragalactic origin of GRBs \citep{Costa1997, vanParadjis1997}. It was then started an unprecedented experimental effort of observing these most unique sources in the vastest range of wavelengths, from the radio to the keV, to the MeV, to the GeV, to the TeV, to the ultrahigh-energy cosmic rays (UHECRs). As a result, we have the opportunity to observe these sources up to high values of the cosmological redshift, e.g. $z=10$, allowing us to witness on a daily basis the birth of NSs and BHs. This new unprecedented condition allows us, for the first time, to submit also our own theoretical analysis to the scrutiny of direct observations and to verify which one of our theoretical assumptions, introduced $49$ years ago, were made just for the sake of simplicity or mathematical convenience.

In parallel, a novel path of theoretical research was gaining attention, diverging from a pure gravitational analysis of an isolated BH, in vacuum and stationary. Interest was called on the electrodynamics and magneto-hydrodynamics around BHs 
and on BH gravito-hydrodynamics \citep[see][and references therein]{Punsly2009}. The BH theory based on the above theoretical developments and on the vastest multiwavelength observations has been started to be constructed,<
with necessary changes of paradigm, and new ones that we infer in this article.

This article is dedicated to the development of the theoretical framework of the ``\textit{inner engine}'' originating the GeV radiation of GRBs within the binary-driven hypernova (BdHN) scenario (see  \citealp{2019ApJ...886...82R}, references therein, and
Sec.~\ref{sec:2} for details). It emphasizes the special role of the interaction of the \textit{gravitomagnetic} field of a Kerr BH with a uniform, asymptotically aligned background test magnetic field, following the mathematical solution of the Einstein-Maxwell equations by \citet{1966AIHPA...4...83P} and \citet{1974PhRvD..10.1680W}, hereafter referred to as the Papapetrou-Wald solution (see Secs.~\ref{sec:3}--\ref{sec:6} for details). As we shall shown, the above leads to a mechanism of particle acceleration and radiation in the BH vicinity, leading to UHECRs along the BH rotation axis, and to GeV radiation at off-axis latitudes within $60^\circ$ of the polar axis, with equatorial symmetry, i.e. there is a double-cone emission structure. The theory equations and their solution describing the particle acceleration and radiation processes, how they extract the rotational energy of the Kerr BH, as well as the consequences for the astrophysics of GRBs and AGN are here outlined. 

The article is organized as follows. In Sec.~\ref{sec:2}, we outline the observational features of the formation of a Kerr BH in a BdHN of type I, using GRB 190114C as a prototype. Section~\ref{sec:3} describes the structure of the electromagnetic field of the Papapetrou-Wald solution. In Sec.~\ref{sec:3}, we present the general relativistic equations of motion for charged particles including radiation losses. We show in Sec.~\ref{sec:4} specific results of the numerical integration of the equations of motion. In Sec.~\ref{sec:5}, we analyze the properties of the photon four-momentum observed at infinity to infer the properties of the observed radiation. Section~\ref{sec:6} is devoted to give quantitative estimates of the radiation power and the spectrum. Finally, we present in Sec.~\ref{sec:7} our conclusions.

\section{Black hole formation in a BdHN I: GRB 190114C as a prototype}\label{sec:2}

Prior to the year $2000$, GRBs were traditionally considered as single objects originating their energy from a Kerr BH \citep[see, e.g.,][]{1992MNRAS.258P..41R, 1993ApJ...405..273W, 1999PhR...314..575P}. Since then, a new scenario has been gradually developing in which GRBs originate in binary systems leading to their classification in nine GRB subclasses \citep[see, e.g.][]{2016ApJ...832..136R, 2019ApJ...874...39W, 2020ApJ...893..148R}. What has become clear is that a BH is generated only in some short GRBs originating in NS-NS mergers, and only in some long GRBs originating in binaries composed of a carbon-oxygen (CO) star and a NS companion in tight orbit.

\begin{figure}
    \centering
    \includegraphics[width=\hsize,clip]{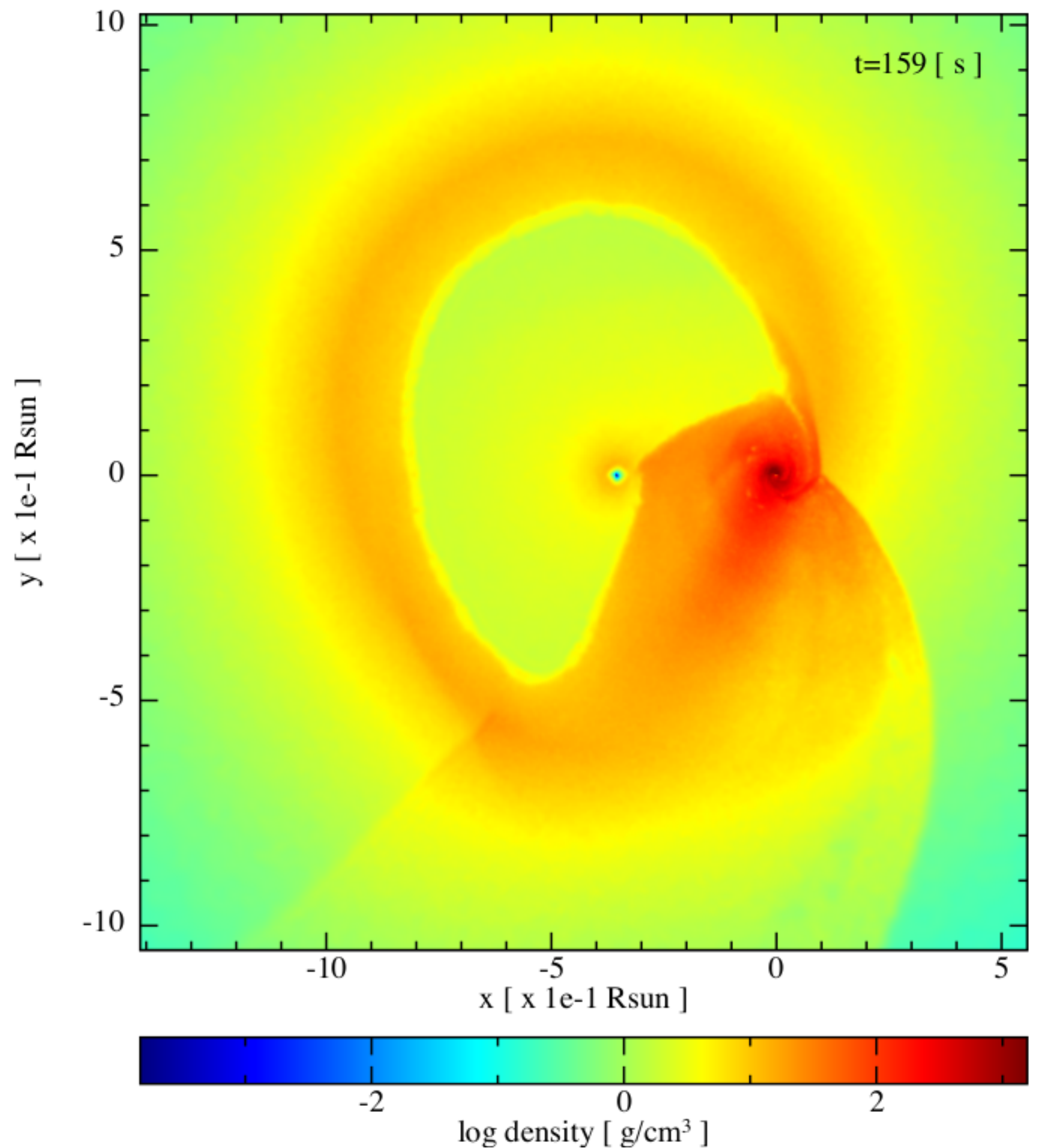}
    \caption{Numerical simulation of a BdHN I from \citet{2019ApJ...871...14B} (Model ``25m1p08E'' from Table~2 therein). The CO star of mass $M_{\rm CO}=6.85~M_\odot$ explodes as SN in presence of a binary companion NS of mass  $M_{\rm NS}=2~M_\odot$. At the center of the SN, it is formed a $\nu$NS of mass $1.85 M_\odot$. The orbital period of the binary system is $4.8$~min. This plot shows a snapshot of the mass density on the binary equatorial plane at a time $159$~s from the SN explosion. The reference system is rotated and translated so that the x-axis is along the line that joins the $\nu$NS and the NS, and the axis origin $(0,0)$ is located at the NS position. In this simulation, the NS reaches the point of gravitational collapse with a mass of $2.26 M_\odot$ and angular momentum $1.24 G M^2_\odot/c$. This binary system kept bound up the final time of the simulation despite the orbit widens, reaching an orbital period of $16.5$~min and an eccentricity of $\epsilon = 0.6$. In this simulation, the collapse of the NS into a BH occurs at $t=21.6$~min.}
    \label{fig:rhoSPH}
\end{figure}

We now focus on the latter special subclass of long GRBs that originate in short-period CO-NS binaries. The GRB ignition works as follows. As the CO star gravitationally collapses, it gives origin to a SN explosion and to a new NS (hereafter $\nu$NS) at its center. The ejected matter in the SN produces a hypercritical (i.e. highly super-Eddington) accretion process both onto the $\nu$NS (via matter fallback) and onto the companion NS. For short orbital periods of the order of a few minutes, the accretion process onto the NS companion is sufficiently massive to lead it to the critical mass, hence forming a BH (see Fig. \ref{fig:rhoSPH}). We have called these systems BdHN of type I \citep[see, e.g.,][]{2019ApJ...874...39W}. Until now, $380$ of BdHN I have been identified, with their evolution following a precise sequence of episodes, which we now indicate using GRB 190114C as a prototype \citep{2021PhRvD.104f3043M, 2021A&A...649A..75M}.

\begin{figure*}
    \centering
    \includegraphics[width=0.49\hsize,clip]{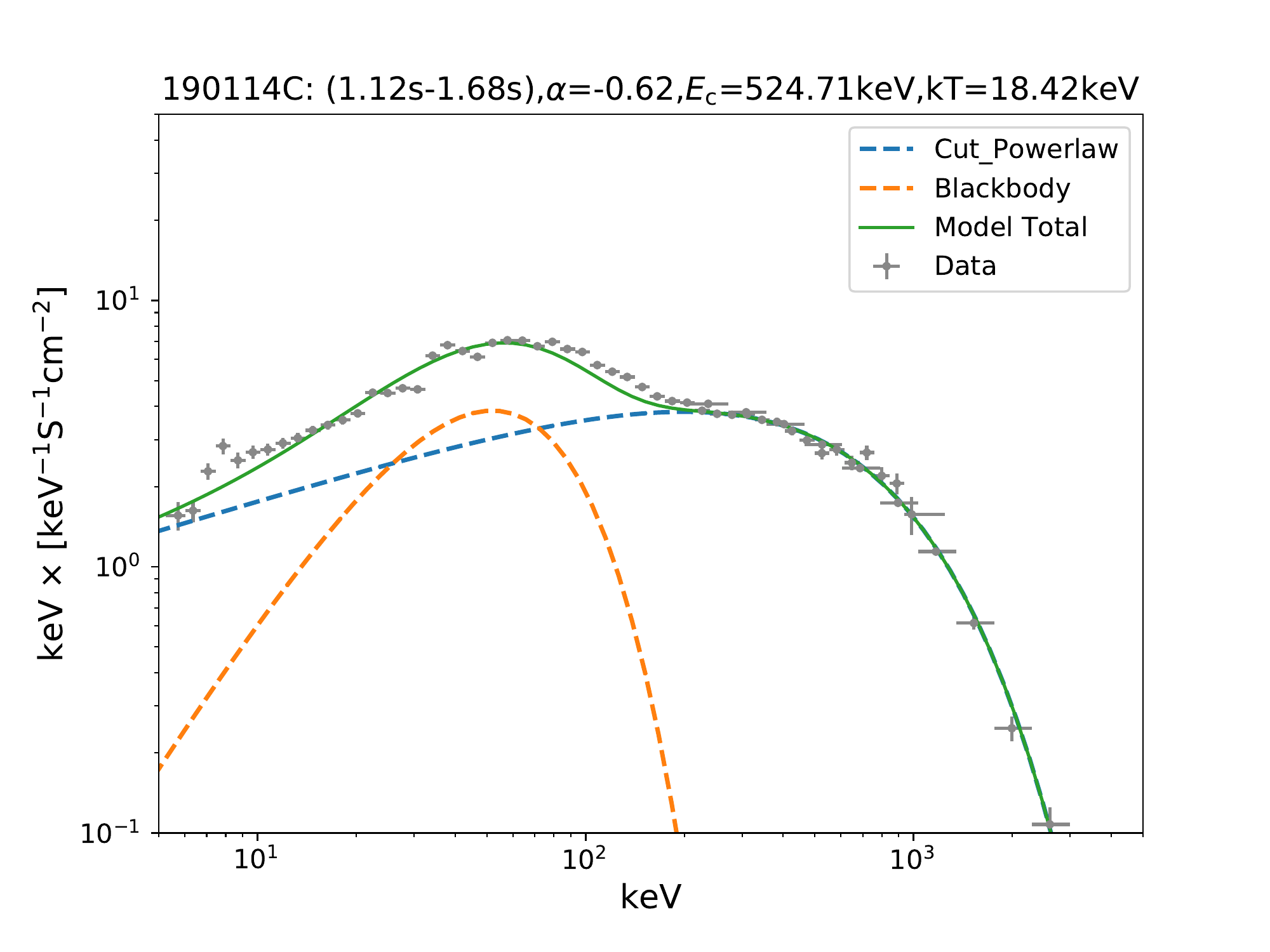}
    \includegraphics[width=0.49\hsize,clip]{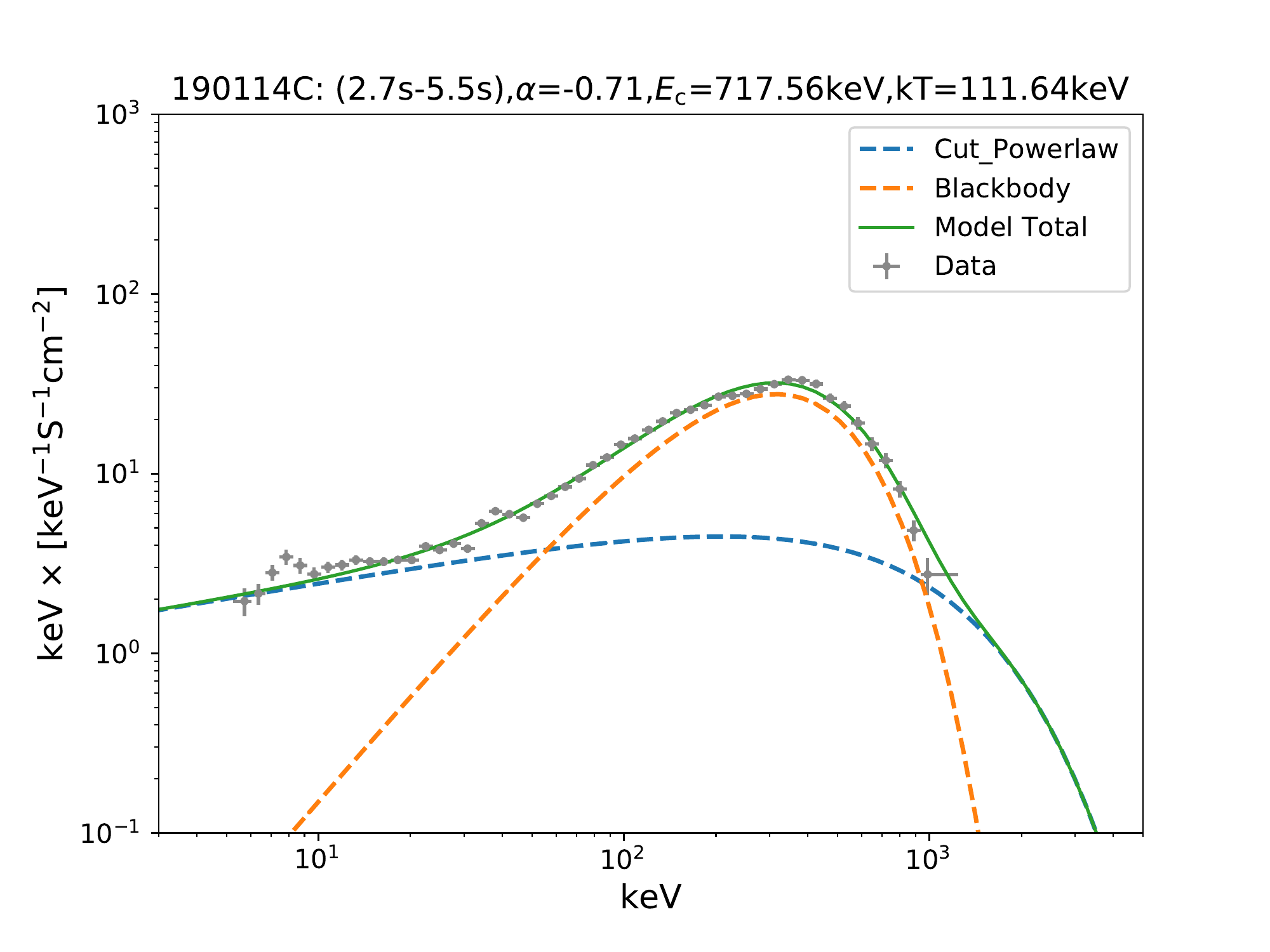} 
    \includegraphics[width=0.49\hsize,clip]{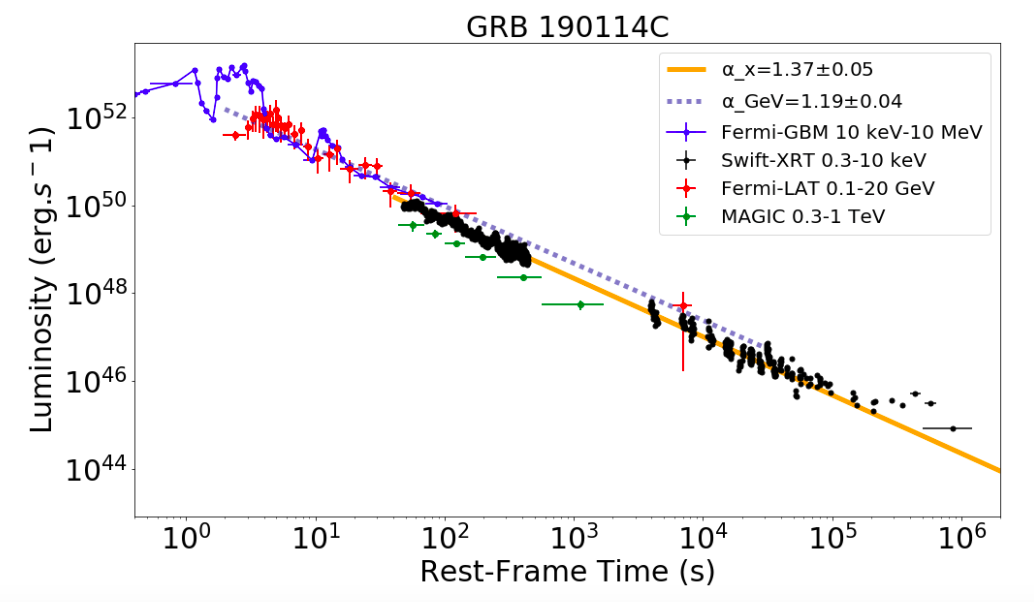}
    \includegraphics[width=0.49\hsize,clip]{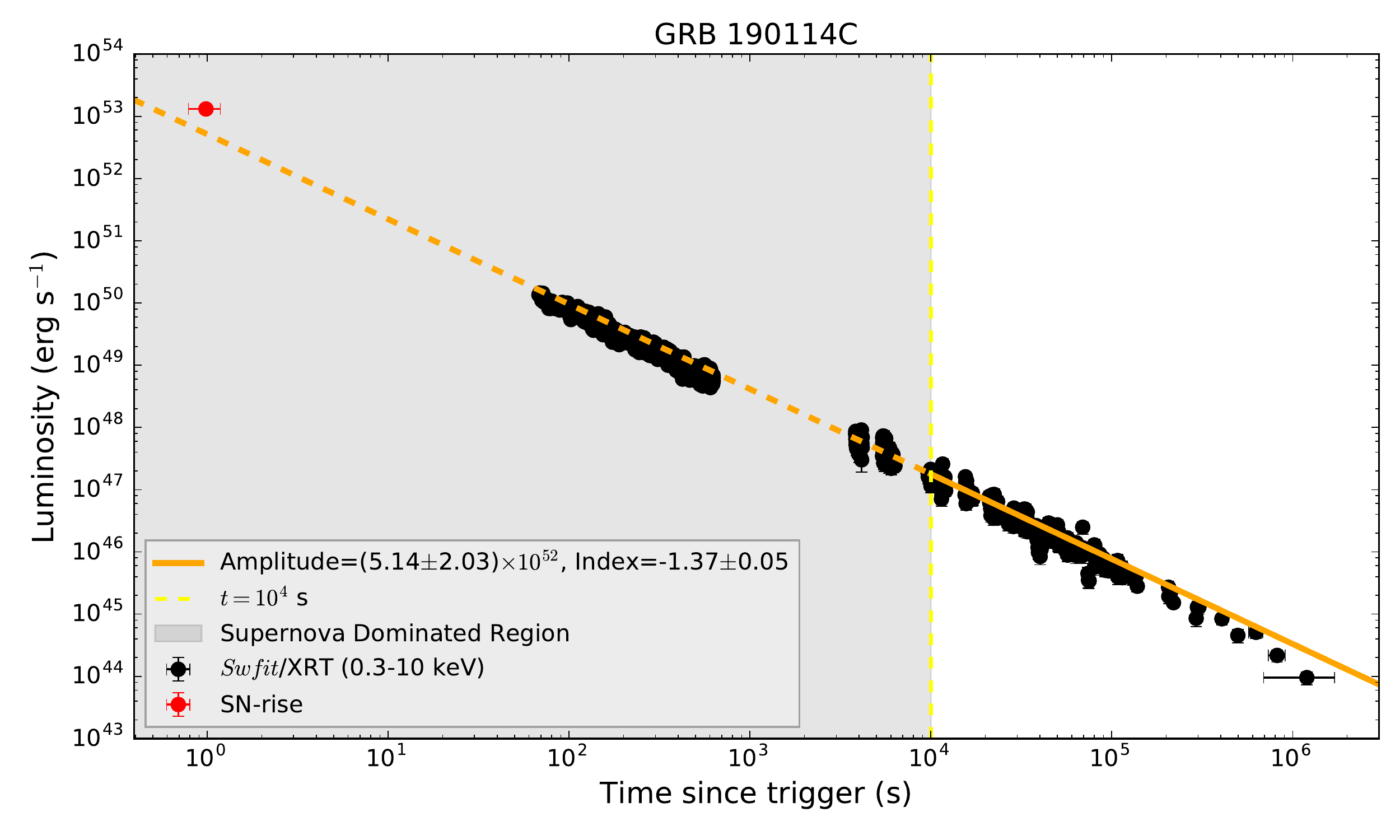}
    \caption{\textbf{\textit{Upper left}:} $\nu$NS-rise spectrum of BdHN I 190114C, corresponding to a time interval from $t = 1.12$~s ($t_{\rm rf} = 0.79s$) to $t=1.68$~s ($t_{\rm rf} = 1.18s$). The best-fit model is a CPL+BB with parameters (in the observer's frame): low-energy photon index $-0.71$, peak energy $E_{\rm c}=524.7$~keV, and blackbody temperature $k T =18.42$~keV. The energy of the $\nu$NS-rise is $E_{\nu\rm NS-rise} = 2.82\times 10^{52}$~erg.
    \textbf{\textit{Upper right}:} Spectral analysis of the UPE phase of GRB 190114C, i.e. in the time interval from $t=2.7$~s ($t_{\rm rf}=1.9$~s) to $t=5.5$~s ($t_{\rm rf}=3.9$~s). The best-fit model is a CPL+BB with the following parameters: power-law index $\alpha=-0.71^{+0.02}_{-0.02}$, cut-off energy $E_c=717.6^{+25.4}_{-25.4}$~keV, temperature $kT= 111.64^{+2.5}_{-2.5}$~keV. The observed energy released in the UPE is $E_{\rm UPE} =1.47\times 10^{53}$~erg \citep{2021PhRvD.104f3043M}.
    \textbf{\textit{Lower left}:} Multiwavelength luminosity of BdHN I 190114C. The blue points are the rest-frame $10$~keV--$10$~MeV luminosity from \textit{Fermi}-GBM. The black points are the rest-frame $0.3$--$10$ keV luminosity from Swift-XRT. The red points are the rest-frame $0.1$--$20$ GeV luminosity from Fermi-LAT. It follows a decaying power-law with amplitude $(4.6 \pm 0.6)\times 10^{52}$ erg s$^{-1}$, and index $\alpha_{\rm GeV} = 1.19 \pm 0.04$. The green points are the rest-frame $0.3$--$1$ TeV luminosity from MAGIC. 
    \textbf{\textit{Lower right}:} Luminosity of GRB 190114C observed by the \textit{Swift}-XRT in the $0.3$--$10$~keV energy band. The luminosity is well fitted by a power-law function given by Eq.~(\ref{eq:LX}), with $A_X = (5.14\pm 2.03)\times 10^{52}$~erg~s$^{-1}$ and $\alpha_X=1.37\pm 0.05$. The observed energy released in the X-ray emission is $E_X=2.11\times10^{52}$~erg \citep{2021PhRvD.104f3043M}.
    }\label{fig:GRB190114C}
\end{figure*}

The time $t=0$ coincides with the trigger of GRB 190114C given by the Fermi satellite. The progenitor is a CO-NS binary with an inferred orbital period of $2.5$~min \citep{2020ApJ...893..148R}. The appearance of the $\nu$NS, hereafter $\nu$NS-rise (formerly called SN-rise), following the CO star core-collapse, occurs between $0.79$~s and $1.18$~s in the rest-frame of the source. The $\nu$NS-rise triggers the entire BdHN I evolution. In the SN explosion, as usual, a $\nu$NS is created, which operates in addition to the binary companion NS (see Fig.~\ref{fig:rhoSPH}). The spectrum of the $\nu$NS-rise, observed in the gamma-rays by the Fermi-GBM detector, is shown in Fig.~\ref{fig:GRB190114C} (upper left panel). The energy of the $\nu$NS-rise in GRB 190114C is $E_{\nu\rm NS-rise} = 2.82\times 10^{52}$~erg \citep{2021MNRAS.504.5301R}.

The SN ejecta now accrete at hypercritical rates onto the binary companion NS. At $1.9$~s, the NS companion reaches its critical mass and collapses creating a Kerr BH, highlighted by the onset of the GeV emission (see Fig.~\ref{fig:GRB190114C}; lower left panel). Between $1.9$~s and $3.9$~s in the source rest-frame, the ultrarelativistic prompt emission (UPE) occurs, characterized by a cutoff power-law plus blackbody (CPL + BB) spectrum and the energetics reported in Fig.~\ref{fig:GRB190114C} (upper right panel). The energy released observed in the UPE phase is $E_{\rm UPE}=1.47\times 10^{53}$~erg \citep{2021PhRvD.104f3043M}.

At $47.83$~s, the turn on of the Neil Gehrels \textit{Swift} satellite occurs and allows to perform the observations of the ongoing emission of the X-ray afterglow. One of the major successes of the BdHN I theory has been the identification in the SN accretion onto the rapidly rotating $\nu$NS, with a spinning period of $2.1$~ms \citep{2020ApJ...893..148R}, the energy source of the afterglow of GRB 190114C (see Fig.~\ref{fig:GRB190114C}; lower right panel). The observations of GRB 190114C and of its ``\textit{twin}'' source GRB 180427A have further allowed to explain the observed spectra of the afterglow characterized by the synchrotron emission originating from the rapidly rotating $\nu$NS, interacting with the SN ejecta. Further examples of afterglows fulfilling this approach and leading to the determination of the $\nu$NS rotation period have been the cases of GRB 180728A with a $\nu$NS spin of $3.5$~ms, GRB 130427A with $0.95$~ms, GRB 160625B with $0.5$~ms, GRB 160509A with $0.75$~ms, and GRB 090926A with $1.1$~ms \citep{2020ApJ...893..148R}.

A most significant result has been that all $378$ identified BdHN I have an afterglow characterised by a decreasing luminosity with time, that expressed in the rest frame of the source is well fitted by a power-law function
\begin{equation}\label{eq:LX}
    L_X = A_X~t^{-\alpha_X},
\end{equation}
which for GRB 190114C $A_X = (5.14\pm 2.03)\times 10^{52}$~erg~s$^{-1}$ and $\alpha_X=1.37\pm 0.05$ \citep{2021MNRAS.504.5301R}. The observed energy released in the X-ray emission is $E_X=2.11\times10^{52}$~erg.

We have identified above the role of the SN in creating a magnetized rapidly rotating $\nu$NS, as well as its role in producing a hypercritical accretion process onto the $\nu$NS which helps to power the X-ray afterglow. 
There are three main conclusions on the afterglow: 
\begin{enumerate}
    \item 
    The initial rotational energy of the $\nu$NS with the observed power-law luminosity given by Eq. (\ref{eq:LX}) may well justify an afterglow emission occurring  for an infinite time.
    \item 
    These results open a new scenario with respect to the usually assumed ultrarelativistic blastwaves in the traditional GRB model, which has been shown to contradict model-independent constraints \citep{2018ApJ...852...53R} and implying unacceptable energetic requirements for the system.
    \item 
    The presence of a rapidly spinning $\nu$NS in the afterglow can lead to the possibility of relating $\nu$NS ``glitches'' to the emission of TeV emission, e.g. in GRB 180720B (Moradi, et al., submitted).
\end{enumerate}

At $3.9$~s, at the end of the UPE phase, the GeV radiation observed by LAT is emitted following a decreasing power-law luminosity (see Fig.~\ref{fig:GRB190114C}; lower left panel)
\begin{equation}\label{eq:Lgev}
    L_{\rm GeV} =A_{\rm GeV}~t^{-\alpha_{\rm GeV}},
\end{equation}
with $A_{\rm GeV} = 7.75 \pm 0.44 \times 10^{52}$~erg~s$^{-1}$ and $\alpha_{\rm GeV} = 1.2\pm 0.04$. The energy of the GeV emission is $E_{\rm GeV} = 8.3\times 10^{52}$~erg \citep{2021A&A...649A..75M}.

The existence of the above mentioned power-laws in the X-ray emission observed by the Swift satellite, originating from the rotational energy extraction from a spinning $\nu$NS, see Eq.~(\ref{eq:LX}), and the analogous one in the GeV observed by the Fermi-LAT satellite, originating from the rotational energy extraction Kerr BH, see Eq.~(\ref{eq:Lgev}), are the two conceptually new observables characterizing GRBs. Every gravitational collapse episode occurring within a GRB, either leading to a rotating NS or to a rotating BH, is not an isolated event in space and time. On the contrary, it is the beginning of a process characterized by a decreasing power-law luminosity that, in principle, may last the entire life of the Universe. What is clear is that, currently, we are observing all \textit{alive} Kerr BHs and not \textit{dead} Schwarzschild BHs. The crucial characterization of this difference resides in the \textit{jetted} emission and the possible presence of trapped surface and of shadows cannot be meaningfully addressed on the ground of this physics.

In order to approach in the following sections the physics of the \textit{inner engine}, we have first  to introduce three main new paradigms:
\begin{enumerate}
    \item 
    The rotational energy extraction from a Kerr BH and the associated acceleration process originates in the \textit{gravitomagnetic} interaction of the Kerr BH with a uniform background magnetic field \citep{2019ApJ...886...82R}. Necessarily, the conditions of stationarity of the Kerr solution, as well as assuming such a solution in vacuum and all the considerations of the geodesics of uncharged particles have to be superseded, see Sec.~\ref{sec:3}.
    \item 
    The geodesic equations of motion of massive particles around a Kerr BH \citep{1971ESRSP..52...45R, 1971PhT....24a..30R} are superseded by the equation of motion of positively and negatively charged particles in the field of a  Papapetrou-Wald solution, taking into due account the radiation reaction forces; see Secs.~\ref{sec:4} and \ref{sec:5}, as well as Appendix \ref{app:A}--\ref{app:C}.
    \item 
    The fundamental role of the reversible transformations stands, indeed, the irreducible mass $M_{\rm irr}$ is the fundamental regulator of the energy extraction process from the Kerr BH and the emission of the radiation power and spectrum; see Secs.~\ref{sec:6} and \ref{sec:7}. 
\end{enumerate}

\section{The electromagnetic field structure}\label{sec:3}

The \textit{inner engine} of the high-energy emission of long GRBs was presented in \citet{2019ApJ...886...82R}, and applied there to GRB 130427A. It is composed of a Kerr BH \citep{doi:10.1063/1.1705193,1968PhRv..174.1559C}, embedded in a test, asymptotically aligned magnetic field, described by the Papapetrou-Wald \citep{1966AIHPA...4...83P,1974PhRvD..10.1680W} solution of the Einstein-Maxwell equations.

We here present a full (numerical) integration of the equations of motion of a charged particle in the electromagnetic field of the Papapetrou-Wald solution, accounting for radiation in the background of the Kerr metric. We use geometric units $c=G=1$, unless otherwise specified.

Denoting by $\vec{\eta} = \partial/\partial \vec{t}$ and $\vec{\psi}= \partial/\partial \vec{\phi}$, respectively the time-like and space-like Killing vectors for the Kerr metric (see Appendix~\ref{app:A}), the electromagnetic four-potential is given by \citep{1974PhRvD..10.1680W}
\begin{equation}\label{eq:Amu}
    A_\mu = \frac{B_0}{2} \,\psi_\mu + a \,B_0\, \eta_\mu,
\end{equation}
where $B_0$ is the asymptotic value of the magnetic field strength. Because of $\eta^\mu = \delta^\mu_{\hphantom{\mu}{t}}$ and $\psi^\mu = \delta^\mu_{\hphantom{\mu}{\phi}}$, the non-vanishing components of the four-potential, in the Boyer-Lindquist coordinate basis of the Kerr metric (see Appendix~\ref{app:A}), read
\begin{align}\label{eq:potential}
    A_t &= -a B_0 \Bigg[ 1 - \frac{M r}{\Sigma} (1+\cos^2\theta) \Bigg],\\
    A_\phi &= \frac{1}{2} B_0\sin^2\theta  \Bigg[ r^2 + a^2 - \frac{2 M r a^2}{\Sigma} (1+\cos^2\theta) \Bigg].
\end{align}

As in \citet{1978PhRvD..17.1518D}, we introduce a local Lorentz observer, specifically a \emph{locally non-rotating} (LNR) observer \citep{1970ApJ...162...71B,1972ApJ...178..347B}, in order to analyze the electromagnetic field properties and the equations of motion.

The LNR observer carries a tetrad basis with vectors $\vec{e}_{\hat a}$; see Appendix~\ref{app:A} for details on the tetrad and related framework of the subsequent calculations. We use a hat to distinguish components (projections) in the LNR frame from the ones in coordinate frame. Latin alphabet (e.g.~$a$) and Greek indexes (e.g.~$\mu$) run over the spacetime coordinates, i.e. from $0$ to $3$ ($t$, $r$, $\theta$, and $\phi$), while Latin indexes (e.g.~$i$) run only over the spatial coordinates.

In the LNR frame, the electric and magnetic field components are given by
\begin{subequations}
\begin{align}
    E_{\hat{i}} & = E_\mu\,\vec{e}^\mu_{\hphantom{\mu}{\hat i}} = F_{\hat{i} \hat{t}}, \\
    B_{\hat{i}} & = B_\mu\,\vec{e}^\mu_{\hphantom{\mu}{\hat i}} = \epsilon_{\hat{i}\hat{j}\hat{k}}F^{\hat{j}\hat{k}},
\end{align}
\end{subequations}
where $F_{\mu \nu}$ is the electromagnetic field tensor in the coordinate basis. Expressed in Boyer-Lindquist coordinates, the electric and magnetic field are \citep{1978PhRvD..17.1518D}
\begin{subequations}\label{eq:Efieldzamo}
\begin{align}
E_{\hat{r}} &= -\frac{B_0 a M}{\Sigma^2 A^{1/2}} \Bigg[(r^2 + a^2)(r^2-a^2\cos^2\theta)(1+\cos^2\theta) \nonumber \\
&- 2 r^2 \sin^2\theta\,\Sigma\Bigg],\\
E_{\hat{\theta}} &= B_0 a M\,\frac{\Delta^{1/2}}{\Sigma^2 A^{1/2}} 2 r a^2 \sin\theta \cos\theta (1+\cos^2\theta),
\end{align}
\end{subequations}
\begin{subequations}\label{eq:Bfieldzamo}
\begin{align}
B_{\hat{r}} &= -\frac{B_0 \cos\theta}{\Sigma^2 A^{1/2}} \Bigg\{2 M r a^2 [2 r^2 \cos^2\theta+a^2(1+\cos^4\theta)] \nonumber \\
&-(r^2+a^2)\Sigma^2\Bigg\},\\
B_{\hat{\theta}}&= -\frac{\Delta^{1/2}B_0 \sin\theta}{\Sigma^2 A^{1/2}} [M a^2 (r^2-a^2\cos^2\theta)(1+\cos^2\theta) \nonumber \\
&+ r \Sigma^2].
\end{align}
\end{subequations}

For completeness, we give the expressions of the electric and magnetic field in the Boyer-Lindquist coordinate basis (see Eqs.~\ref{eq:ZAMOtetrad}) 
\begin{subequations}
\begin{align}
    \vec{E} &= E_{\hat{r}}\,\vec{e}_{\hat{r}} + E_{\hat{\theta}}\,\vec{e}_{\hat{\theta}} = E_{\hat{r}} \sqrt{\frac{\Delta}{\Sigma}}\vec{e}_r + E_{\hat{\theta}}\frac{1}{\sqrt{\Sigma}}\vec{e}_\theta,\\
    \vec{B} &= B_{\hat{r}}\,\vec{e}_{\hat{r}} + B_{\hat{\theta}}\,\vec{e}_{\hat{\theta}} = B_{\hat{r}} \sqrt{\frac{\Delta}{\Sigma}}\vec{e}_r + B_{\hat{\theta}}\frac{1}{\sqrt{\Sigma}}\vec{e}_\theta.
\end{align}
\end{subequations}

\begin{figure*}
    \centering
    \includegraphics[width=0.32\hsize,clip]{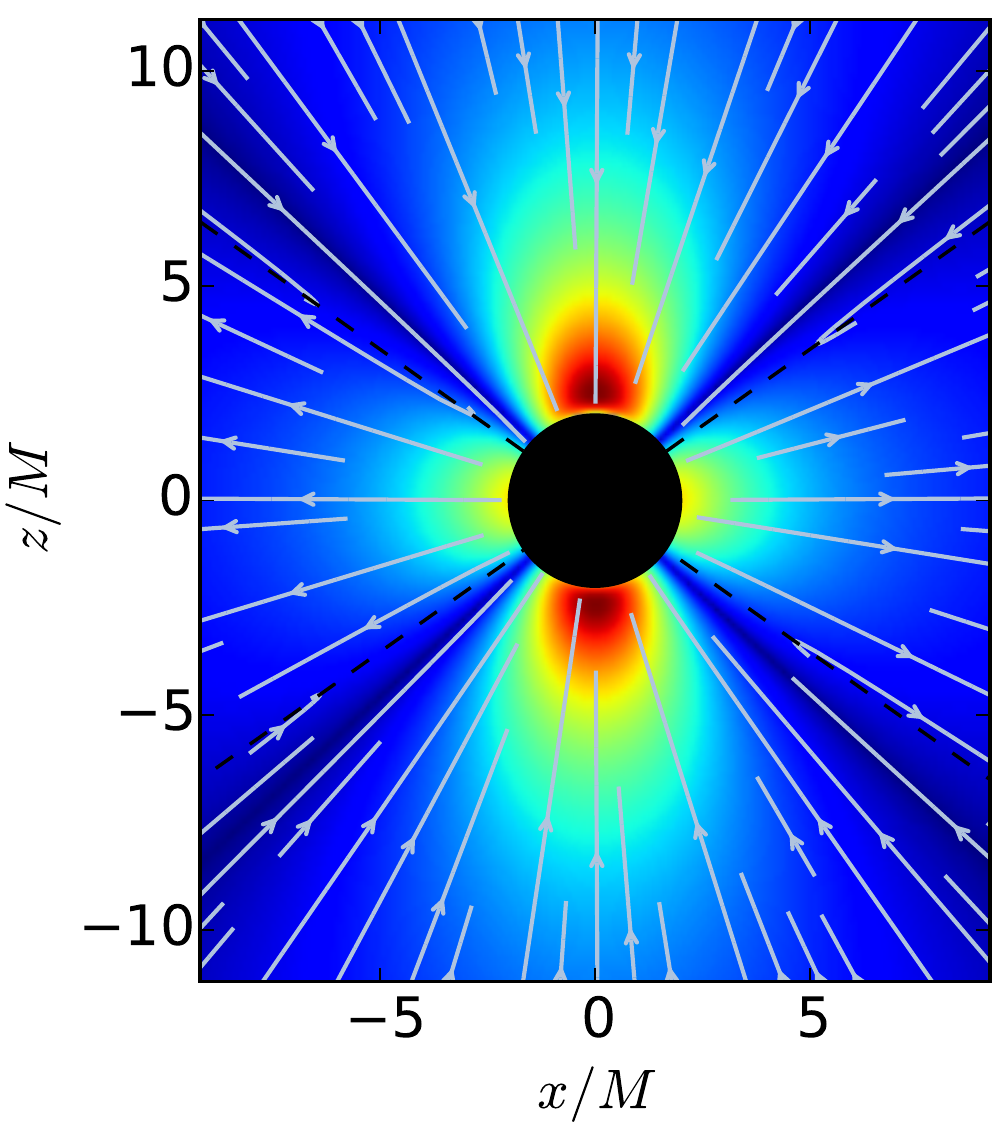}
    \includegraphics[width=0.32\hsize,clip]{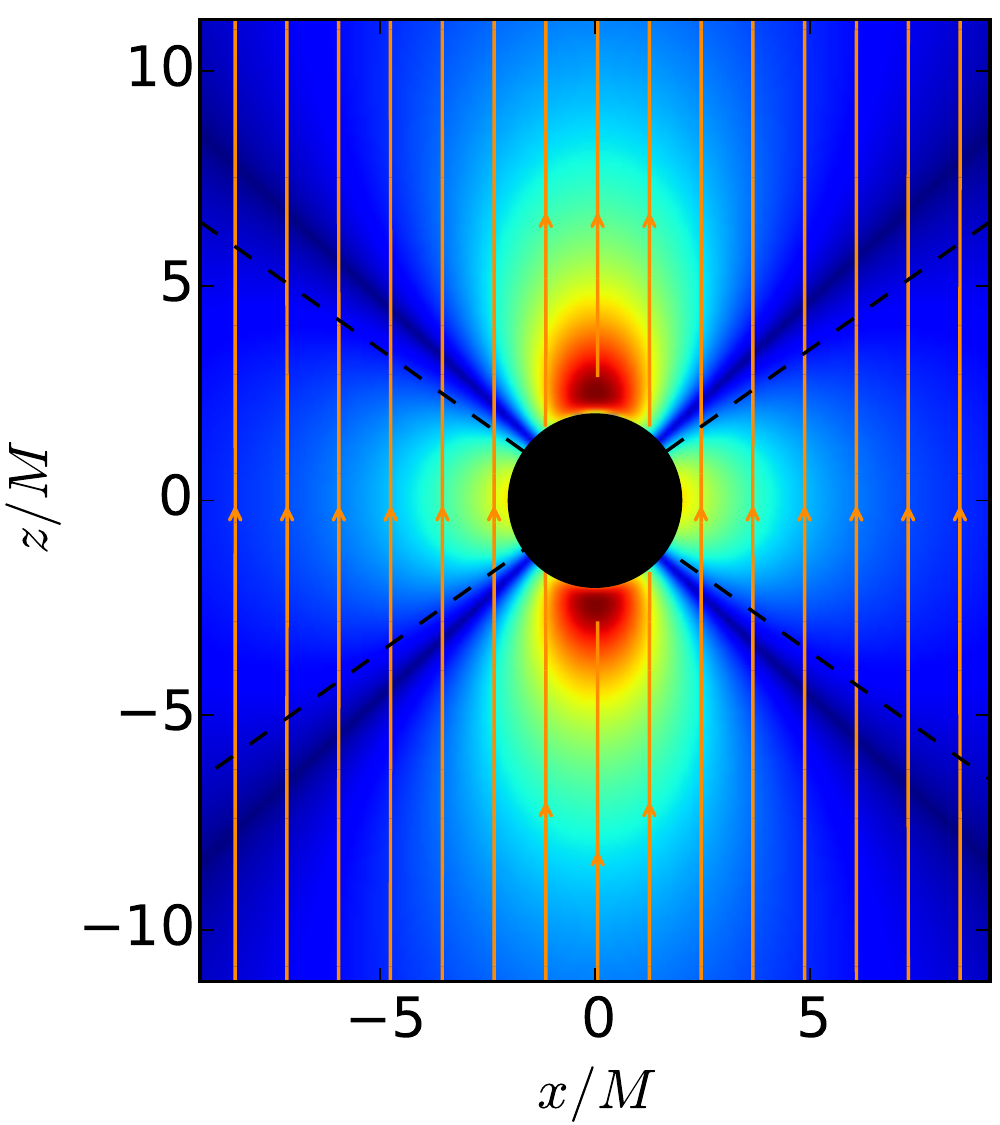}
    \includegraphics[width=0.32\hsize,clip]{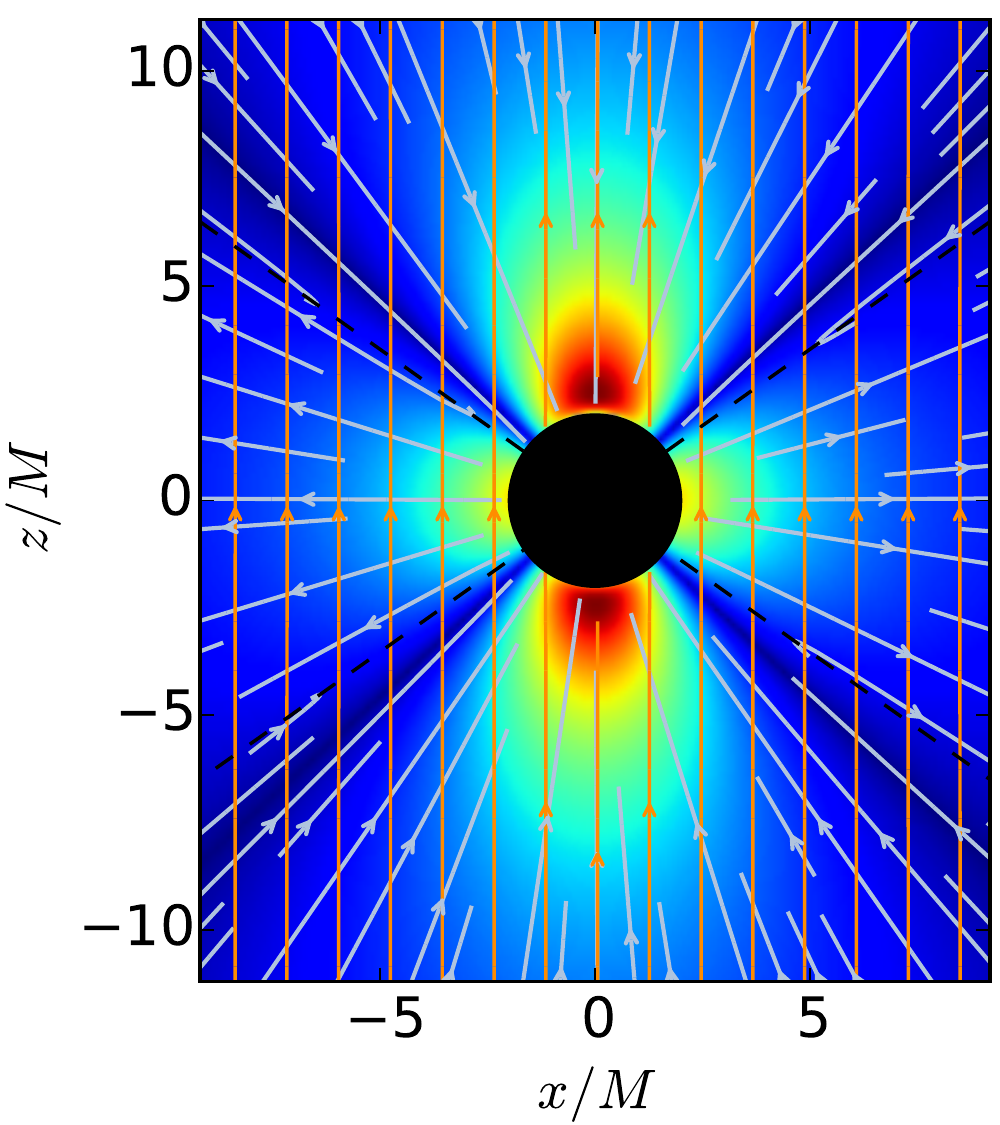}
    \caption{Electromagnetic field configuration for a \textit{polar electronic jet}. Left: Electric field lines (blue-colored lines) of the Papapetrou-Wald solution for a BH spin parameter $a/M = 0.3$, in the xz plane in Kerr-Schild  coordinates. Center: magnetic field lines (gold-colored lines). Right: Electric and magnetic field lines together. It can be directly checked from the slow-rotation expressions of the fields, Eqs.~(\ref{eq:EMslow}), that the electric field intensity decreases from the polar axis to the equator along a contour of constant radial distance. The magnetic field is asymptotically (at infinity) parallel to the BH spin. The colored background is a density plot of the electric field energy density which decreases from red to blue. The BH horizon is represented by the black filled disk. Distances are in units of $M$ and the fields in units of $B_0$. Electron acceleration (outward photon emission) occurs in the region where the electric field is inwardly directed. This region, limited by the dashed-black lines, is defined in the northern hemisphere by the spherical polar angles (measured clockwise from the rotation axis) $-\theta_\pm<\theta<\theta_\pm$, and by equatorial symmetry, in the southern hemisphere by $\pi-\theta_\pm<\theta<\pi+\theta_\pm$. At the angle $\theta_\pm$, the electric field reverses direction, i.e. $\vec{E}\cdot \vec{B}=0$. For the above parameters, $\theta_{\pm} \approx 55^\circ$. We have used the general transformation from Boyer-Lindquist to Kerr-Schild  coordinates (\ref{eq:xyz}); see Appendix~\ref{app:A} for details.}
    \label{fig:fieldlines}
\end{figure*}

\begin{figure*}
    \centering
    \includegraphics[width=0.32\hsize,clip]{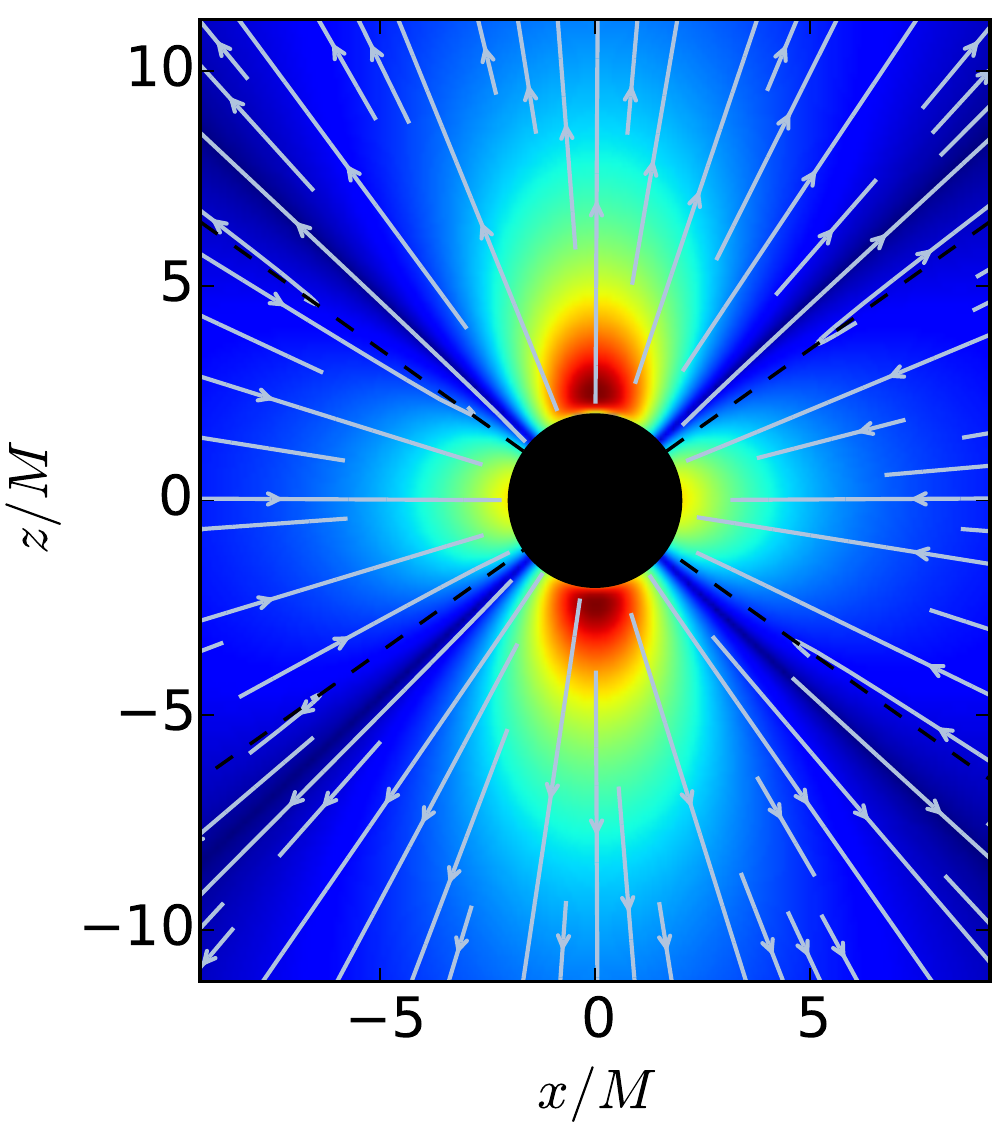}
    \includegraphics[width=0.32\hsize,clip]{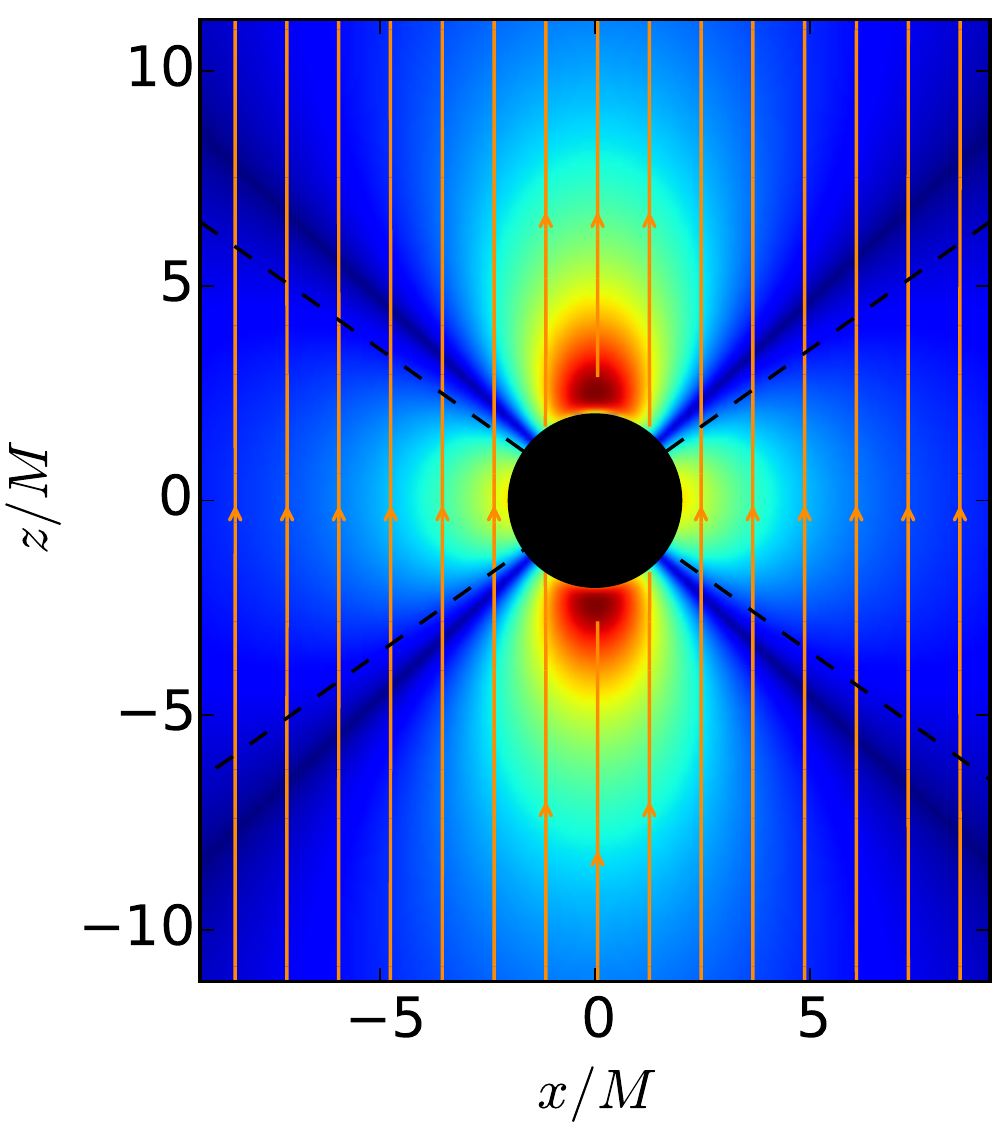}
    \includegraphics[width=0.32\hsize,clip]{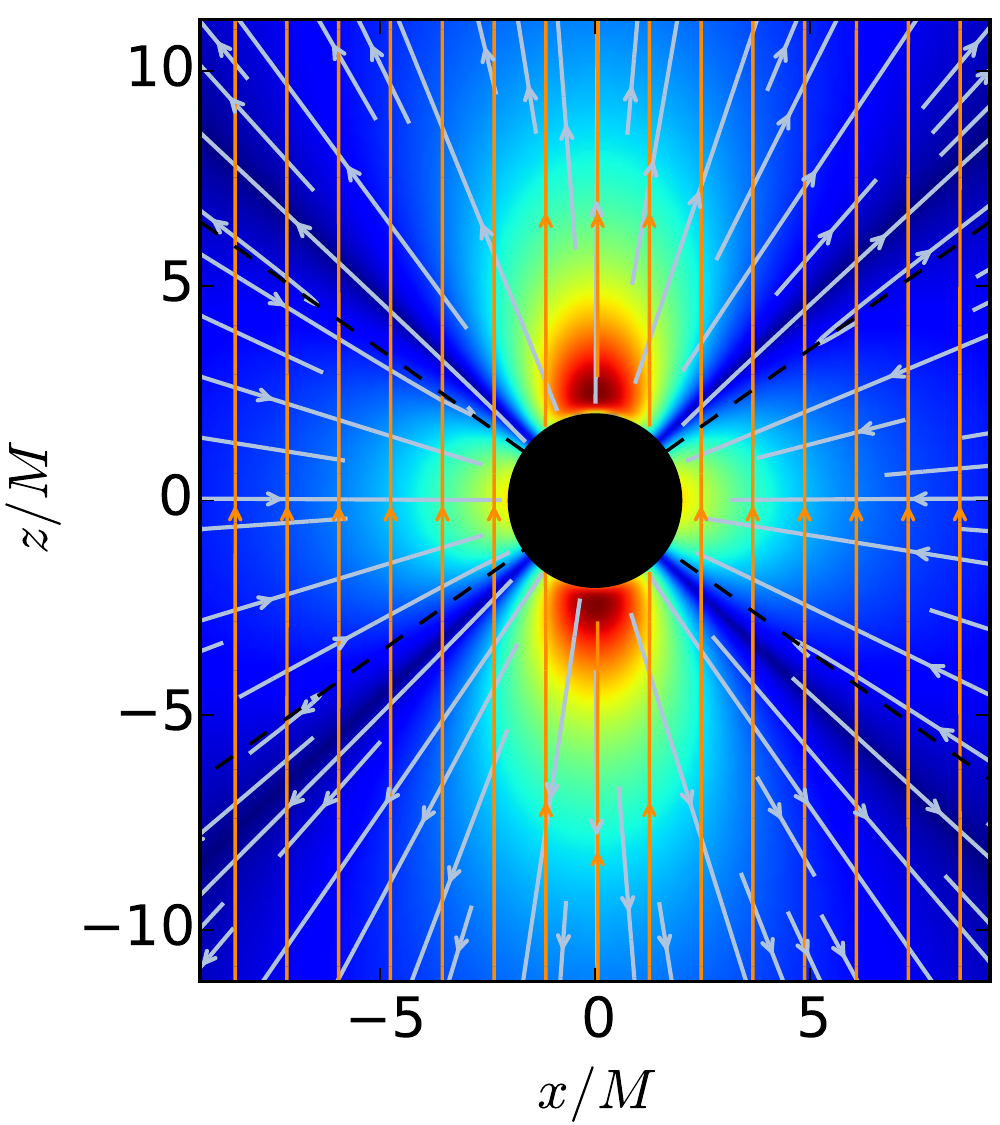}
    \caption{Electromagnetic field configuration for a \textit{polar protonic jet}. Similar to Fig. \ref{fig:fieldlines} but for a magnetic field asymptotically (at infinity) antiparallel to the BH spin. In this case, protons move outward where electrons do in Fig. \ref{fig:fieldlines}, namely, in the double-cone of angle $\theta_\pm \approx 55^\circ$ from the polar axis.}
    \label{fig:fieldlines2}
\end{figure*}

Figures~\ref{fig:fieldlines} and \ref{fig:fieldlines2} show the electric and magnetic field lines as seen by the LNR observer, in the $x$-$z$ plane in Kerr-Schild coordinates (see Appendix~\ref{app:A}). 
We describe now the physical situation in the northern hemisphere since because of the equatorial symmetry, the situation in southern hemisphere will be exactly the same. Electrons move outwardly where the electric field is inwardly directed. Clearly, the opposite happens for protons, they move outwardly where the electric field is outwardly directed. In Fig.~\ref{fig:fieldlines}, we show the case of an asymptotically parallel magnetic field to the BH spin. The electric field is inwardly directed in the cone of semi-aperture angle $\theta_{\pm} \approx 55^\circ$ from the polar axis. Therefore, in this situation we have a \textit{polar electronic jet}. At $\theta = \theta_{\pm}$, the electric field vanishes and reverse direction, i.e. the electric field becomes outwardly directed. In Fig.~\ref{fig:fieldlines2}, we show the case of an asymptotically antiparallel magnetic field to the BH spin. In this case, the electric field is outwardly directed in the cone of semi-aperture angle $\theta_{\pm} \approx 55^\circ$ from the polar axis, creating a \textit{polar electronic jet}.

For moderate spin values (i.e. $a/M\lesssim 0.7$), the electric and magnetic fields are accurately represented by the lowest order of their expansions in the dimensionless spin parameter $a/M$, i.e. up to first order we have
\begin{subequations}\label{eq:EMslow}
\begin{align}
    E_{\hat{r}} &\approx -\frac{B_0 a M}{r^2}(3 \cos^2\theta-1),\\
    E_{\hat{\theta}} &\approx 0,\\
    B_{\hat{r}} &\approx B_0 \cos\theta,\\
    B_{\hat{\theta}} &\approx -B_0 \sqrt{1-\frac{2 M}{r}} \sin\theta.
\end{align}
\end{subequations}
We can see that within this approximation, the electric field reverses sign at $\cos\theta_\pm = \sqrt{3}/3$, i.e. $\theta_\pm \approx 54.74^\circ$, in agreement with the full numerical result. 
It can be checked that the magnetic field is asymptotically directed along the $z$-direction. Evaluating the Kerr-Schild  components (see Appendix~\ref{app:B}), e.g. on the plane $\phi=0$, we obtain
\begin{subequations}\label{eq:EMslowKS}
\begin{align}
    B_{\hat{x}} &= B_{\hat{r}}\sin\theta + B_{\hat{\theta}}\cos\theta \nonumber \\
    &= B_0 \sin\theta\cos\theta \left(1-\sqrt{1-\frac{2 M}{r}}\right),\\
    B_{\hat{y}} &= B_{\hat{\theta}}\frac{a}{r}\cos\theta = -B_0 \sin\theta\cos\theta\, \frac{a}{r}\sqrt{1-\frac{2 M}{r}},\\
    B_{\hat{z}} &= B_{\hat{r}}\cos\theta - B_{\hat{\theta}}\sin\theta \nonumber \\
    &= B_0\left(\cos^2\theta + \sin^2\theta\,\sqrt{1-\frac{2 M}{r}} \right).
\end{align}
\end{subequations}

It is then clear that for radial distances $r\gg 2M$, we have $B_{\hat{x}} \to 0$, $B_{\hat{y}}\to 0$, and $B_{\hat{z}}\to B_0$. The dominance of the z-component actually occurs everywhere and for any value of the BH spin parameter $a$, as it can be seen from Fig.~\ref{fig:fieldlines} which shows the electric and magnetic field lines without any approximation.

The electric field (\ref{eq:Efieldzamo}) is induced by interaction of the magnetic field and the BH gravitomagnetic field. In fact, it is easy to check that it vanishes for $a = 0$. As we have seen in Figs. \ref{fig:fieldlines} and \ref{fig:fieldlines2}, the electric field has a quadrupolar nature. Thus, although the BH has a zero net charge, we can think of this field as produced by a quadrupolar distribution of charges of surface density \citep{1986bhmp.book.....T}
\begin{equation}\label{eq:sigma}
    \sigma = \frac{1}{4\pi} B_0 a r_+ (r_+ - M)\frac{r_+\sin^4\theta - M \cos^2\theta (1+\cos^2\theta)}{(r_+^2 + a^2 \cos^2\theta)},
\end{equation}
that in the slow rotation approximation becomes
\begin{equation}\label{eq:sigmaapp}
    \sigma \approx \frac{B_0}{16\pi}\frac{a}{M}(1-3 \cos^2\theta).
\end{equation}
From this expression it is clear that it equals $E_{\hat{r}}/(4\pi)$, where the electric field is given by Eq.~(\ref{eq:EMslow}), as expected. From this expression, we have that the surface charge vanishes at $\cos\theta_\pm = \sqrt{3}/3$, which leads to $\theta_\pm = 54.74^\circ$, consistent with the value already obtained directly from the electric field expression and shown in Figs. \ref{fig:fieldlines} and \ref{fig:fieldlines2}.

\begin{figure}
    \centering
    \includegraphics[width=\hsize,clip]{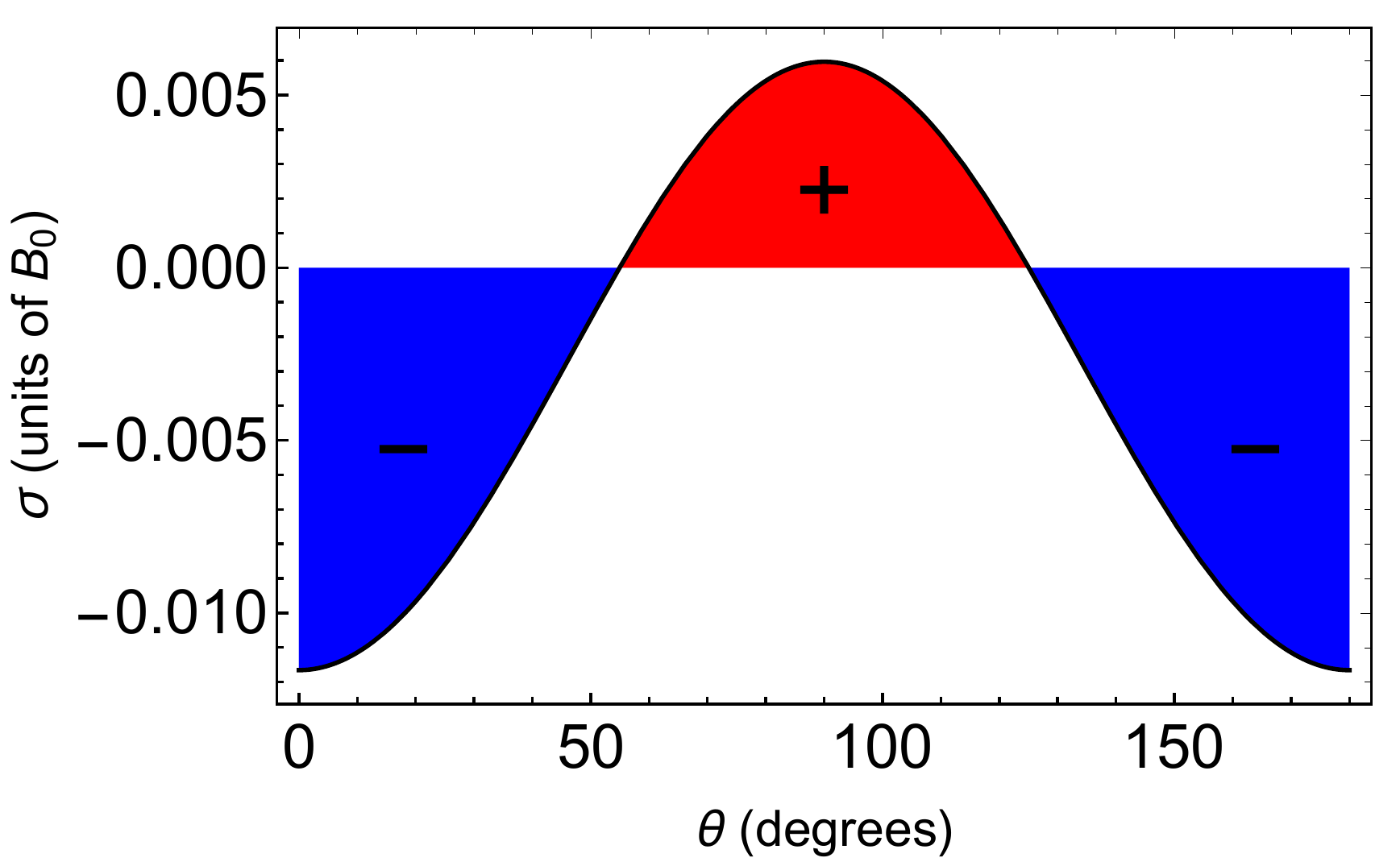}
    \caption{Surface charge (\ref{eq:sigma}) in units of $B_0$ as a function of the spherical polar angle $\theta$, and for a BH spin parameter $a/M = 0.3$. The blue and red colors indicate, respectively, the regions with negative and positive surface charge density.}
    \label{fig:surfacecharge}
\end{figure}

By introducing the dimensionless radius and the spin parameter in units of $M$, it can be seen that the surface charge $\sigma$ does not depend explicitly on $M$, but only on $B_0$ and $a/M$. Figure~\ref{fig:surfacecharge} shows the surface charge (\ref{eq:sigma}) as a function of $\theta$, for a BH spin parameter $a/M= 0.3$, the same used in Fig.~\ref{fig:fieldlines}. This surface charge $\sigma$ in fact agrees with the electric field lines in the upper row plots of Fig.~\ref{fig:fieldlines}. In the first patch, $\theta = [0,\theta_\pm]$ (the angle $\theta$ is measured clockwise from the polar axis), the electric field lines are inwardly directed as expected from $\sigma<0$; the electric field lines in the region $\theta = [\theta_\pm,\pi - \theta_\pm]$ point outward in agreement with $\sigma >0$. A similar analysis for the other patches confirm the self-consistency of the scenario. The angle $\theta_\pm$ is the one for which $\sigma$ vanishes. For this numerical example, $\theta_\pm  \approx 54.95^\circ$.

We can compute the charge induced on a surface patch of the horizon as
\begin{equation}\label{eq:Qtheta}
    Q_{\rm patch} = \iint \sigma \sqrt{|h_{i j}|}\,dx^i dx^j = \iint \sigma \sqrt{|h_{\theta \theta}h_{\phi \phi}|}\,d\theta \,d\phi,
\end{equation}
where $h_{i j}$ is the induced metric on the horizon, $ds^2 = h_{i j} dx^i dx^j = h_{\theta \theta} d\theta^2 + h_{\phi \phi} d\phi^2$, being $h_{\theta \theta} = \Sigma_+ = r_+^2 + a^2 \cos^2\theta$, and $h_{\phi \phi} = (2 M r_+)^2 \sin^2\theta/\Sigma_+$, obtained by taking constant slices $t$ and $r = r_+$ in the Kerr metric (\ref{eq:Kmetric}). Therefore, Eq.~(\ref{eq:Qtheta}) becomes
\begin{equation}\label{eq:Qtheta2}
    Q_{\rm patch} = 4\pi M r_+ \int_{\Delta\theta} \sigma(\theta) \sin\theta d\theta,
\end{equation}
where we have already performed the integral over $\phi$ taking advantage of the axial symmetry and $\Delta\theta$ denotes the angular region subtended by two spherical polar angles. The integral in Eq.~(\ref{eq:Qtheta2}) from $0$ to $\pi$ vanishes, so the net charge of the BH is zero. In the example of Fig.~\ref{fig:surfacecharge}, the charge of the negatively charged patches is obtained by summing the integrals in the regions $\theta= [0,\theta_\pm]$ and $\theta = [\pi - \theta_\pm,\pi]$, while the positively charged one is obtained from the integral in the region $\theta = [\theta_\pm, \pi - \theta_\pm]$. For the case of $a/M = 0.3$, we obtain
\begin{equation}\label{eq:Qexample}
    Q_\pm = \pm 0.1119 \times M^2 B_0,
\end{equation}
and therefore, $Q_{-} + Q_{+} = 0$, as expected.

We can obtain an analytic expression of the patch charge in the \textit{slow-rotation} regime. By replacing $\sigma$ given by Eq.~(\ref{eq:sigmaapp}) into Eq.~(\ref{eq:Qtheta2}), we obtain the charge on any patch
\begin{equation}\label{eq:Qapp}
    Q_{\rm patch} \approx \frac{1}{2} M^2 B_0 \frac{a}{M} \cos\theta (\cos^2\theta-1)\Big|_{\Delta\theta}.
\end{equation}
We can now evaluate the total negative and positive charge as we proceeded before, which leads us to
\begin{equation}\label{eq:Qpm}
    Q_\pm = \pm\frac{2\sqrt{3}}{9} M^2 B_0 \frac{a}{M} = \pm\frac{2\sqrt{3}}{9} B_0 J,
\end{equation}
where in the last equality we have used the BH angular momentum via the relation $a = J/M$. Turning to the previous numerical example, i.e. $a/M = 0.3$, Eq.~(\ref{eq:Qpm}) gives $Q_\pm = \pm 0.1155 M^2 B_0$, which is pretty close to the value given by Eq.~(\ref{eq:Qexample}), validating once more the accuracy of the \textit{slow-rotation} approximation.

\section{Equations of motion}\label{sec:4}

In the LNR frame, the particle equations of motion can be written as 
\begin{equation}\label{eq:EOM}
    \frac{D u^{\hat{a}}}{d\tau} = \frac{d u^{\hat{a}}}{d \tau} + \omega_{\hat{c}\,\,\,\hat{b}}^{\hphantom{\mu}{\hat{a}}}u^{\hat{b}}u^{\hat{c}} = \frac{q}{m} F^{\hat{a}}_{\hphantom{\hat{a}}{\hat{b}}}u^{\hat{b}} - {\cal F}^{\hat{a}},
\end{equation}
where $u^{\hat{a}}$ and $F^{\hat{a}}_{\hphantom{\hat{a}}{\hat{b}}}$ and ${\cal F}^{\hat{a}}$ are, respectively, the components of the particle's four-velocity and the electromagnetic field tensor, projected onto the observer's tetrad, see Eqs.~(\ref{eq:uontotetrad}) and (\ref{eq:projections}), $\tau$ is the particle's proper time along its worldline, and $\omega_{\hat{c}\,\,\,\hat{b}}^{\hphantom{\mu}{\hat{a}}}$ are the \emph{spin coefficients} given by Eqs.~(\ref{eq:spincoefficients}) The last term on the right-hand side are the components of the radiation-reaction ``force'' \citep[see, e.g.,][]{1975ctf..book.....L} per unit mass, projected onto the LNR frame, i.e.
\begin{align}\label{eq:Frad}
    &{\cal F}^{\hat{a}} = \frac{2}{3}\left(\frac{q}{m}\right)^2 \frac{q^2}{m}\left( F_{\hat{c}\hat{d}}F^{\hat{d}}_{\hphantom{\hat{d}}{\hat{e}}}u^{\hat{c}}u^{\hat{e}}\right) u^{\hat{a}} \nonumber \\
    &+ \frac{2}{3}\left(\frac{q}{m}\right)^2\frac{q^2}{m} F^{\hat{a}}_{\hphantom{\hat{a}}{\hat{b}}}F^{\hat{b}}_{\hphantom{\hat{b}}{\hat{c}}}u^{\hat{c}} + \frac{2}{3}\left(\frac{q}{m}\right)^2 q\frac{D F^{\hat{a}}_{\hphantom{\hat{a}}{\hat{b}}}}{d x^{\hat{c}}}u^{\hat{b}}u^{\hat{c}},
\end{align}
where in the last term the covariant derivative of the electromagnetic tensor in the LNR frame can be computed via  Eq.~(\ref{eq:covariantD}).

The radiation reaction force (\ref{eq:Frad}) is largely dominated by the first term, hence
\begin{equation}
    {\cal F}^{\hat{a}}\approx \frac{2}{3}\left(\frac{q}{m}\right)^2 \frac{q^2}{m}\left( F_{\hat{c}\hat{d}}F^{\hat{d}}_{\hphantom{\hat{d}}{\hat{e}}}u^{\hat{c}}u^{\hat{e}}\right) u^{\hat{a}} = -\frac{{\cal P}}{m} v^{\hat{a}},
\end{equation}
where we have defined the radiated off power
\begin{equation}\label{eq:power}
     {\cal P} \equiv \frac{2}{3}\left(\frac{q}{m}\right)^2 q^2 \hat{\gamma}^3 \left[ (\vec{E} + \vec{v}\times \vec{B})^2 - (\vec{v}\cdot \vec{E})^2 \right].
\end{equation}

We can work in the slow-rotation regime which provides sufficient accuracy for our purpose. Within this approximation, Eqs.~(\ref{eq:EOM}) become
\begin{align}\label{eq:EOMslowa}
    \frac{d\hat{\gamma}}{d\tau} &= -\frac{e}{m}E^{\hat{r}}v^{\hat{r}} \hat{\gamma} - \frac{{\cal P}}{m}\nonumber \\
    &+\left[\frac{M}{r^2\sqrt{1-2 M/r}} -\frac{6 M a \sin\theta}{r^3} v^{\hat{\theta}}\right]\hat{\gamma} v^{\hat{r}},\\
\end{align}
and 
\begin{align}\label{eq:EOMslowb}
    &\frac{d{v}^{\hat{i}}}{d\tau} = -\frac{e}{m} \left[(E^{\hat{r}}-v^{\hat{\phi}}B^{\hat{\theta}})\delta^{\hat{i}}_{\hphantom{\hat{i}}{\hat{r}}} 
    + v^{\hat{\phi}}B^{\hat{r}}\delta^{\hat{i}}_{\hphantom{\hat{i}}{\hat{\theta}}} + (v^{\hat{r}}B^{\hat{\theta}}-v^{\hat{\theta}}B^{\hat{r}})\delta^{\hat{i}}_{\hphantom{\hat{i}}{\hat{\phi}}} \right. \nonumber \\
    & \left.- E^{\hat{r}} v^{\hat{r}} v^{\hat{i}} \right] - \frac{{\cal P}v^{\hat{i}}}{m \hat{\gamma}}- \left(\frac{M\,r}{\sqrt{1-2 M/r}} - 6 M a \sin\theta v^{\hat{\theta}}
    \right)\frac{\hat{\gamma} v^{\hat{r}}v^{\hat{i}}}{r^3}\nonumber \\
    &- \left[ 6 M a \sin\theta v^{\hat{\phi}}
    + \frac{M\,r\,(v^{\hat{\theta}})^2}{\left(1-\frac{2 M}{r}\right)^{5/2}} +  r^2\sqrt{1-\frac{2 M}{r}} (v^{\hat{\phi}})^2 
    \right]\frac{\hat{\gamma}\,\delta^{\hat{i}}_{\hphantom{\hat{i}}{\hat{r}}}}{r^3}
    \nonumber \\ 
    &+ \left[
    \frac{M}{(1-2 M/r)^{5/2}}v^{\hat{r}} v^{\hat{\theta}}
    - \frac{r \cos\theta}{\sin\theta} (v^{\hat{\phi}})^2 
    \right]\frac{\hat{\gamma}\,\delta^{\hat{i}}_{\hphantom{\hat{i}}{\hat{\theta}}}}{r^2}\nonumber \\
    &- \left[
    6 M a \sin\theta v^{\hat{r}}
    - r^2\sqrt{1-\frac{2 M}{r}} v^{\hat{r}} v^{\hat{\phi}} 
    - \frac{r^2 \cos\theta}{\sin\theta} v^{\hat{\theta}} v^{\hat{\phi}}
    \right]\frac{\hat{\gamma}\,\delta^{\hat{i}}_{\hphantom{\hat{i}}{\hat{\phi}}}}{r^3},
\end{align}
where $E^{\hat{r}}$, $B^{\hat{r}}$ and $B^{\hat{\theta}}$ are given by their corresponding slow-regime expressions given by Eqs.~(\ref{eq:EMslow}).

It is worth to assess the relative importance of the purely gravitational terms in the dynamics of the particle. It is clear than the contribution of the square brackets on the right-hand side of Eq.~(\ref{eq:EOMslowa}) is dominated by the first term inside them. Then, because of $E^{\hat{r}}\propto 1/r^2$, the order of magnitude of the ratio between the electric field term and the gravitational one is given by $\omega_B a/M$, where $ \omega_B \equiv e B_0/m$ is the so-called gyration angular frequency. Therefore, since the largest value of the spin parameter is $a = M$, the largest value that the electric to gravitational contributions ratio can attain is $\omega_B$, that becomes of the order of unity only for an extremely weak magnetic field of the order of $\mu$G. This implies that, for a magnetic field $10^{11}$~G, the electric field term is $10^{17}$ bigger than the gravitational one. Only in the limit $r\to r_+$, the gravitational attraction can play some role against such a huge electric force. Therefore, we expect the evolution of $\gamma$ to be dictated only by the first term on the right-hand side of Eq.~(\ref{eq:EOMslowa}). A similar analysis can be done for the evolution of the spatial velocity, which we therefore expect to be fully dominated by the terms in the first row of Eq.~(\ref{eq:EOMslowb}).

\subsection{A specific numerical solution}

We turn now to show specific examples of numerical solutions to the equations of motion (\ref{eq:EOMslowa}) and (\ref{eq:EOMslowb}), for \emph{inner engine} parameters: $M=4.4 M_\odot$, $a/M = 0.3$, $B_0 = 10^{11}$~G (so $\omega_B\approx 1.8\times 10^{18}$~rad~s$^{-1}$).

\begin{figure}
    \centering
    \includegraphics[width=\hsize,clip]{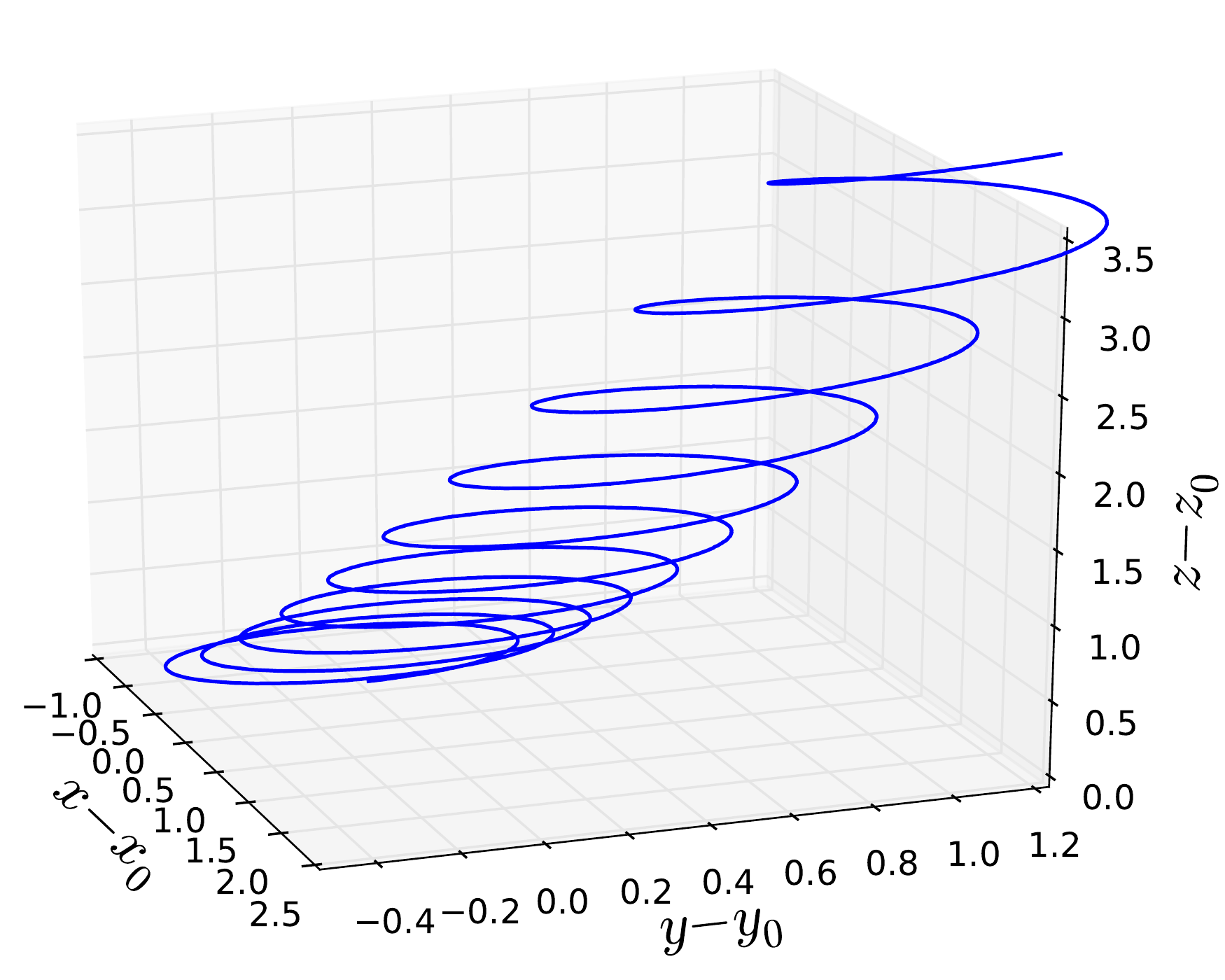}
    \caption{Example of numerical solution of the equations of motion in the slow-rotation regime, given by Eqs.~(\ref{eq:EOMslowa}) and (\ref{eq:EOMslowb}). The figure shows the electron's trajectory in Kerr-Schild Cartesian coordinates, for which we have used the transformation from Boyer-Lindquist coordinates given by Eq.~(\ref{eq:BLto}). The \emph{inner engine} parameters in this case are: $M=4.4 M_\odot$, $a/M = 0.3$, $B_0 = 10^{11}$~G (so $\omega_B\approx 1.8\times 10^{18}$~rad~s$^{-1}$). We have set as initial conditions: the location $r_0 = 2 r_+ = 4 M$, $\theta(0) = 20^\circ$, $\phi(0) = 0$, and velocity components $v^{\hat{r}}(0) = v^{\hat{\theta}}(0) = 0$, and $v^{\hat{\phi}}(0) = \sqrt{\hat{\gamma}_0^2-1}/\hat{\gamma}_0$, with $\hat{\gamma}_0 = 2$. For the sake of illustration, the trajectory is here shown at times $0\leq \omega_B \tau \leq 65$, in which the spiraling behavior is better appreciated. The displacements in the x and y directions are in units of $10^{-13} M$ and, the one in the z-direction, in units of $10^{-12} M$.
    }
    \label{fig:trajectory}
\end{figure}

\begin{figure*}
    \centering
    \includegraphics[width=0.49\hsize,clip]{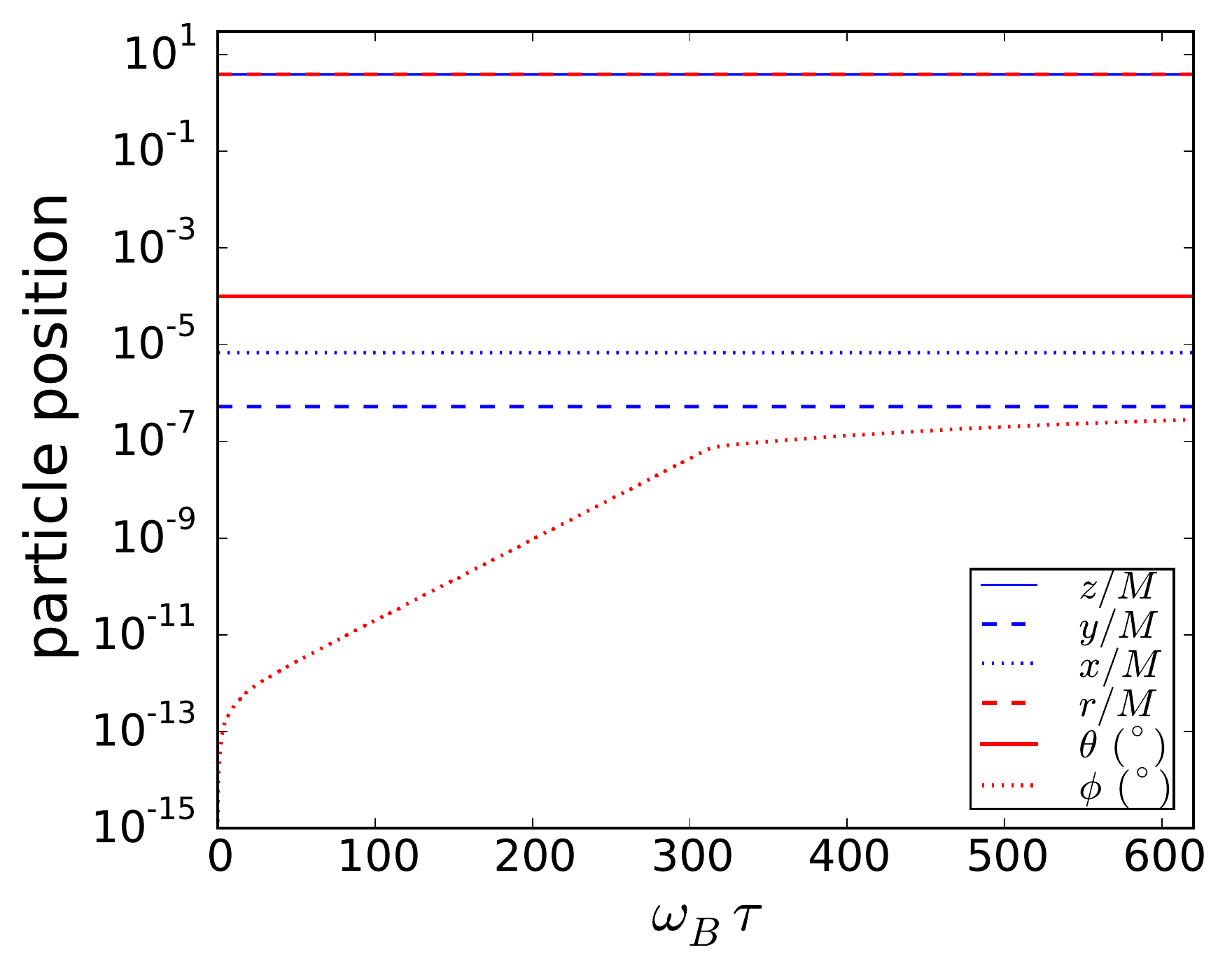}
    \includegraphics[width=0.49\hsize,clip]{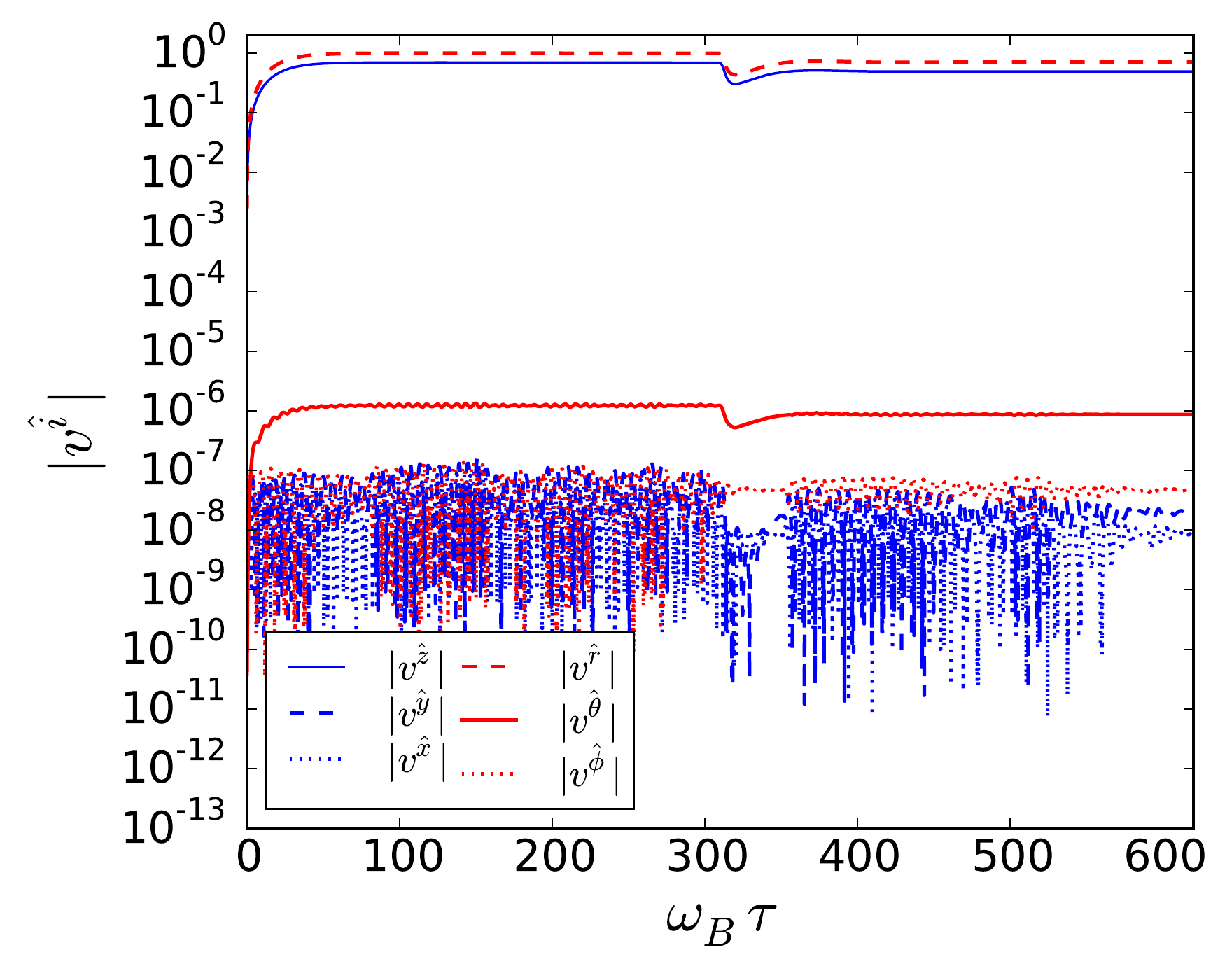}\\
    \includegraphics[width=0.49\hsize,clip]{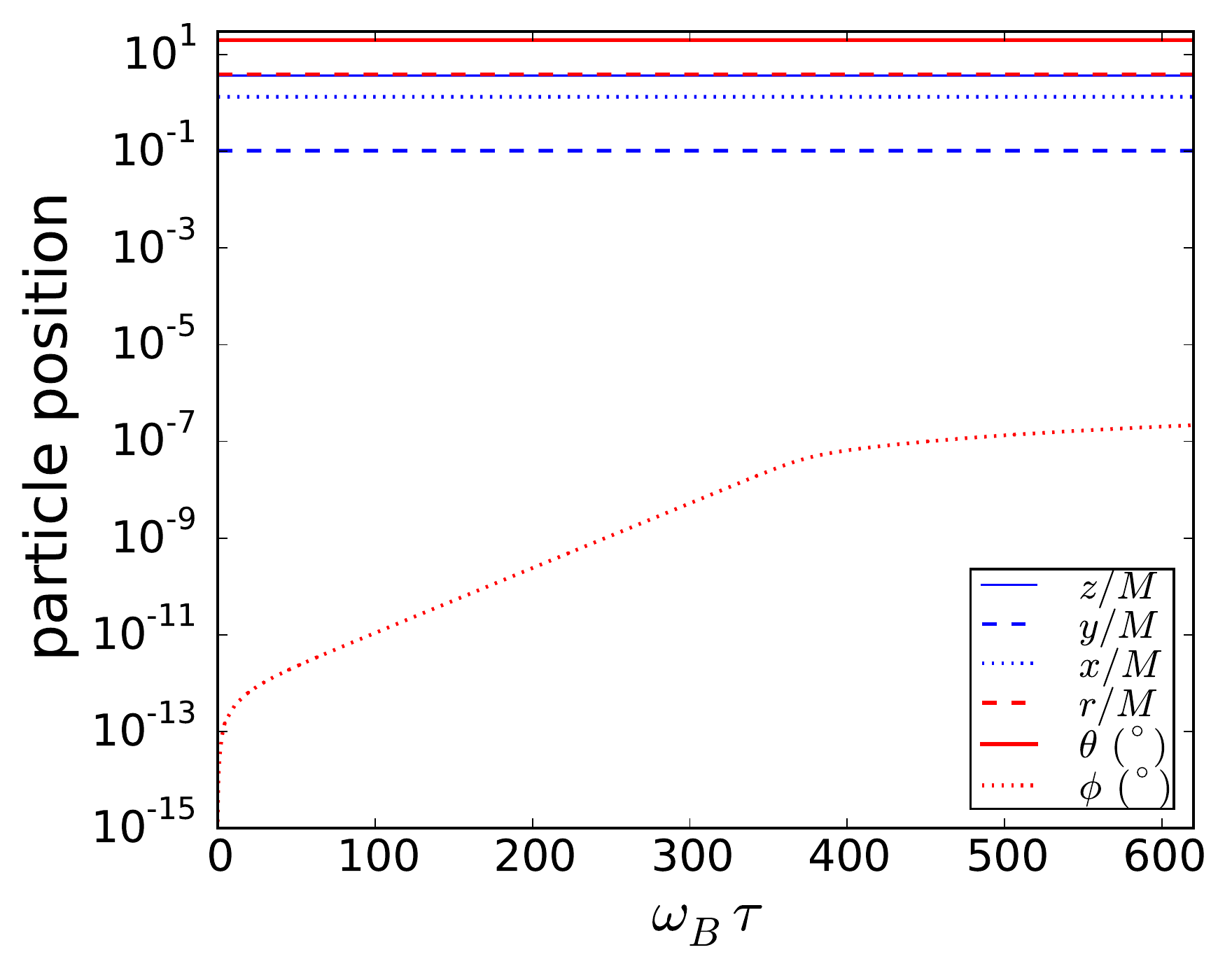}
    \includegraphics[width=0.49\hsize,clip]{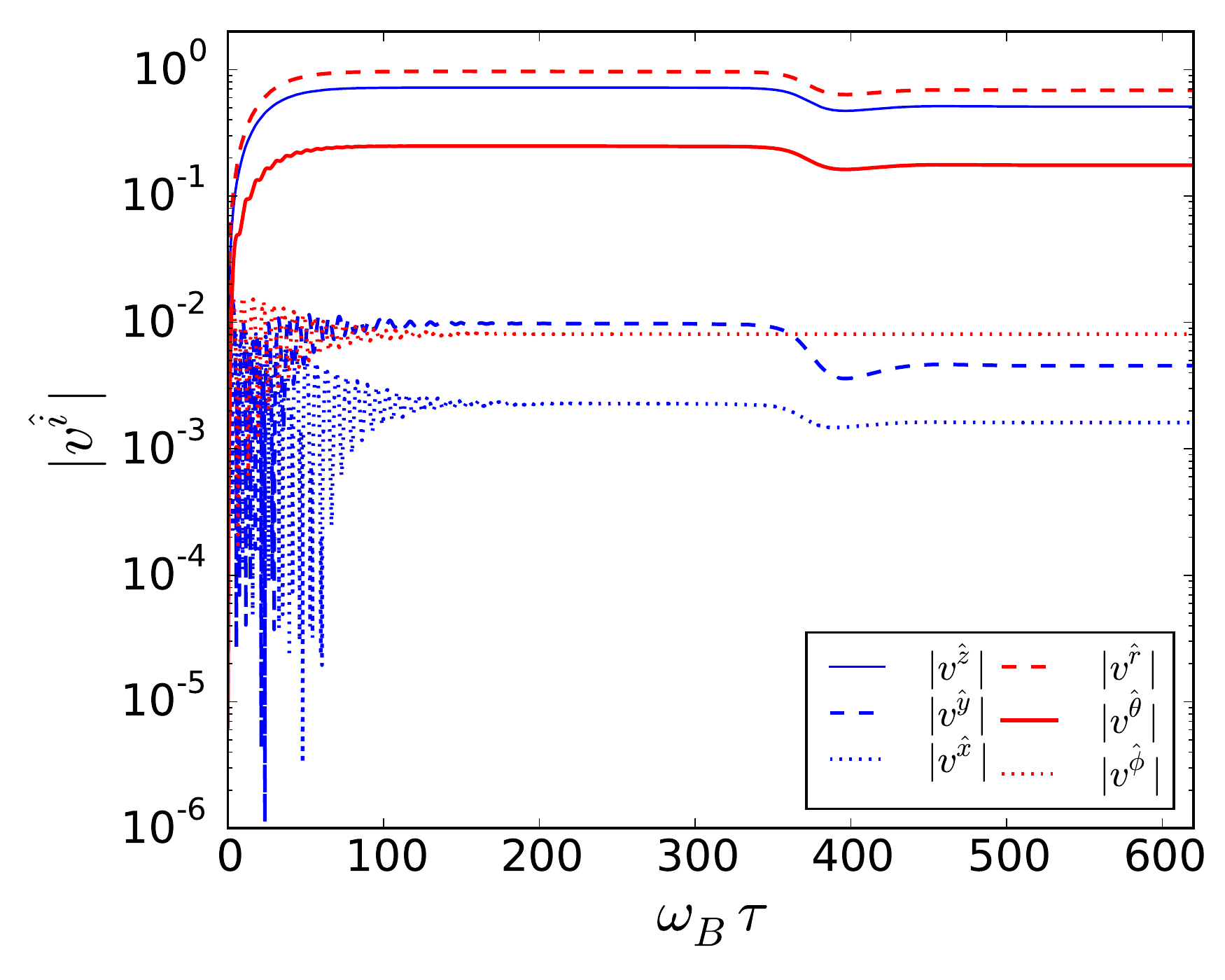}\\
    \caption{Electron's position (coordinate frame) and velocity (LNR frame), as a function of dimensionless proper time, $\omega_B \tau$. The electron is initially located at Boyer-Lindquist coordinates $r(0) = 2 r_+$, $\theta(0) = 0.0001^\circ$ (upper row), $\theta(0) = 20^\circ$ (lower row) and $\phi(0) = 0$. The electron is initially at rest, i.e. $v^{\hat{r}}(0) = v^{\hat{\theta}}(0) = v^{\hat{\phi}}(0) = 0$, i.e. $\hat{\gamma}(0) = 1$. The particle position and the velocity components are shown in both Boyer-Lindquist and (Kerr-Schild) coordinates. The \textit{inner engine} parameters are the same of Fig.~\ref{fig:trajectory}. Notice that in Fig.~\ref{fig:trajectory} we represent displacements, namely the relative particle position with respect to the initial position, while here we represent the actual position of the particle, namely the position with respect to the coordinates origin. In addition, in Fig.~\ref{fig:trajectory} the displacement in the $x$ and $y$ directions is in units of $10^{-13}M$ and the one in the $z$ direction is in units $10^{-12}M$, while in this figure the $x$, $y$, and $z$ position are in units of $M$.}
    \label{fig:xyz}
\end{figure*}

Figure~\ref{fig:trajectory} shows an example of the trajectory of an electron initially located at $r_0 = 2 r_+ \approx 4 M$, $\theta(0) = 20^\circ$, $\phi(0) = 0$. The initial values of the electron's velocity components were, in this specific example, were set to have a spiraling electron motion, that is setting up an initially dominating azimuthal velocity, in particular we have chosen $v^{\hat{r}}(0) = v^{\hat{\theta}}(0) = 0$, and $v^{\hat{\phi}}(0) = \sqrt{\hat{\gamma}_0^2-1}/\hat{\gamma}_0$, with $\hat{\gamma}_0 = 2$. The trajectory is shown at times $\omega_B\,\tau \leq 65$, when the spiraling behavior is better appreciated. At longer times, the electron follows closely the magnetic-field lines which are nearly $z$-directed (see Fig.~\ref{fig:fieldlines}), so the z-component of the velocity becomes dominant (see also Fig.~\ref{fig:xyz}).

In Fig.~\ref{fig:trajectory}, the position displacements in the $x$ and $y$ direction are in units of $10^{-13} M$ and the one in the $z$ direction in units of $10^{-12} M$. Indeed, a dimensionless proper time $\omega_B\,\tau = 65$ corresponds to $\tau \approx 3.7\times 10^{-17}$~s, so the observer measures a displacement $\Delta z~\sim \hat{\gamma} c \tau \approx 2.2 \times 10^{-6} $~cm$\approx 4\times 10^{-12} M$, for the chosen BH mass. The displacements in the $x$-$y$ plane are $\sim 1/\hat{\gamma}$ smaller since the particle's velocity is dominated by the $z$-component.

Figure~\ref{fig:xyz} shows the evolution of the position and velocity components $v^{\hat{i}}$ of an electron located at $r(0) = 2 r_+$, $\theta(0) = 0.0001^\circ$ (upper row), $\theta(0) = 20^\circ$ (lower row) and $\phi(0) = 0$. The electron has been set initially at rest, i.e. $v^{\hat{r}}(0) = v^{\hat{\theta}}(0) = v^{\hat{\phi}}(0) = 0$, so $\hat{\gamma}(0) = 1$. This figure shows that, in general, the motion along the $z$-direction is dominant, and it is even more pronounced for small spherical polar angles (motion nearly attached to the BH rotation axis).

\section{Photon four-momentum measured at infinity }\label{sec:5}

We refer the reader to App.~\ref{app:B} for the general equations between the four-momentum measured by the observer at rest at infinity, the LNR observer and the comoving observer. We can gain some (analytical) insight into the features of the photon emission by assuming $v^{\hat{r}}\approx 1$, $v^{\hat{\theta}}\approx 0$ and $v^{\hat{\phi}}\approx 0$. In this one-dimensional approximation of motion, Eqs.~(\ref{eq:kmuka}) reduce to
\begin{subequations}\label{eq:kmukaapp}
\begin{align}
    k^0 &= k^{(0)}\,\hat{\gamma} e^{-\nu} [1 + v^{\hat{r}} \,n^{(1)}],\quad n^r = e^{\nu-\mu_1} \frac{v^{\hat{r}} + n^{(1)}}{1 + v^{\hat{r}} \,n^{(1)}},\\
    n^\theta &= \frac{e^{\nu-\mu_2} n^{(2)}}{\hat{\gamma} [1 + v^{\hat{r}} \,n^{(1)}]},\quad n^\phi = \omega +  \frac{e^{\nu-\Psi} n^{(3)}}{\hat{\gamma} [1 + v^{\hat{r}} \,n^{(1)}]},
\end{align}
\end{subequations}
which in the slow-rotation regime can be written as
\begin{subequations}\label{eq:kmukaslow}
\begin{align}
    k^0 &= \frac{k^{(0)}}{\sqrt{1-2 M/r}}\,\hat{\gamma}\,(1 + \hat{\beta} \cos\Theta),\\
    n^r &= \left(1-\frac{2 M}{r}\right)\frac{\cos\Theta + \hat{\beta}}{1 + \hat{\beta} \cos\Theta},\\
    r\,n^\theta &= \sqrt{1-\frac{2 M}{r}}\left(\frac{\sin\Theta \cos\Phi}{1 + \hat{\beta} \cos\Theta}\right)\frac{1}{\hat{\gamma}},\\
    r\sin\theta\,n^\phi &= \frac{2 M a \sin\theta}{r^2} + \sqrt{1-\frac{2 M}{r}}\left(\frac{\sin\Theta \sin\Phi}{1 + \hat{\beta} \cos\Theta}\right) \frac{1}{\hat{\gamma}},
\end{align}
\end{subequations}
where we have introduced the notation $\hat{\beta} = v^{\hat{r}}$. We have also parametrized the spatial components of the photon four-momentum in the orthonormal comoving frame as
\begin{equation}
    n^{(1)} = \cos\Theta,\quad  n^{(2)} = \sin\Theta \cos\Phi,\quad n^{(3)} = \sin\Theta \sin\Phi,
\end{equation}
where $\Theta$ and $\Phi$ are spherical polar and azimuth angles measured by this local observer. This choice satisfies $n^{(i)}n_{(i)} = 1$, which derives from $k^{(a)}k_{(a)} = 0$.

\begin{figure*}
    \centering
    \includegraphics[width=0.49\hsize,clip]{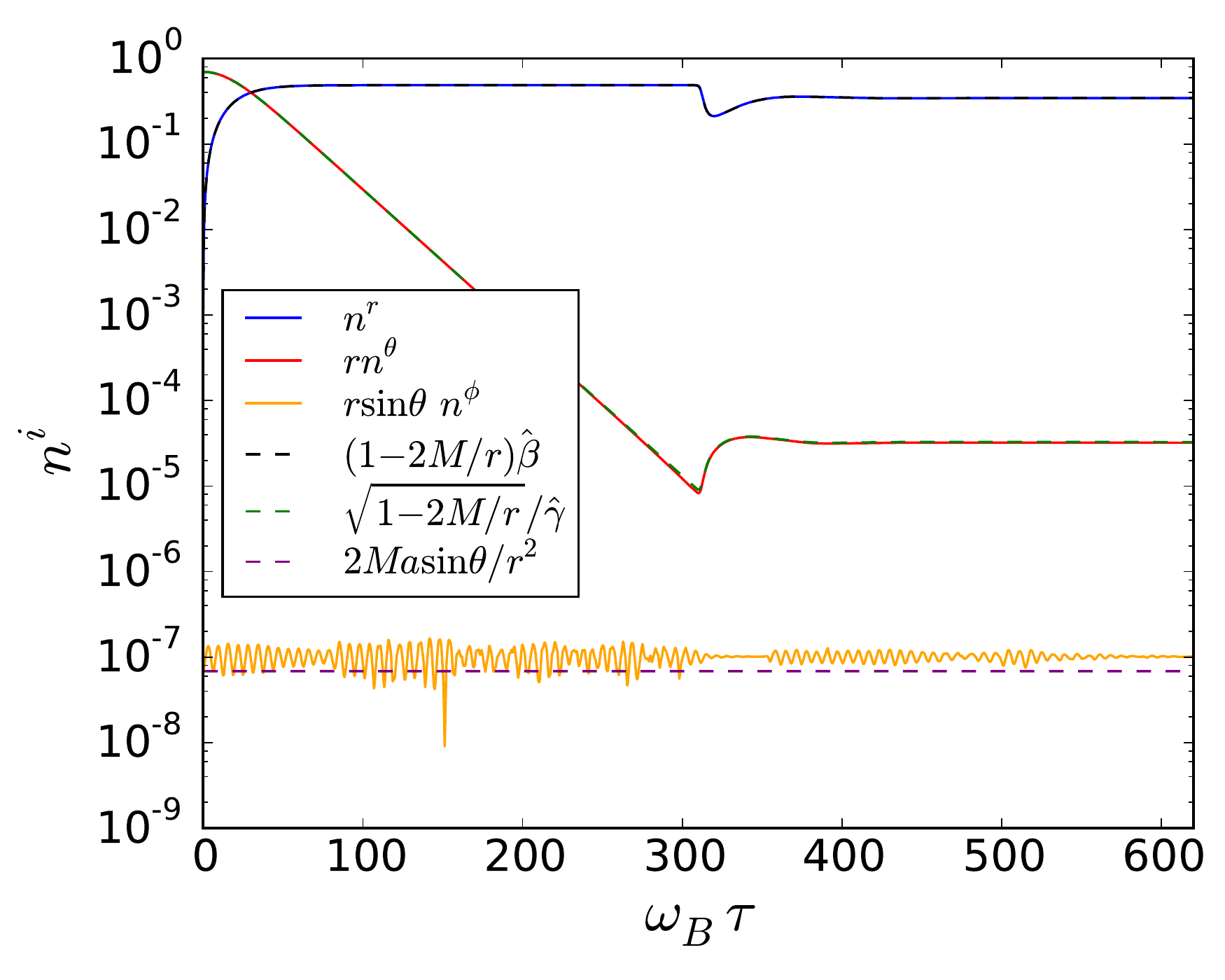}
    \includegraphics[width=0.49\hsize,clip]{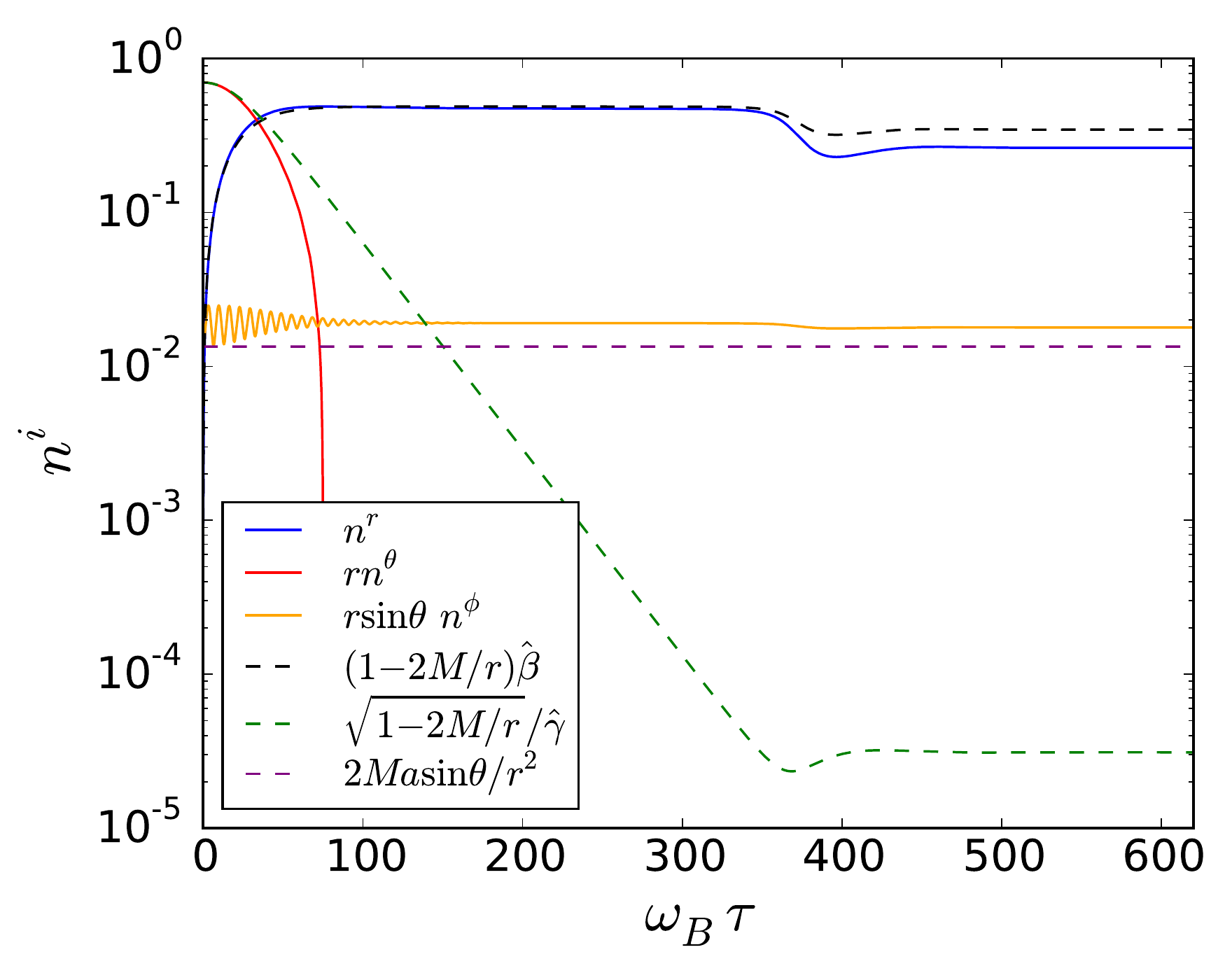}
    \caption{Spatial components of the photon four-momentum measured by an observer at rest at infinity. The emitter (electron) is initially located at Boyer-Lindquist coordinates $r(0) = 2 r_+$, $\theta(0) = 0.0001^\circ$ (left) and $\theta(0) = 20^\circ$ (right), and $\phi(0) = 0$. The comoving frame emission angles have been set to $\Phi = 0$ and $\Theta = \pi/2$. The solid curves are obtained from the full expressions given by Eqs.~    (\ref{eq:kmuka}), while the dashed curves are the corresponding components of the four-momentum given by the approximate expressions given by Eqs.~(\ref{eq:kmukaapp}); see also Table~\ref{tab:ks}. The approximate expressions remain accurate for the entire evolution only for an emitter moving along the BH rotation axis, or for small departures of the spherical polar angle. The azimuth component of the four-momentum remains nearly constant and is a purely general relativistic effect of dragging of inertial frames.}
    \label{fig:pmutheta0}
\end{figure*}

Thus, by assigning values to these angles in their corresponding range, i.e. $\Theta \in [0,\pi]$ and $\Phi \in  [0,2\pi]$, we can compute from Eqs.~(\ref{eq:kmukaslow}) the corresponding normalized components of the photon four-momentum in the coordinate frame. Figure~\ref{fig:pmutheta0} compare and contrast the photon four-momentum components calculated with the general equations~(\ref{eq:kmuka}) with the ones obtained from the approximate equations~(\ref{eq:kmukaslow}). In this example, the photons are emitted from the two positions $(r,\theta)$ of Fig.~\ref{fig:xyz}, and make angles $\Phi = 0$ and $\Theta = \pi/2$ as measured by the comoving observer.

Table~\ref{tab:ks} shows the photon four-momentum components as measured by the observer at rest at infinity, i.e. $k^\mu$, for photons emitted in the plane $\Phi = 0$ of the comoving observer (i.e. $\vec{e}_{(z)}$-$\vec{e}_{(x)}$), and in the plane $\Phi = \pi/2$ (i.e. $\vec{e}_{(z)}$-$\vec{e}_{(y)}$), in the directions $\Theta = 0$, $\pi/2$, and $\pi$.

\begin{table*}
    \centering
    \begin{tabular}{c|c|c|c|c|c|c}
    & \multicolumn{3}{c|}{$\Phi = 0$: plane  $\vec{e}_{(z)}$-$\vec{e}_{(x)}$} & \multicolumn{3}{c}{$\Phi = \pi/2$: plane $\vec{e}_{(z)}$-$\vec{e}_{(y)}$} \\
    \hline
         & $\Theta = 0$ & $\Theta = \pi/2$ & $\Theta = \pi$
         & $\Theta = 0$ & $\Theta = \pi/2$ & $\Theta = \pi$ \\
        \hline
        $k^0$ & $\frac{k^{(0)}}{\sqrt{1-2 M/r}}\sqrt{\frac{1 + \hat{\beta}}{1 - \hat{\beta}}}$ & $k^{(0)} \gamma$ & $\frac{k^{(0)}}{\sqrt{1-2 M/r}} \sqrt{\frac{1 - \hat{\beta}}{1 + \hat{\beta}}}$
        & $\frac{k^{(0)}}{\sqrt{1-2 M/r}}\sqrt{\frac{1 + \hat{\beta}}{1 - \hat{\beta}}}$ & $k^{(0)} \gamma$ & $\frac{k^{(0)}}{\sqrt{1-2 M/r}} \sqrt{\frac{1 - \hat{\beta}}{1 + \hat{\beta}}}$\\
        $n^r$ & $1-2 M/r$ & $\left(1-2 M/r\right)\hat{\beta}$ & $-\left(1-2 M/r\right)$ 
        & $1-2 M/r$ & $\left(1-2 M/r\right)\hat{\beta}$ & $-\left(1-2 M/r\right)$\\
        $n^\theta$ & $0$ & $(1/\gamma) (1/r)$ & $0$ 
        & 0 & 0 & 0 \\
        $n^\phi$ & $2 M a/r^3$ & $2 M a/r^3$ & $2 M a/r^3$
        & $2 M a/r^3$ & $2 M a/r^3$ + $(1/\gamma) (1/r \sin\theta)$ & $2 M a/r^3$\\
        \hline
    \end{tabular}
    \caption{Photon four-momentum components calculated with Eqs.~(\ref{eq:kmukaslow}), for specific angles of emission, $\Theta$ and $\Phi$, as measured by the comoving observer. We have used the relation between the Lorentz factor $\hat{\gamma}$ and $\gamma$ given by Eq ~(\ref{eq:gammas}).}
    \label{tab:ks}
\end{table*}

It can be seen from the above equations (see also their summary in Table~\ref{tab:ks}) that
\begin{enumerate}
    \item
    There is a purely general relativistic effect: the azimuthal photon four-momentum component, $n^\phi$, has a contribution from the angular velocity $\omega$ associated with the \emph{frame-dragging} effect. This contribution vanishes only for photons emitted by electrons accelerated along the BH rotation axis ($\theta = 0$). For off-axis motion, it does not vanish and it can be even larger than the polar component, $n^\theta$ (see Table~\ref{tab:ks} and Fig.~\ref{fig:pmutheta0}).
    \item
    The above implies that \emph{relativistic aberration} by which photons are measured at infinity as confined within an angle $1/\gamma$ around the direction of motion of the emitter \citep[see e.g.][]{1975ctf..book.....L}, strictly holds only along the BH rotation axis. However, the radial component of the four-momentum is still dominant and the traditional relativistic aberration can be assumed for practical purposes.
\end{enumerate}

\section{Radiation power and spectrum}\label{sec:6}

The above result allows us to estimate the radiation by acceleration using known results from special relativity. Most of the synchrotron power is radiated around the peak of the spectrum which occurs near the characteristic angular frequency \citep{1975ctf..book.....L}
\begin{equation}\label{eq:egamma}
    \omega_c = \frac{c}{\rho}\gamma^3,
\end{equation}
where $\rho$ is the relativistic curvature radius
\begin{equation}
    \rho = \frac{m c^2 \gamma}{|q|}\left[ (\vec{E} + \vec{v}\times \vec{B})^2 - (\vec{v}\cdot \vec{E})^2 \right]^{-1/2}.
\end{equation}

\begin{figure*}
    \centering
    \includegraphics[width=0.49\hsize,clip]{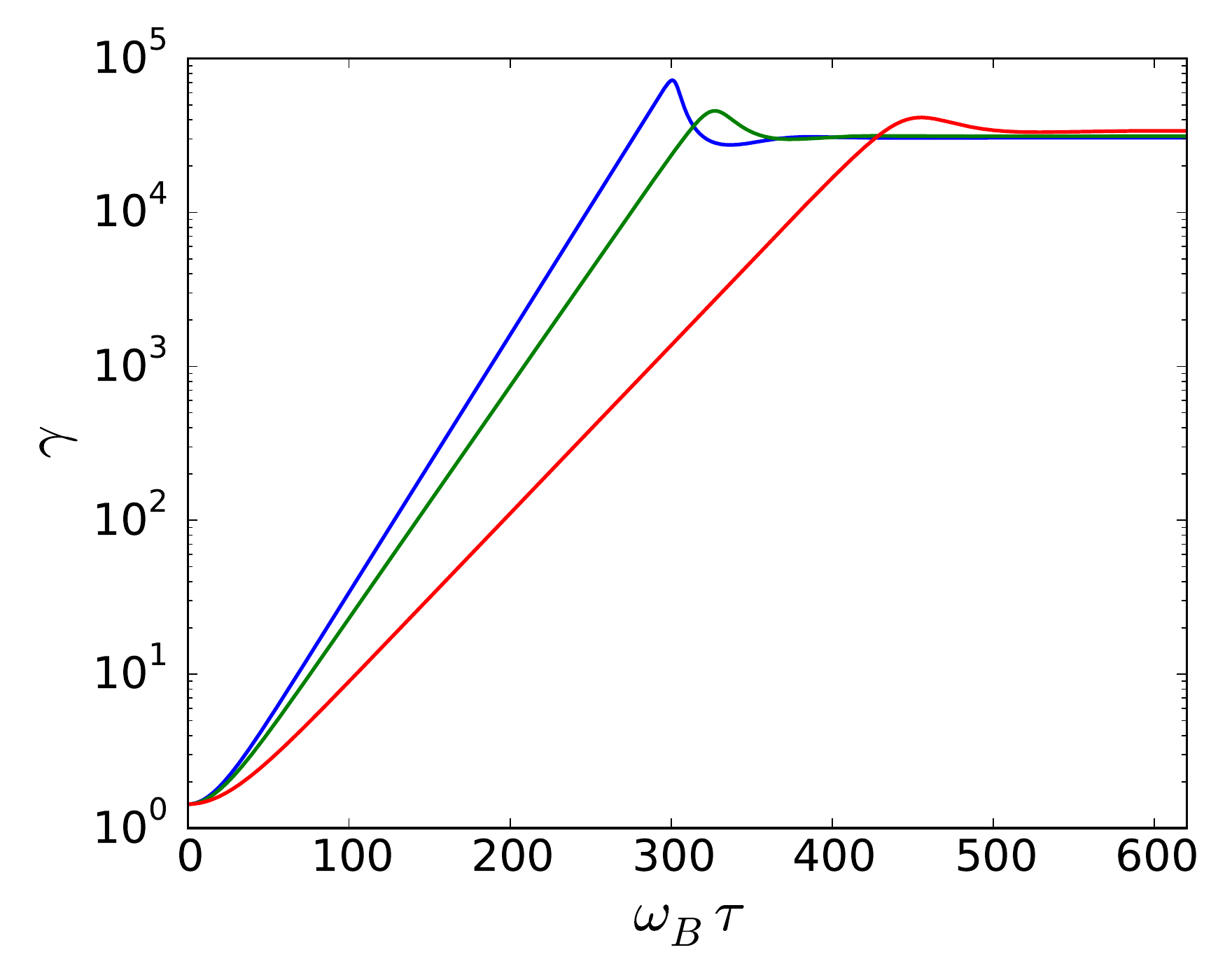}
    \includegraphics[width=0.49\hsize,clip]{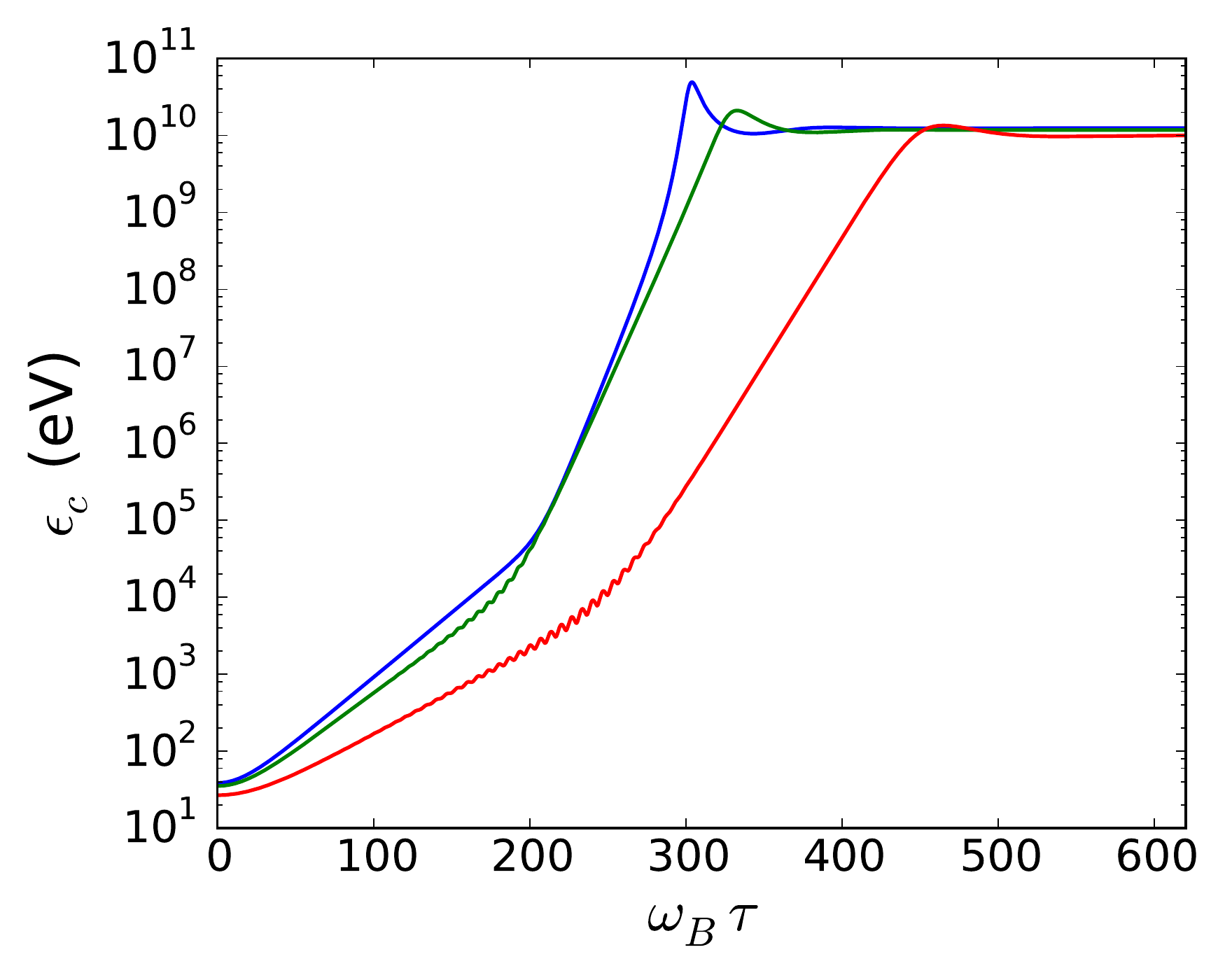}
    \caption{Electron Lorentz factor $\gamma$, given by Eq.~(\ref{eq:gammas}), and photon critical energy $\epsilon_c$, given by Eq.~(\ref{eq:egamma}), as a function of the emitter (electron) proper time, $\omega_B \tau$. In this example, the electron starts is motion at rest at the position $r(0) = 2 r_+$, $\phi(0) = 0$, and $\theta(0) = 1^\circ$ (blue), $14^\circ$ (green), and $27^\circ$ (red).}
    \label{fig:gammaEdotephoton}
\end{figure*}

The left panel of Fig.~\ref{fig:gammaEdotephoton} shows electron Lorentz factor as a function of the dimensionless proper time, and the right panel the characteristic photon energy, $\epsilon_c = \hbar \omega_c$. We show the results for three different initial conditions; the electron starts its motion at rest at $[r(0),\theta(0),\phi(0)]$, where $r(0) = 2 r_+$, $\phi(0) = 0$, and $\theta(0) = 1^\circ$ (blue curves), $\theta(0) = 14^\circ$ (green curves), and $\theta(0) = 27^\circ$.

The electron reaches an asymptotic value when acceleration and radiation losses balance each other. At small values of the spherical polar angle, the electron experiences less radiation losses reaching a higher Lorentz factor and higher photon energy. Furthermore, the photon characteristic energy falls in the GeV regime at the asymptotic Lorentz factor. Therefore, most of the radiation is emitted at GeV energies.

We now turn to the energy distribution of the radiation. The synchrotron power radiated by an electron, per unit angular frequency $\omega$, and integrated over the solid angle is \citep[see, e.g.,][]{1975ctf..book.....L}
\begin{equation}
    P_\omega= \frac{\sqrt{3}}{2\pi}\frac{e^2}{\rho}\gamma F_\omega(x), \qquad F_\omega(x) \equiv x \int_x^\infty K_{5/3}(y)dy,
\end{equation}
where $x \equiv \omega/\omega_c$, and $K_{5/3}$ is the modified Bessel function of the second kind.

\begin{figure*}
    \centering
    \includegraphics[width=0.49\hsize,clip]{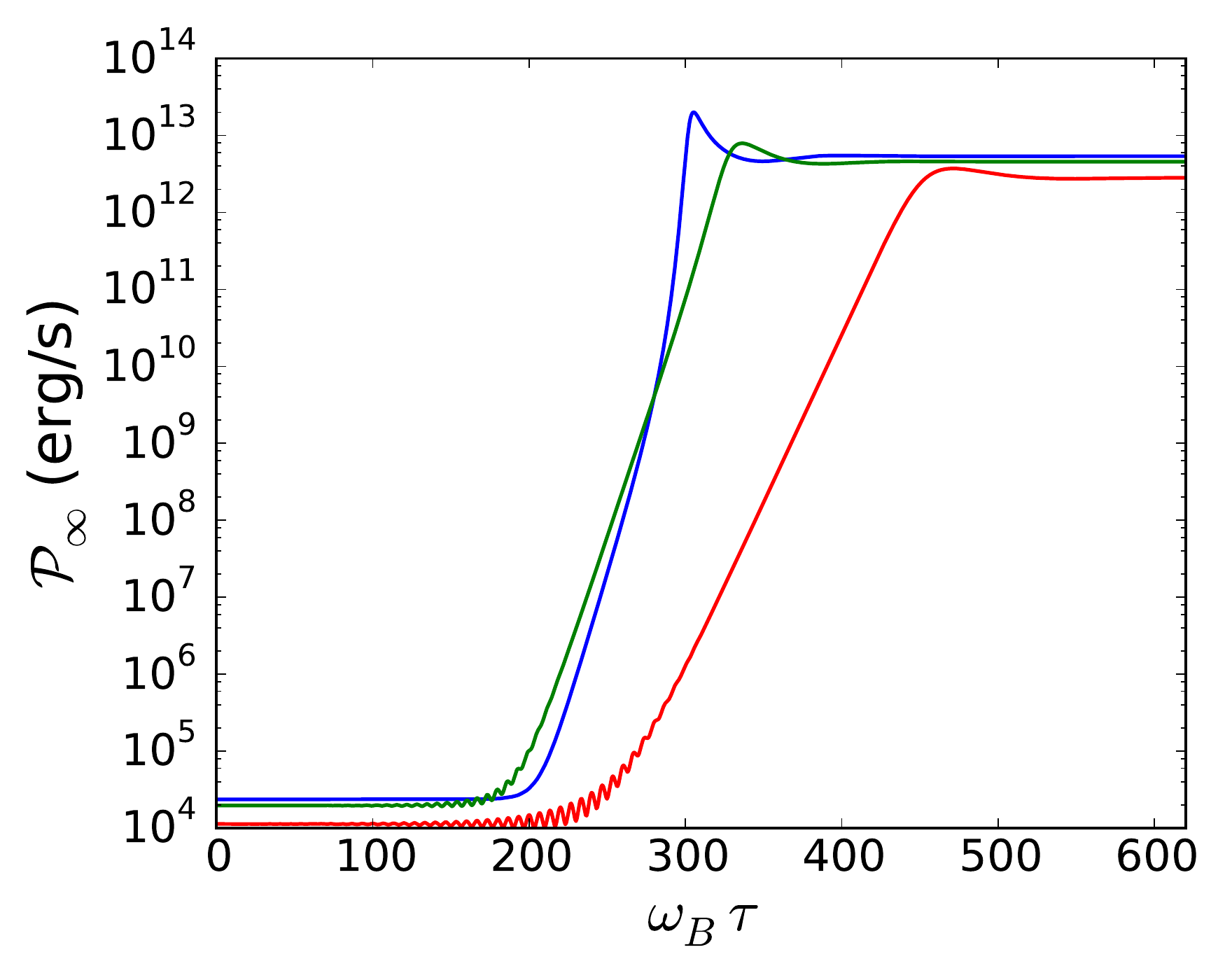}
    \includegraphics[width=0.49\hsize,clip]{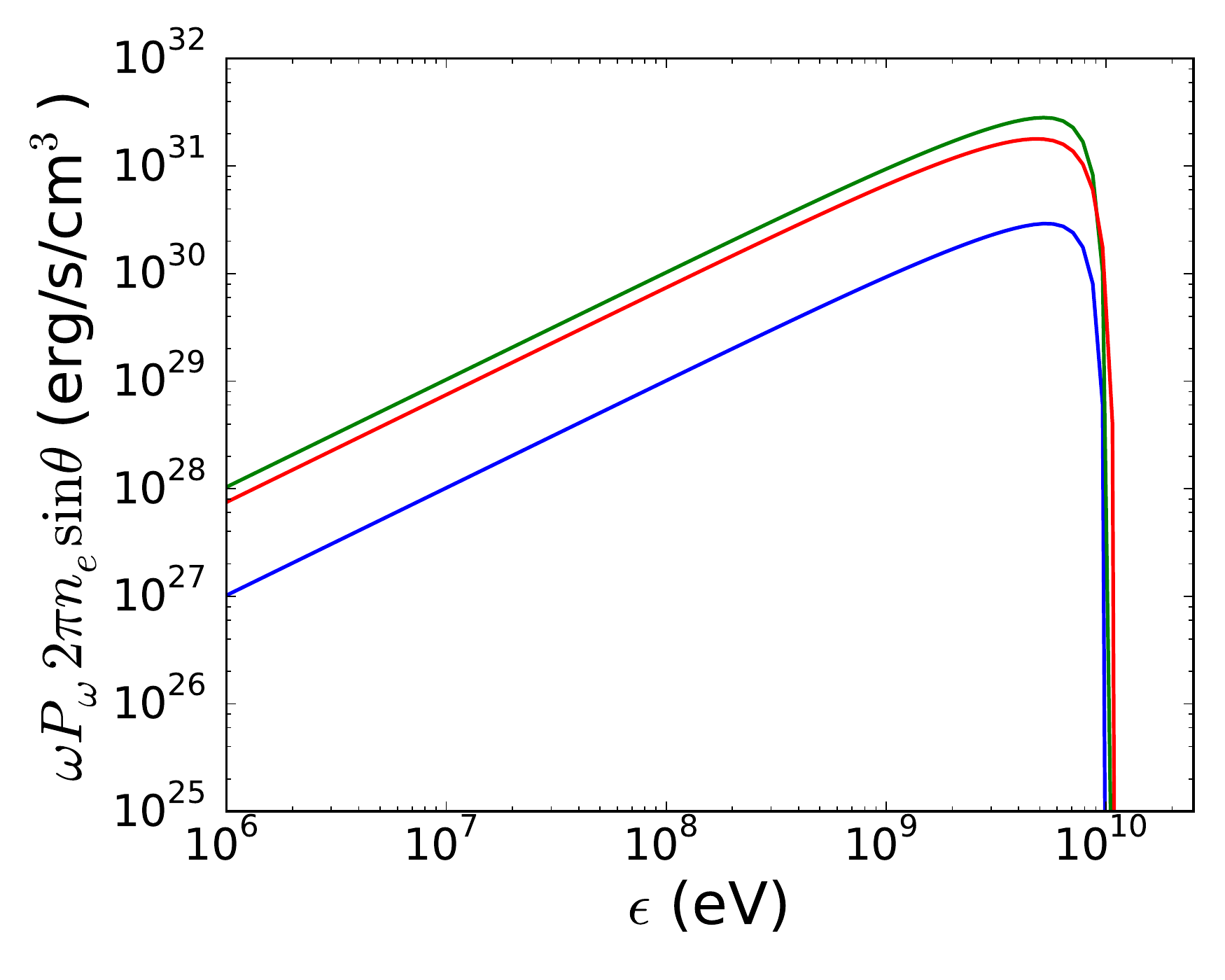}
    \caption{Left: power radiated off to infinity as a function the dimensionless proper time. Right: energy distribution of the synchrotron power per unit volume, $\omega P_\omega n_e\sin\theta$, where $n_e$ is given by Eq. (\ref{eq:ne}). This gives an estimate of the energy radiated to infinity per unit time, per unit volume. In this example, the electron starts is motion at rest at three different selected positions $(r,\theta,\phi)$: $r(0) = 2 r_+$, $\phi(0) = 0$, and $\theta(0) = 1^\circ$ (blue), $14^\circ$ (green), and $27^\circ$ (red).}
    \label{fig:nuFnu}
\end{figure*}

We plot in Fig.~\ref{fig:nuFnu} the (bolometric) power  radiated off at infinity (left panel), ${\cal P}_\infty$, as a function of dimensionless proper time. The obtained power per electrons shows that to emit, for instance, a luminosity of $L = 10^{51}$~erg~s$^{-1}$, we must accelerate about $N_e \sim L/{\cal P}_\infty \sim 10^{38}$ electrons. Since the acceleration occurs in the proximity of the BH horizon, these electrons occupy a volume ${\cal V}_e \sim 4\pi r_+^3/3 \sim 10^{19}$~cm$^3$, implying a number density $n_e \sim N_e/{\cal V}_e\sim 10^{38}/10^{19} = 10^{19}$~cm$^{-3}$. In fact, for the LNR observer the \textit{inner engine} can accelerate the following number of electrons per unit volume
\begin{equation}\label{eq:ne}
    n_e = \frac{E^2}{8\pi\,\hat\gamma m_e c^2},
\end{equation}
where $E$ is given by Eqs.~(\ref{eq:Efieldzamo}). The electric field near the horizon is $E \approx \alpha B_0/2$, where $\alpha = a/M$, so Eq. (\ref{eq:ne}) gives $n_e \approx 3 (\alpha B_0)^2/(32\pi \hat\gamma m_e c^2)$. For the present quantitative example, so using $\hat\gamma \approx 10^4$ (see Fig. \ref{fig:gammaEdotephoton}), Eq. (\ref{eq:ne}) gives $n_e \approx 10^{19}$ cm$^{-3}$, as expected from our previous estimate. The number of electrons around the Kerr BH can be either lower and higher than this value, but the \textit{inner engine} can accelerate at most the above number of electrons. The density in the \textit{cavity} created around the Kerr BH formed from the gravitational collapse of the NS in a BdHN is of the same order as the above value \citep[see][for numerical simulations]{2019ApJ...871...14B, 2019ApJ...883..191R}. This implies that the \textit{inner engine} in BdHNe works efficiently using its full energy reservoir.

The right panel of Fig.~\ref{fig:nuFnu} shows the spectrum density, i.e. the radiated power per unit volume, $\omega P_\omega 2\pi n_e\sin\theta$, as a function of the photon energy (calculated as $\epsilon = \hbar \omega$). Here, $n_e$ is given by Eq. (\ref{eq:ne}). This plot confirms that most of the energy is emitted in the GeV domain, and the above-mentioned value of the electron density of particles that can be accelerated, in fact ${\cal P}_\infty \times n_e\sim $ $10^{13}\times 10^{18} = 10^{31}$~erg~s$^{-1}$~cm$^{-3}$.

Summarizing, the gravitomagnetic interaction of the Kerr BH with the surrounding magnetic field efficiently accelerate electrons from the ionized environment in the BH vicinity. This acceleration and radiation process emits photons in the GeV regime for magnetic fields in the $10^{10}$--$10^{11}$~G and stellar-mass Kerr BHs. The emission originates in proximity of the BH horizon and within $60^\circ$ from the BH rotation axis with equatorial symmetry, hence generating a double-cone ``\textit{jetted}'' high-energy ($\gtrsim$GeV) emission.

\section{The black hole energy extraction}\label{sec:7}

It is clear that the energy radiated off to infinity must necessarily come from the BH extractable energy
\begin{equation}\label{eq:Eext}
    E_{\rm extr}\equiv (M-M_{\rm irr}) c^2,
\end{equation}
where the total mass $M$ is given by the BH mass-energy formula \citep{1970PhRvL..25.1596C,1971PhRvD...4.3552C,1971PhRvL..26.1344H}
\begin{equation}
\label{eq:Mbh}
M^2 = \frac{c^2 J^2}{4 G^2 M^2_{\rm irr}}+M_{\rm irr}^2,
\end{equation}
where $J$, $M$, are respectively the angular momentum and the mass of the BH. 

In order to understand better how the BH rotational energy is extracted, it is useful to recall the \emph{inner engine} operation:
\begin{enumerate}
    \item 
    The magnetic field and the BH rotation induce an electric field as given by the Papapetrou-Wald solution (see Sec.~\ref{sec:2}). For aligned and parallel magnetic field to the BH spin, the electric field is nearly radial and inwardly around the BH rotation axis within an angle $\theta_\pm \approx 60^\circ$ (see Fig.~\ref{fig:fieldlines}).
    \item
    The induced electric field accelerates electrons outwardly. The number of electrons that can be accelerated is set by the energy of the electric field
\begin{equation}\label{eq:Equantum}
    {\cal E} \approx \frac{1}{2}E_{\hat{r}}^2 r_+^3 \approx \frac{G}{c^4} \frac{B_0^2 J^2}{M},
\end{equation}
where we have used Eq. (\ref{eq:EMslow}). 
    %
    \item
    The maximum possible electron acceleration/energy is set by the electric potential energy difference from the horizon to infinity
    %
    \begin{equation}\label{eq:deltaphi}
    \Delta \Phi = \frac{1}{c}e\,a\,B_0.
\end{equation}
%
    \item 
    Along the polar axis, radiation losses by acceleration are absent because the electric and magnetic fields are parallel. Therefore, electrons accelerated on the BH rotation axis can gain the full potential energy difference (\ref{eq:deltaphi}).
    \item 
    At off-axis latitudes, the accelerated electrons emit synchrotron radiation. In order to explain the observed luminosity, $L_{\rm GeV}$, the radiation timescale, $t_{\rm rad}$, must fulfill
    \begin{equation}\label{eq:taurad}
        L_{\rm GeV} = \frac{d E_{\rm GeV}}{dt} \leq \frac{dE_{\rm extr}}{dt} \approx \frac{{\cal E}}{t_{\rm rad}},
    \end{equation}
    where ${\cal E}$ is given by Eq.~(\ref{eq:Equantum}), and we assume most of this energy is radiated off in high-energy photons, as we have shown in Sec.~\ref{sec:6}. We take here into account that the energy reservoir is the rotational energy of the BH, therefore, the extractable energy must satisfy
\begin{equation}\label{eq:EextEgev}
    E_{\rm extr} \geq E_{\rm GeV},
\end{equation}
where $E_{\rm extr}$ is given by Eq.~(\ref{eq:Eext}) and $E_{\rm GeV}$ is the energy observed at high-energies, i.e.:
\begin{equation}\label{eq:LgeV}
    E_{\rm GeV} = \int L_{\rm GeV}\,dt.
\end{equation}
    \item
    Once the energy ${\cal E}$ has been emitted, the BH is left with new values of mass and angular momentum which have been reduced by amounts $dM$ and $dJ$, respectively. The change in the BH mass is 
    \begin{equation}\label{eq:dM}
        c^2 dM = dE_{\rm extr} \approx dE_{\rm GeV} \approx {\cal E}.
    \end{equation}
According to the BH mass-energy formula (\ref{eq:Mbh}), if in the energy extraction process the irreducible mass is kept constant, the change in the BH angular momentum is
    \begin{equation}\label{eq:dJ}
        dJ = \frac{c^2 dM}{\Omega_+}.
    \end{equation}
    \item
    The above steps are repeated, with the same efficiency, if the density of plasma is sufficient, namely if the number of the particles is enough to cover the new electric energy. Therefore, the \emph{inner engine} evolves in a sequence of \emph{elementary processes}, each emitting a well-defined, precise amount of BH rotational energy.
\end{enumerate}
Therefore, the total fractional changes of mass and angular momentum in the whole emission process are
\begin{align}\label{eq:fractionaldMdJ}
    \frac{\Delta M}{M} &= 1-\frac{M_{\rm irr}}{M},\\
    \frac{\Delta J}{J} &= \left(1+\frac{M_{\rm irr}}{M}\right)^{-1},
\end{align}
where we have used Eqs.~(\ref{eq:Mbh}), (\ref{eq:dM}), (\ref{eq:dJ}), and the relation between the BH irreducible mass and the horizon: 
\begin{equation}\label{eq:rhMirr}
    r_+ = \frac{G}{c^2}\frac{2 M_{\rm irr}^2}{M}.
\end{equation}

\begin{figure}
    \centering
    \includegraphics[width=\hsize,clip]{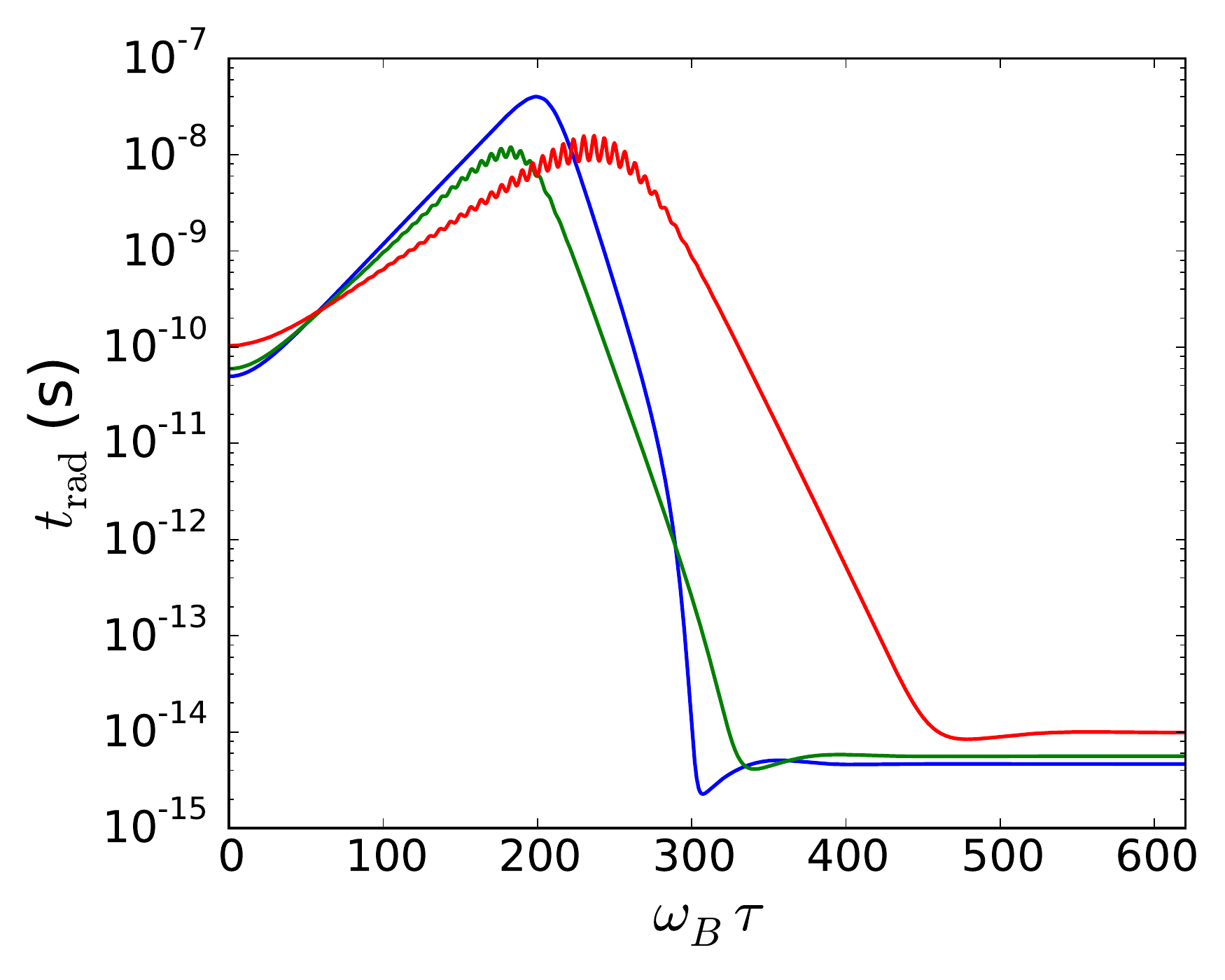}
    \caption{Radiation timescale $t_{\rm rad}$ given by Eq.~(\ref{eq:trad}), as a function of the dimensionless proper time $\omega_B \tau$. It can be seen how $t_{\rm rad}$ becomes as short as $10^{-14}$~s when the electrons reach the peak radiation power at $\omega_B \tau \sim 300$ (see Fig.~\ref{fig:gammaEdotephoton}).}
    \label{fig:trad}
\end{figure}

As a quantitative example, let us use our present fiducial parameters $B_0=10^{11}$~G, $M=4.4 M_\odot$ and $\alpha=0.3$. In this case, the electric energy is ${\cal E}\approx 2.04\times 10^{37}$~erg. The radiated luminosity is set by the timescale at which ${\cal E}$ is radiated off. Figures~\ref{fig:gammaEdotephoton} and \ref{fig:nuFnu} show that electrons reach a Lorentz factor $\gamma\approx 3\times 10^4$, so an energy of nearly $\epsilon_e = \gamma m_e c^2 \approx 10^{11}\,{\rm eV} \approx 10^{-1}$~erg, which is then radiated off at a rate ${\cal P}_\infty\approx 10^{13}$~erg~s$^{-1}$, leading to a radiation timescale
\begin{equation}\label{eq:trad}
    t_{\rm rad} = -\frac{p_\mu \eta^\mu}{{\cal P}_\infty},
\end{equation}
of the order of $\approx 10^{-14}$~s. We recall that $p_\mu$ is the four-momentum of the particle measure by an observer at rest at infinity and $\eta^\mu$ is the time-like Killing vector. This estimate of the timescale is confirmed by Fig.~\ref{fig:trad} which shows the time evolution of $t_{\rm rad}$. This radiation timescale implies that the system radiates off with a power ${\cal E}/t_{\rm rad}\approx 10^{51}$~erg~s$^{-1}$. This energy is mainly radiated at photon of energies $\epsilon_c$ of the order of GeV (see right panels of Figs.~\ref{fig:gammaEdotephoton} and \ref{fig:nuFnu}).

For these parameters, in each elementary emission, the BH mass and angular momentum experience fractional changes $dM/M\approx 2.59\times 10^{-18}$, and $dJ/J \approx 1.12\times 10^{-16}$, respectively. The BH irreducible mass can be readily obtained from Eq.~(\ref{eq:rhMirr}), i.e. $M_{\rm irr} \approx 4.35\,M_\odot$. Therefore, the total extractable energy is $E_{\rm extr} = \Delta M c^2 \approx 0.012 M c^2 \approx 9.12\times 10^{52}$~erg. If $E_{\rm extr}$ is emitted, the total fractional changes in the BH mass and angular momentum will be, respectively, $\Delta M/M \approx 0.012$ and $\Delta J/J\approx 0.50$.

\section{Discussion and Conclusions}\label{sec:8}

Summarizing, all BdHN I are powered by three independent sources of energy:
\begin{enumerate}
    \item
    The BdHN I is triggered by the $\nu$NS-rise produced in the core-collapse of the CO star generating a $\nu$NS at its center. The $\nu$NS rotational energy powers the synchrotron emission from the expanding SN that originates the X-ray afterglow \citep[see][and references therein]{2020ApJ...893..148R}.
    \item 
    The hypercritical accretion of the SN material onto the binary companion NS leads to the formation of the BH when it reaches the critical mass for gravitational collapse \citep[see, e.g.][]{2016ApJ...833..107B, 2019ApJ...871...14B}. This ``smooth'' accretion leading to BH formation does not emit any detectable signal of gravitational waves and is alternative to the direct gravitational collapse of a massive star (\emph{collapsar}).
    \item
    The gravitomagnetic interaction of the Kerr BH with the surrounding magnetic field, in presence of ionized matter of the SN ejecta, leads to the process of extraction of the BH rotational energy, generating the ``\textit{jetted}'' high-energy ($\gtrsim$GeV) emission. We have here shown that this radiation is emitted in the vicinity of the BH horizon within $60^\circ$ from the BH rotation axis (see Secs.~\ref{sec:2} and \ref{sec:6}).
\end{enumerate}

We have focused in this article on the \emph{inner engine} of the high-energy emission, which as we have shown drives a new paradigm in the theory of GRBs:
\begin{enumerate}
    \item
    \emph{There is no need for bulk motion}. The traditional GRB model uses the gravitational pull to accelerate matter in bulk up to very large distances $\sim 10^{16}$--$10^{18}$~cm, where the ultrarelativistic blastwave becomes transparent to high-energy photons. The \textit{inner engine}, instead, radiates at horizon scales the kinetic energy rapidly gained by single-particle acceleration (Secs.~\ref{sec:3}--\ref{sec:6}). As shown in \citet{2021A&A...649A..75M, 2019ApJ...886...82R}, the radiated high-energy photons are transparent to magnetic pair production, likely the most important opacity source of this system.
    \item 
    \emph{There is no need of massive accretion}. The density of ionized matter needed for the \emph{inner engine} to explain the GeV emission of a long GRB is much lower than the corresponding one requested by traditional matter accretion. For example, an accretion disk produces a luminosity $L_{\rm disk}=\eta \dot{M} c^2$, where $\eta$ is a parameter accounting for the efficiency in converting gravitational energy into radiation. Adopting a fiducial value $\eta = 0.1$, in order to get a luminosity of $10^{50}$~erg~s$^{-1}$, the BH must accrete matter at a rate $\dot{M} = 10^{-3}~M_\odot$~s$^{-1}$. This implies a consumption of protons by the BH at a rate of $\dot{N}_p \sim \dot{M}/m_p\sim 10^{51}$~s$^{-1}$, or a proton number density rate $\dot{n}_p =\dot{N}_p/(4\pi r_+^3/3)\sim 10^{33}$~cm$^{-3}$~s$^{-1}$. Clearly, smaller values of $\eta$ require larger accretion rates. The \emph{inner engine}, instead, produces the same luminosity by accelerating electrons with a rate $\dot{N}_e\sim N_e/\tau_{\rm rad} = {\cal E}/(\Delta \Phi\,\tau_{\rm rad}) \sim 10^{47}$~s$^{-1}$, i.e. $\dot{M}\sim m_p N_e \sim 10^{-7}~M_\odot$~s$^{-1}$ (for fully ionized matter), or an electron density of $\dot{n}_e\sim 10^{29}$~cm$^{-3}$~s$^{-1}$; see discussion of Eq.~(\ref{eq:ne}) in Sec.~\ref{sec:6} for further details. This implies that the geodesic equations of motion of massive particles around a Kerr BH, traditionally applied to the problem of gravitational matter accretion, are superseded by the equation of motion of charged particles, accounting for the radiation reaction, accelerated in the electromagnetic field of the Papapetrou-Wald solution (see Secs.~\ref{sec:4}--\ref{sec:6}).
    \item
    \emph{The BH is responsible only for the high-energy radiation}. During the acceleration region, the electrons radiate mainly at high-energies, e.g.~$\gtrsim 0.1$~GeV (see  Sec.~\ref{sec:6}). The fundamental role of the reversible transformations holds. Indeed, the energy budget is paid by the BH extractable energy \citep{2020EPJC...80..300R}, and the irreducible mass $M_{\rm irr}$ of the BH mediates the energy extraction process (see Sec.~\ref{sec:7}).
\end{enumerate}

\begin{acknowledgments}
In occasion of this fifty anniversary of \textit{Introducing the black hole}, we would like to thank all the people who have contributed in these fifty years to finally identify the energy extraction process of a Kerr BH. We are most grateful to Prof. Robert Jantzen for insightful discussions on the general relativistic framework presented in this article.
\end{acknowledgments}


\begin{thebibliography}{}
\expandafter\ifx\csname natexlab\endcsname\relax\def\natexlab#1{#1}\fi
\providecommand{\url}[1]{\href{#1}{#1}}
\providecommand{\dodoi}[1]{doi:~\href{http://doi.org/#1}{\nolinkurl{#1}}}
\providecommand{\doeprint}[1]{\href{http://ascl.net/#1}{\nolinkurl{http://ascl.net/#1}}}
\providecommand{\doarXiv}[1]{\href{https://arxiv.org/abs/#1}{\nolinkurl{https://arxiv.org/abs/#1}}}

\bibitem[{{Baade} \& {Zwicky}(1934)}]{1934PhRv...46...76B}
{Baade}, W., \& {Zwicky}, F. 1934, Physical Review, 46, 76,
  \dodoi{10.1103/PhysRev.46.76.2}

\bibitem[{{Bardeen}(1970)}]{1970ApJ...162...71B}
{Bardeen}, J.~M. 1970, \apj, 162, 71, \dodoi{10.1086/150635}

\bibitem[{{Bardeen} {et~al.}(1972){Bardeen}, {Press}, \&
  {Teukolsky}}]{1972ApJ...178..347B}
{Bardeen}, J.~M., {Press}, W.~H., \& {Teukolsky}, S.~A. 1972, \apj, 178, 347,
  \dodoi{10.1086/151796}

\bibitem[{{Becerra} {et~al.}(2016){Becerra}, {Bianco}, {Fryer}, {Rueda}, \&
  {Ruffini}}]{2016ApJ...833..107B}
{Becerra}, L., {Bianco}, C.~L., {Fryer}, C.~L., {Rueda}, J.~A., \& {Ruffini},
  R. 2016, \apj, 833, 107, \dodoi{10.3847/1538-4357/833/1/107}

\bibitem[{{Becerra} {et~al.}(2019){Becerra}, {Ellinger}, {Fryer}, {Rueda}, \&
  {Ruffini}}]{2019ApJ...871...14B}
{Becerra}, L., {Ellinger}, C.~L., {Fryer}, C.~L., {Rueda}, J.~A., \& {Ruffini},
  R. 2019, \apj, 871, 14, \dodoi{10.3847/1538-4357/aaf6b3}

\bibitem[{{Bernardi} {et~al.}(2002){Bernardi}, {Alonso}, {da Costa}, {Willmer},
  {Wegner}, {Pellegrini}, {Rit{\'e}}, \& {Maia}}]{2002AJ....123.2990B}
{Bernardi}, M., {Alonso}, M.~V., {da Costa}, L.~N., {et~al.} 2002, \aj, 123,
  2990, \dodoi{10.1086/340463}

\bibitem[{{Bird} {et~al.}(2010){Bird}, {Harris}, {Blakeslee}, \&
  {Flynn}}]{2010A&A...524A..71B}
{Bird}, S., {Harris}, W.~E., {Blakeslee}, J.~P., \& {Flynn}, C. 2010, \aap,
  524, A71, \dodoi{10.1051/0004-6361/201014876}

\bibitem[{Boyer \& Lindquist(1967)}]{doi:10.1063/1.1705193}
Boyer, R.~H., \& Lindquist, R.~W. 1967, Journal of Mathematical Physics, 8,
  265, \dodoi{10.1063/1.1705193}

\bibitem[{{Carter}(1968)}]{1968PhRv..174.1559C}
{Carter}, B. 1968, Physical Review, 174, 1559, \dodoi{10.1103/PhysRev.174.1559}

\bibitem[{{Christodoulou}(1970)}]{1970PhRvL..25.1596C}
{Christodoulou}, D. 1970, \prl, 25, 1596, \dodoi{10.1103/PhysRevLett.25.1596}

\bibitem[{{Christodoulou} \& {Ruffini}(1971)}]{1971PhRvD...4.3552C}
{Christodoulou}, D., \& {Ruffini}, R. 1971, \prd, 4, 3552,
  \dodoi{10.1103/PhysRevD.4.3552}

\bibitem[{{Costa} {et~al.}(1997){Costa}, {Frontera}, {Heise}, {Feroci}, {in't
  Zand}, {Fiore}, {Cinti}, {Dal Fiume}, {Nicastro}, {Orlandini}, {Palazzi},
  {Rapisarda}, {Zavattini}, {Jager}, {Parmar}, {Owens}, {Molendi}, {Cusumano},
  {Maccarone}, {Giarrusso}, {Coletta}, {Antonelli}, {Giommi}, {Muller}, {Piro},
  \& {Butler}}]{Costa1997}
{Costa}, E., {Frontera}, F., {Heise}, J., {et~al.} 1997, \nat, 387, 783,
  \dodoi{10.1038/42885}

\bibitem[{{Damour} {et~al.}(1978){Damour}, {Hanni}, {Ruffini}, \&
  {Wilson}}]{1978PhRvD..17.1518D}
{Damour}, T., {Hanni}, R.~S., {Ruffini}, R., \& {Wilson}, J.~R. 1978, \prd, 17,
  1518, \dodoi{10.1103/PhysRevD.17.1518}

\bibitem[{{Finkelstein}(1958)}]{1958PhRv..110..965F}
{Finkelstein}, D. 1958, Physical Review, 110, 965,
  \dodoi{10.1103/PhysRev.110.965}

\bibitem[{{Finzi} \& {Wolf}(1969)}]{1969ApJ...155L.107F}
{Finzi}, A., \& {Wolf}, R.~A. 1969, \apjl, 155, L107, \dodoi{10.1086/180312}

\bibitem[{{Giacconi}(2003)}]{2003RvMP...75..995G}
{Giacconi}, R. 2003, Reviews of Modern Physics, 75, 995,
  \dodoi{10.1103/RevModPhys.75.995}

\bibitem[{Giacconi {et~al.}(1962)Giacconi, Gursky, Paolini, \&
  Rossi}]{PhysRevLett.9.439}
Giacconi, R., Gursky, H., Paolini, F.~R., \& Rossi, B.~B. 1962, Phys. Rev.
  Lett., 9, 439, \dodoi{10.1103/PhysRevLett.9.439}

\bibitem[{{Giacconi} \& {Ruffini}(1978)}]{1978pans.proc.....G}
{Giacconi}, R., \& {Ruffini}, R., eds. 1978, {Physics and astrophysics of
  neutron stars and black holes}

\bibitem[{{Gold}(1968)}]{1968Natur.218..731G}
{Gold}, T. 1968, \nat, 218, 731, \dodoi{10.1038/218731a0}

\bibitem[{{Hawking}(1971)}]{1971PhRvL..26.1344H}
{Hawking}, S.~W. 1971, Physical Review Letters, 26, 1344,
  \dodoi{10.1103/PhysRevLett.26.1344}

\bibitem[{{Hewish} {et~al.}(1968){Hewish}, {Bell}, {Pilkington}, {Scott}, \&
  {Collins}}]{1968Natur.217..709H}
{Hewish}, A., {Bell}, S.~J., {Pilkington}, J.~D.~H., {Scott}, P.~F., \&
  {Collins}, R.~A. 1968, \nat, 217, 709, \dodoi{10.1038/217709a0}

\bibitem[{{Jia} {et~al.}(2020){Jia}, {Bu}, {Qu}, {Lu}, {Zhang}, {Huang}, {Ma},
  {Tao}, {Xiao}, {Zhang}, {Chen}, {Song}, {Zhang}, {Li}, {Xu}, {Cao}, {Chen},
  {Liu}, {Cai}, {Chang}, {Chen}, {Chen}, {Chen}, {Chen}, {Cui}, {Cui}, {Deng},
  {Dong}, {Du}, {Fu}, {Gao}, {Gao}, {Gao}, {Ge}, {Gu}, {Guan}, {Guo}, {Han},
  {Huo}, {Jiang}, {Jiang}, {Jin}, {Jin}, {Kong}, {Li}, {Li}, {Li}, {Li}, {Li},
  {Li}, {Li}, {Li}, {Li}, {Li}, {Liang}, {Liao}, {Liu}, {Liu}, {Liu}, {Liu},
  {Lu}, {Lu}, {Luo}, {Luo}, {Meng}, {Nang}, {Nie}, {Ou}, {Sai}, {Shang},
  {Song}, {Sun}, {Tan}, {Tuo}, {Wang}, {Wang}, {Wang}, {Wang}, {Wang}, {Wen},
  {Wu}, {Wu}, {Wu}, {Xiao}, {Xiong}, {Yang}, {Yang}, {Yang}, {Yin}, {Yi},
  {You}, {Zhang}, {Zhang}, {Zhang}, {Zhang}, {Zhang}, {Zhang}, {Zhang},
  {Zhang}, {Zhang}, {Zhang}, {Zhang}, {Zhang}, {Zhang}, {Zhang}, {Zhao},
  {Zhao}, {Zheng}, {Zhou}, {Zhou}, {Zhu}, {Zhu}, {Zhuang}, \& {Insight-HXMT
  Collaboration}}]{2020JHEAp..25....1J}
{Jia}, S.~M., {Bu}, Q.~C., {Qu}, J.~L., {et~al.} 2020, Journal of High Energy
  Astrophysics, 25, 1, \dodoi{10.1016/j.jheap.2019.11.001}

\bibitem[{{Kerr}(1963)}]{1963PhRvL..11..237K}
{Kerr}, R.~P. 1963, \prl, 11, 237, \dodoi{10.1103/PhysRevLett.11.237}

\bibitem[{{Kerr}(2007)}]{2007arXiv0706.1109K}
---. 2007, arXiv e-prints, arXiv:0706.1109.
\newblock \doarXiv{0706.1109}

\bibitem[{{Kruskal}(1960)}]{1960PhRv..119.1743K}
{Kruskal}, M.~D. 1960, Physical Review, 119, 1743,
  \dodoi{10.1103/PhysRev.119.1743}

\bibitem[{{Landau} \& {Lifshitz}(1975)}]{1975ctf..book.....L}
{Landau}, L.~D., \& {Lifshitz}, E.~M. 1975, {The classical theory of fields}

\bibitem[{{Leach} \& {Ruffini}(1973)}]{1973ApJ...180L..15L}
{Leach}, R.~W., \& {Ruffini}, R. 1973, \apjl, 180, L15, \dodoi{10.1086/181143}

\bibitem[{{Moradi} {et~al.}(2021{\natexlab{a}}){Moradi}, {Rueda}, {Ruffini}, \&
  {Wang}}]{2021A&A...649A..75M}
{Moradi}, R., {Rueda}, J.~A., {Ruffini}, R., \& {Wang}, Y. 2021{\natexlab{a}},
  \aap, 649, A75, \dodoi{10.1051/0004-6361/201937135}

\bibitem[{{Moradi} {et~al.}(2021{\natexlab{b}}){Moradi}, {Rueda}, {Ruffini},
  {Li}, {Bianco}, {Campion}, {Cherubini}, {Filippi}, {Wang}, \&
  {Xue}}]{2021PhRvD.104f3043M}
{Moradi}, R., {Rueda}, J.~A., {Ruffini}, R., {et~al.} 2021{\natexlab{b}}, \prd,
  104, 063043, \dodoi{10.1103/PhysRevD.104.063043}

\bibitem[{{Oppenheimer} \& {Snyder}(1939)}]{1939PhRv...56..455O}
{Oppenheimer}, J.~R., \& {Snyder}, H. 1939, Physical Review, 56, 455,
  \dodoi{10.1103/PhysRev.56.455}

\bibitem[{{Oppenheimer} \& {Volkoff}(1939)}]{1939PhRv...55..374O}
{Oppenheimer}, J.~R., \& {Volkoff}, G.~M. 1939, Physical Review, 55, 374,
  \dodoi{10.1103/PhysRev.55.374}

\bibitem[{{Pacini}(1968)}]{1968Natur.219..145P}
{Pacini}, F. 1968, \nat, 219, 145, \dodoi{10.1038/219145a0}

\bibitem[{{Papapetrou}(1966)}]{1966AIHPA...4...83P}
{Papapetrou}, A. 1966, Annales de L'Institut Henri Poincare Section (A)
  Physique Theorique, 4, 83

\bibitem[{{Penrose} \& {Floyd}(1971)}]{1971NPhS..229..177P}
{Penrose}, R., \& {Floyd}, R.~M. 1971, Nature Physical Science, 229, 177,
  \dodoi{10.1038/physci229177a0}

\bibitem[{{Piran}(1999)}]{1999PhR...314..575P}
{Piran}, T. 1999, \physrep, 314, 575, \dodoi{10.1016/S0370-1573(98)00127-6}

\bibitem[{Punsly(2009)}]{Punsly2009}
Punsly, B. 2009, Black Hole Gravitohydromagnetics (Springer Berlin Heidelberg),
  \dodoi{10.1007/978-3-540-76957-6}.
\newblock \url{https://doi.org/10.1007%2F978-3-540-76957-6}

\bibitem[{{Rees} \& {Meszaros}(1992)}]{1992MNRAS.258P..41R}
{Rees}, M.~J., \& {Meszaros}, P. 1992, \mnras, 258, 41P,
  \dodoi{10.1093/mnras/258.1.41P}

\bibitem[{{Rhoades} \& {Ruffini}(1974)}]{1974PhRvL..32..324R}
{Rhoades}, C.~E., \& {Ruffini}, R. 1974, \prl, 32, 324,
  \dodoi{10.1103/PhysRevLett.32.324}

\bibitem[{{Rueda} \& {Ruffini}(2020)}]{2020EPJC...80..300R}
{Rueda}, J.~A., \& {Ruffini}, R. 2020, European Physical Journal C, 80, 300,
  \dodoi{10.1140/epjc/s10052-020-7868-z}

\bibitem[{{Rueda} {et~al.}(2020){Rueda}, {Ruffini}, {Karlica}, {Moradi}, \&
  {Wang}}]{2020ApJ...893..148R}
{Rueda}, J.~A., {Ruffini}, R., {Karlica}, M., {Moradi}, R., \& {Wang}, Y. 2020,
  \apj, 893, 148, \dodoi{10.3847/1538-4357/ab80b9}

\bibitem[{{Ruffini}(1974)}]{1974asgr.proc..349R}
{Ruffini}, R. 1974, in Astrophysics and Gravitation, 349--424

\bibitem[{{Ruffini} {et~al.}(2019{\natexlab{a}}){Ruffini}, {Melon Fuksman}, \&
  {Vereshchagin}}]{2019ApJ...883..191R}
{Ruffini}, R., {Melon Fuksman}, J.~D., \& {Vereshchagin}, G.~V.
  2019{\natexlab{a}}, \apj, 883, 191, \dodoi{10.3847/1538-4357/ab3c51}

\bibitem[{{Ruffini} \& {Wheeler}(1971{\natexlab{a}})}]{1971PhT....24a..30R}
{Ruffini}, R., \& {Wheeler}, J.~A. 1971{\natexlab{a}}, Physics Today, 24, 30,
  \dodoi{10.1063/1.3022513}

\bibitem[{{Ruffini} \& {Wheeler}(1971{\natexlab{b}})}]{1971ESRSP..52...45R}
---. 1971{\natexlab{b}}, ESRO, 52, 45

\bibitem[{{Ruffini} {et~al.}(2016){Ruffini}, {Rueda}, {Muccino}, {Aimuratov},
  {Becerra}, {Bianco}, {Kovacevic}, {Moradi}, {Oliveira}, {Pisani}, \&
  {Wang}}]{2016ApJ...832..136R}
{Ruffini}, R., {Rueda}, J.~A., {Muccino}, M., {et~al.} 2016, \apj, 832, 136,
  \dodoi{10.3847/0004-637X/832/2/136}

\bibitem[{{Ruffini} {et~al.}(2018){Ruffini}, {Wang}, {Aimuratov}, {Barres de
  Almeida}, {Becerra}, {Bianco}, {Chen}, {Karlica}, {Kovacevic}, {Li}, {Melon
  Fuksman}, {Moradi}, {Muccino}, {Penacchioni}, {Pisani}, {Primorac}, {Rueda},
  {Shakeri}, {Vereshchagin}, \& {Xue}}]{2018ApJ...852...53R}
{Ruffini}, R., {Wang}, Y., {Aimuratov}, Y., {et~al.} 2018, \apj, 852, 53,
  \dodoi{10.3847/1538-4357/aa9e8b}

\bibitem[{{Ruffini} {et~al.}(2019{\natexlab{b}}){Ruffini}, {Moradi}, {Rueda},
  {Becerra}, {Bianco}, {Cherubini}, {Filippi}, {Chen}, {Karlica}, {Sahakyan},
  {Wang}, \& {Xue}}]{2019ApJ...886...82R}
{Ruffini}, R., {Moradi}, R., {Rueda}, J.~A., {et~al.} 2019{\natexlab{b}}, \apj,
  886, 82, \dodoi{10.3847/1538-4357/ab4ce6}

\bibitem[{{Ruffini} {et~al.}(2021){Ruffini}, {Moradi}, {Rueda}, {Li},
  {Sahakyan}, {Chen}, {Wang}, {Aimuratov}, {Becerra}, {Bianco}, {Cherubini},
  {Filippi}, {Karlica}, {Mathews}, {Muccino}, {Pisani}, \&
  {Xue}}]{2021MNRAS.504.5301R}
---. 2021, \mnras, 504, 5301, \dodoi{10.1093/mnras/stab724}

\bibitem[{{Schmidt}(1963)}]{1963Natur.197.1040S}
{Schmidt}, M. 1963, \nat, 197, 1040, \dodoi{10.1038/1971040a0}

\bibitem[{{Shklovskii}(1953)}]{1953DoSSR..90..983S}
{Shklovskii}, I.~S. 1953, Akademiia Nauk SSSR Doklady, 90, 983

\bibitem[{{Shklovskii}(1968)}]{1968SvA....11..749S}
---. 1968, \sovast, 11, 749

\bibitem[{{Shklovskij}(1969)}]{1969supe.book.....S}
{Shklovskij}, I.~S. 1969, {Supernovae.}

\bibitem[{{Thorne} {et~al.}(1986){Thorne}, {Price}, \&
  {MacDonald}}]{1986bhmp.book.....T}
{Thorne}, K.~S., {Price}, R.~H., \& {MacDonald}, D.~A. 1986, {Black holes: The
  membrane paradigm}

\bibitem[{{van Paradijs} {et~al.}(1997){van Paradijs}, {Groot}, {Galama},
  {Kouveliotou}, {Strom}, {Telting}, {Rutten}, {Fishman}, {Meegan}, {Pettini},
  {Tanvir}, {Bloom}, {Pedersen}, {N{\o}rdgaard-Nielsen}, {Linden-V{\o}rnle},
  {Melnick}, {van der Steene}, {Bremer}, {Naber}, {Heise}, {in't Zand},
  {Costa}, {Feroci}, {Piro}, {Frontera}, {Zavattini}, {Nicastro}, {Palazzi},
  {Bennett}, {Hanlon}, \& {Parmar}}]{vanParadjis1997}
{van Paradijs}, J., {Groot}, P.~J., {Galama}, T., {et~al.} 1997, \nat, 386,
  686, \dodoi{10.1038/386686a0}

\bibitem[{{Wald}(1974)}]{1974PhRvD..10.1680W}
{Wald}, R.~M. 1974, \prd, 10, 1680, \dodoi{10.1103/PhysRevD.10.1680}

\bibitem[{{Wang} {et~al.}(2019){Wang}, {Rueda}, {Ruffini}, {Becerra}, {Bianco},
  {Becerra}, {Li}, \& {Karlica}}]{2019ApJ...874...39W}
{Wang}, Y., {Rueda}, J.~A., {Ruffini}, R., {et~al.} 2019, \apj, 874, 39,
  \dodoi{10.3847/1538-4357/ab04f8}

\bibitem[{{Wiltshire} {et~al.}(2009){Wiltshire}, {Visser}, \&
  {Scott}}]{2009kesp.book.....W}
{Wiltshire}, D.~L., {Visser}, M., \& {Scott}, S.~M. 2009, {The Kerr Spacetime}

\bibitem[{{Woosley}(1993)}]{1993ApJ...405..273W}
{Woosley}, S.~E. 1993, \apj, 405, 273, \dodoi{10.1086/172359}

\end{thebibliography}


\appendix

\section{The Kerr metric}\label{app:A}

\subsection{The original Kerr metric and the Boyer-Lindquist coordinates form}\label{app:A1}

The Kerr spacetime metric, which is stationary and axially symmetric, describes the exterior field of a rotating BH. In Eddington-Finkelstein-like coordinates $(u, r, \theta, \phi)$, it reads
\begin{equation}\label{eq:Kmetricu}
    ds^2 = -\left( 1- \frac{2 M r}{\Sigma} \right)du^2 + 2 du\,dr + \Sigma\,d\theta^2 + \frac{A}{\Sigma}\sin^2\theta\,d\phi^2-2 a \sin^2\theta\,dr\,d\phi -\frac{ 4 a \,M\,r\sin^2\theta }{\Sigma} du\,d\phi,
\end{equation}
where $\Sigma=r^2+a^2\cos^2\theta$, $\Delta=r^2-2 M r+ a^2$, and $A = (r^2+a^2)^2-\Delta a^2 \sin^2\theta$, with $M$ and $a=J/M$, respectively, the BH mass and angular momentum per unit mass. This form of the Kerr metric differs from the original one presented by \citet{1963PhRvL..11..237K}, in the signs of the $u$ coordinate and of the BH spin parameter $a$ (see \citealp{2007arXiv0706.1109K} in \citealp{2009kesp.book.....W}, for details).

The current most used form of the Kerr metric uses the ``spheroidal'' Boyer-Lindquist coordinates \citep{doi:10.1063/1.1705193}, which is obtained via the coordinate transformation \citep{1968PhRv..174.1559C, 2007arXiv0706.1109K}
\begin{equation}
    du \to dt + \frac{r^2 + a^2}{\Delta} dr,\quad d\phi \to d\phi + \frac{a}{\Delta} dr, 
\end{equation}
while $r$ and $\theta$ hold the same. In these coordinates, the Kerr metric reads \citep{1968PhRv..174.1559C,2007arXiv0706.1109K}:
\begin{equation}\label{eq:Kmetric}
ds^2 = -\left( 1- \frac{2 M r}{\Sigma} \right)dt^2 +\frac{\Sigma}{\Delta} dr^2+ \Sigma d\theta^2+ \frac{A}{\Sigma}\sin^2\theta d\phi^2-\frac{ 4 a \,M\,r\sin^2\theta }{\Sigma} dt d\phi,
\end{equation}
and the (outer) event horizon is located at $r_+=M+\sqrt{M^2-a^2}$.

Computations are facilitated by writing the metric in the general form of an asymptotically flat, stationary, axisymmetric metric, which using the notation of \citet{1970ApJ...162...71B,1972ApJ...178..347B} reads
\begin{equation}\label{eq:Kerr2}
    ds^2 = -e^{2\nu}dt^2 + e^{2\Psi} (d\phi - \omega dt)^2 + e^{2\mu_1} dr^2 + e^{2\mu_2} d\theta^2, 
\end{equation}
where 
\begin{equation}\label{eq:defsKerr2}
    e^{2\nu} = \frac{\Sigma \Delta}{A},\quad e^{2\Psi} = \frac{A}{\Sigma}\sin^2\theta,\quad e^{2\mu_1} = \frac{\Sigma}{\Delta}, \quad
    e^{2\mu_2} = \Sigma,\quad \omega = \frac{2 M a r}{A}.
\end{equation}
%

\subsection{Kerr-Schild  coordinates}\label{app:A2}

The Kerr metric was also presented by \citet{1963PhRvL..11..237K} in Kerr-Schild  spacetime coordinates $(t,x,y,z)$ which are related to the Boyer-Lindquist ones by \citep[see][for details]{1963PhRvL..11..237K,2007arXiv0706.1109K}
%
\begin{equation}\label{eq:xyz}
    x = (r \cos\phi - a\sin\phi) \sin\theta,\quad
    y = (r \sin\phi + a\cos\phi) \sin\theta,\quad
    z = r \cos\theta,
\end{equation}
%
whose inverse transformation can be written as
%
\begin{equation}\label{eq:xyzinv}
    r^2 = \frac{r_f^2 -a^2 +\sqrt{(r_f^2 - a^2)^2 + 4 a^2 z^2}}{2},\quad
    \theta  = \arctan{\left(\frac{r}{z}\sqrt{\frac{x^2 + y^2}{r^2 + a^2}}\right)},\quad
    \phi = \arctan{\left(\frac{y}{x}\right)} + \arctan{\left(\frac{a}{r}\right)},
\end{equation}
%
where $r_f = \sqrt{x^2 + y^2 + z^2}$, and we have given the expressions for $\theta$ and $\phi$ in terms of $\arctan$ functions which give the right angle sign on any quadrant. In these coordinates, the Kerr metric reads \citep{1963PhRvL..11..237K}
\begin{equation}
    ds^2 = -dt^2 + dx^2 + dy^2 + dz^2 + \frac{2 M r^3}{r^4 + a^2 z^2} \Bigg[ dt + \frac{z}{r} dz + \frac{r}{r^2 + a^2} (x\,dx + y\, dy) - \frac{a}{r^2 + a^2} (x\,dy - y\, dx) \Bigg]^2,
\end{equation}
which explicitly show the ``quasi-Minkowskian'' character of the metric \citep[see][for details]{2007arXiv0706.1109K}.

\section{Locally non-rotating observers}\label{app:B}

\subsection{Observer's frame using the Boyer-Lindquist coordinate basis}\label{app:B1}

We denote coordinate basis vectors by $\vec{e}_a = \partial/\partial \vec{x^a}$, and dual one-form basis vectors by $\vec{e}^a = \vec{d}x^a$. Thus, the tetrad carried by the LNR observer \citep{1970ApJ...162...71B,1972ApJ...178..347B} has the following dual 1-form basis vectors
%
\begin{equation}\label{eq:ZAMOoneforms}
    \vec{e}^{\hat{t}} = \sqrt{\frac{\Sigma\,\Delta}{A}}\, \vec{e}^t,\quad
    \vec{e}^{\hat{r}} = \sqrt{\frac{\Sigma}{\Delta}}\, \vec{e}^r, \quad
    \vec{e}^{\hat{\theta}} = \sqrt{\Sigma}\, \vec{e}^\theta,\quad
    \vec{e}^{\hat{\phi}} = - \frac{2 M r a \sin\theta}{\sqrt{\Sigma\, A}}\, \vec{e}^t +  \sqrt{\frac{A}{\Sigma}} \sin\theta\, \vec{e}^\phi,
\end{equation}
%
which are naturally identified from the metric (\ref{eq:Kerr2}), and tetrad vectors given by
%
\begin{equation}\label{eq:ZAMOtetrad}
    \vec{e}_{\hat{t}}  = \sqrt{\frac{A}{\Sigma\,\Delta}}\, \vec{e}_t + \frac{2 M a r}{\sqrt{\Sigma\, \Delta\, A}}\, \vec{e}_\phi,\quad
    \vec{e}_{\hat{r}} =  \sqrt{\frac{\Delta}{\Sigma}}\, \vec{e}_r,\quad
    \vec{e}_{\hat{\theta}} =  \frac{1}{\sqrt{\Sigma}}\,\vec{e}_\theta,\quad
    \vec{e}_{\hat{\phi}} =  \sqrt{\frac{\Sigma}{A}}\frac{1}{\sin\theta}\,\vec{e}_\phi,
\end{equation}
%
where in the second equality of each covector and vector, we have used the definitions (\ref{eq:defsKerr2}).

It is also useful to write the coordinate basis 1-forms in terms of the observer's basis ones, i.e.
%
\begin{equation}\label{eq:BLtoZAMOoneforms}
    \vec{e}^t =  \sqrt{\frac{A}{\Sigma\,\Delta}}\, \vec{e}^{\hat{t}},\quad
    \vec{e}^r = \sqrt{\frac{\Delta}{\Sigma}}\, \vec{e}^{\hat{r}},\quad
    \vec{e}^\theta =  \frac{1}{\sqrt{\Sigma}}\, \vec{e}^{\hat{\theta}},\quad
    \vec{e}^\phi = \frac{2 M a r}{\sqrt{\Sigma\,\Delta\,A}}\frac{1}{\sin\theta}\, \vec{e}^{\hat{t}} + \sqrt{\frac{\Sigma}{A}}\frac{1}{\sin\theta}\, \vec{e}^{\hat{\phi}},
\end{equation}
%
and the corresponding relation among basis vectors
%
\begin{equation}\label{eq:BLtoZAMOtetrad}
    \vec{e}_t =  \sqrt{\frac{\Sigma\,\Delta}{A}}\, \vec{e}_{\hat{t}} - \frac{2 M a r}{\sqrt{\Sigma\,A}}\, \vec{e}_{\hat{\phi}},\quad
    \vec{e}_r = \sqrt{\frac{\Sigma}{\Delta}}\, \vec{e}_{\hat{r}},\quad
    \vec{e}_\theta = \sqrt{\Sigma}\, \vec{e}_{\hat{\theta}},\quad
    \vec{e}_\phi =  \sqrt{\frac{A}{\Sigma}}\sin\theta\, \vec{e}_{\hat{\phi}}.
\end{equation}

For the formulation of the equations of motion in the LNR frame, we must know the components of the involved four-vectors and tensors projected onto the observer's tetrad. The covariant (contravariant) components of a vector and a tensor in the observer's frame are given by
%
\begin{equation}\label{eq:projections}
    A_{\hat{a}} = e^\mu_{\hphantom{\mu}{\hat{a}}} A_\mu,\quad A^{\hat{a}} = e_\mu^{\hphantom{\mu}{\hat{a}}} A^\mu\quad B_{\hat{a} \hat{b}}= e^\mu_{\hphantom{\mu}{\hat{a}}}e^\nu_{\hphantom{\mu}{\hat{b}}} B_{\mu \nu},\quad B^{\hat{a} \hat{b}} = e_\mu^{\hphantom{\mu}{\hat{a}}}e_\nu^{\hphantom{\mu}{\hat{b}}} B^{\mu \nu},
\end{equation}
%
where we recall the Greek indexes label the components in the coordinate frame. Thus, the four-velocity velocity of a particle with respect to the observer's frame is
\begin{equation}\label{eq:uontotetrad}
    u^{\hat{a}} = u^\mu\,e_\mu^{\hphantom{\mu}{\hat{a}}},
\end{equation}
and the spatial components
\begin{equation}
    v^{\hat{i}} = \frac{u^\mu\,e_\mu^{\hphantom{\mu}{\hat{i}}}}{u^\mu\,e_\mu^{\hphantom{\mu}{\hat{t}}}},
\end{equation}
so we can write the velocity four-vector of the particle as $\vec{\hat{u}} = (\hat{\gamma}, \hat{\gamma} \vec{v})$, where $\hat{\gamma} = (1-v^{\hat{i}} v_{\hat{i}})^{-1/2}$ is the Lorentz factor of the particle measured by the LNR observer. The four-velocity of the LNR observer, $e_{\hat{t}}$, with respect to an \emph{observer at rest at infinity} can be written as
\begin{equation}\label{eq:vzamo}
    u^{\hat{a}}_{\rm lnr} = \Gamma \vec{e}_t + \Gamma V \vec{e}_\phi,
\end{equation}
where
\begin{equation}\label{eq:vzamob}
    \Gamma = e^{-\nu} = \sqrt{\frac{A}{\Sigma \Delta}},\qquad V = \omega = \frac{2 M a r}{A}. 
\end{equation}
In the \emph{slow rotation} regime, at first-order in the spin parameter, the spatial velocity of the observer becomes
\begin{equation}\label{eq:vzamolow}
      V = \frac{2 M a}{r^3}.
\end{equation}
This expression is much accurate than expected, indeed, for a spin parameter $a\lesssim 0.7$, it overestimates the full expression by less than $5\%$ for radial distances $r\gtrsim 2 r_+$ and for any polar angle. Of course, the expression increases the accuracy for the lower spin values and for the larger distances. 

The spatial velocity of the LNR observer, as seen by a \emph{locally static observer}, is instead given by
\begin{equation}\label{eq:vzamo2}
     V_{\rm lnr,s} = \frac{2 M a r \sin\theta}{\Sigma \sqrt{\Delta}} \approx \frac{2 M a \sin\theta}{r^2 \sqrt{1-2 M/r}},
\end{equation}
where the last approximation corresponds to the slow rotation regime. The tetrad basis carried by this observer is not used in this article, hence it is not here presented.

In a non-coordinate frame, the covariant derivative of a tensor is given by
\begin{equation}\label{eq:covariantD}
    \frac{D B^{\hat{a}}_{\hphantom{\hat{a}}{\hat{b}}}}{d x^{\hat{c}}} = \frac{\partial B^{\hat{a}}_{\hphantom{\hat{a}}{\hat{b}}}}{\partial x^{\hat{c}}} + \omega_{\hat{c}\,\,\,\hat{d}}^{\hphantom{\mu}{\hat{a}}}\,B^{\hat{d}}_{\hphantom{\hat{d}}{\hat{b}}} - \omega_{\hat{c}\,\,\,\hat{b}}^{\hphantom{\mu}{\hat{d}}}\,B^{\hat{a}}_{\hphantom{\hat{a}}{\hat{d}}},
\end{equation}
where the \emph{spin connection} coefficients, $\omega_{\hat{c}\,\,\,\hat{b}}^{\hphantom{\mu}{\hat{a}}}$, play the role of the \emph{affine connection} coefficients.  For the present LNR observer, the spin coefficients can be obtained directly from the rotation vectors \citep{1972ApJ...178..347B}
\begin{equation}
    \vec{\omega}_{\hat{a}\hat{b}} = \omega_{\hat{a}\hat{b}\hat{c}}\,\vec{e}^{\hat{c}},
\end{equation}
where
\begin{subequations}\label{eq:spincoefficients}
\begin{align}
    &\omega_{\hat{t}\hat{r}\hat{t}} = -\omega_{\hat{r}\hat{t}\hat{t}} = -\nu_{,r}\,e^{-\mu_1},
    \quad
     \omega_{\hat{t}\hat{r}\hat{\phi}} = -\omega_{\hat{r}\hat{t}\hat{\phi}} = -\frac{\omega_{,r}}{2}e^{\Psi-\nu-\mu_1},\\
    &\omega_{\hat{t}\hat{\theta}\hat{t}} = -\omega_{\hat{\theta}\hat{t}\hat{t}} = -\nu_{,\theta}\,e^{-\mu_2},
    \quad
     \omega_{\hat{t}\hat{\theta}\hat{\phi}} = -\omega_{\hat{\theta}\hat{t}\hat{\phi}} = -\frac{\omega_{,\theta}}{2}e^{\Psi-\nu-\mu_2},\\
     &\omega_{\hat{t}\hat{\phi}\hat{r}} = -\omega_{\hat{\phi}\hat{t}\hat{r}} = -\frac{\omega_{,r}}{2}e^{\Psi-\nu-\mu_1},
     \\
     &\omega_{\hat{t}\hat{\phi}\hat{\theta}} = -\omega_{\hat{\phi}\hat{t}\hat{\theta}} =  -\frac{\omega_{,\theta}}{2}e^{\Psi-\nu-\mu_2},\\
     &\omega_{\hat{r}\hat{\theta}\hat{r}} = -\omega_{\hat{\theta}\hat{r}\hat{r}} = \mu_{1,\theta}\,e^{-\mu_2},
     \quad
     \omega_{\hat{r}\hat{\theta}\hat{\theta}} = -\omega_{\hat{\theta}\hat{r}\hat{\theta}} =  -\mu_{2,r}\,e^{-\mu_1},\\
     &\omega_{\hat{r}\hat{\phi}\hat{\phi}} = -\omega_{\hat{\phi}\hat{r}\hat{\phi}} = -\Psi_{,r}\,e^{-\mu_1},
     \quad
     \omega_{\hat{r}\hat{\phi}\hat{t}} = -\omega_{\hat{\phi}\hat{r}\hat{t}} = \frac{1}{2}\omega_{,r}\,e^{\Psi-\nu-\mu_1},\\
     &\omega_{\hat{\theta}\hat{\phi}\hat{\phi}} = -\omega_{\hat{\phi}\hat{\theta}\hat{\phi}} = -\Psi_{,\theta}\,e^{-\mu_2},
     \quad
     \omega_{\hat{\theta}\hat{\phi}\hat{t}} = -\omega_{\hat{\phi}\hat{\theta}\hat{t}} = \frac{1}{2}\omega_{,\theta}\,e^{\Psi-\nu-\mu_2},
\end{align}
\end{subequations}
with the colon notation $f_{,\sigma}\equiv \partial f/\partial x^\sigma$, and we have used the antisymmetric property $\vec{\omega}_{\hat{a}\hat{b}} = -\vec{\omega}_{\hat{b}\hat{a}}$.

\subsection{Observer's frame using the Kerr-Schild  coordinate basis}\label{app:B2}

The relation between the spatial coordinate basis vectors in Boyer-Lindquist and  coordinates is therefore given by
\begin{subequations}\label{eq:BLto}
\begin{align}
    \vec{e}_r &=  \sin\theta \cos\phi\,\vec{e}_x + \sin\theta \sin\phi\,\vec{e}_y + \cos\theta \,\vec{e}_z,\\
    \vec{e}_\theta &= (r\cos\theta \cos\phi -a\cos\theta\sin\phi)\,\vec{e}_x + (r\cos\theta \sin\phi +a\cos\theta\cos\phi)\,\vec{e}_y -r \sin\theta\,\vec{e}_z,\\
    \vec{e}_\phi &= -(r \sin\theta\sin\phi + a\sin\theta\cos\phi)\,\vec{e}_x + (r \sin\theta\cos\phi - a\sin\theta\sin\phi)\,\vec{e}_y.
\end{align}
\end{subequations}
while the vectors of the inverse transformation are
\begin{subequations}\label{eq:toBL}
\begin{align}
    \vec{e}_x & = \left(\frac{r^2}{\Sigma} \sin\theta \cos\phi - \frac{a\,r}{\Sigma}\sin\theta\sin\phi\right)\,\vec{e}_r + \left(\frac{r}{\Sigma} \cos\theta \cos\phi - \frac{a}{\Sigma}\cos\theta\sin\phi \right)\,\vec{e}_\theta - \left(\frac{r}{\Sigma \sin\theta}\sin\phi + \frac{a}{\Sigma \sin\theta} \cos^2\theta\cos\phi\right)\,\vec{e}_\phi,\\
    \vec{e}_y & = \left(\frac{r^2}{\Sigma} \sin\theta \sin\phi + \frac{a\,r}{\Sigma}\sin\theta\cos\phi\right)\,\vec{e}_r + \left(\frac{r}{\Sigma} \cos\theta \sin\phi + \frac{a}{\Sigma}\cos\theta\cos\phi \right)\,\vec{e}_\theta + \left(\frac{r}{\Sigma \sin\theta}\cos\phi- \frac{a}{\Sigma \sin\theta} \cos^2\theta\sin\phi\right)\,\vec{e}_\phi,\\
    \vec{e}_z & = \frac{r^2+a^2}{\Sigma}\cos\theta\,\vec{e}_r - \frac{r}{\Sigma}\sin\theta\,\vec{e}_\theta + \frac{a}{\Sigma}\cos\theta\,\vec{e}_\phi.
\end{align}
\end{subequations}

In the limit $a\to 0$, or in the weak-field limit $r\to \infty$ (so $a/r\to 0$), the above transformation reduce to the ones from traditional spherical to  coordinates:
\begin{subequations}\label{eq:Sphto}
\begin{align}
    \vec{e}_r &=  \sin\theta \cos\phi\,\vec{e}_x + \sin\theta \sin\phi\,\vec{e}_y + \cos\theta \,\vec{e}_z,\\
    \vec{e}_\theta &= r\cos\theta \cos\phi\,\vec{e}_x + r\cos\theta \sin\phi\,\vec{e}_y -r \sin\theta\,\vec{e}_z,\\
    \vec{e}_\phi &= -r \sin\theta\sin\phi\,\vec{e}_x + r \sin\theta\cos\phi\,\vec{e}_y.
\end{align}
\end{subequations}
and
\begin{subequations}\label{eq:toSph}
\begin{align}
    \vec{e}_x & = \sin\theta \cos\phi \,\vec{e}_r + \frac{1}{r}\cos\theta \cos\phi\,\vec{e}_\theta - \frac{1}{r\sin\theta} \sin\phi\,\vec{e}_\phi,\\
    \vec{e}_y & = \sin\theta \sin\phi\,\vec{e}_r  + \frac{1}{r}\cos\theta \sin\phi\,\vec{e}_\theta + \frac{1}{r\sin\theta}\cos\phi\,\vec{e}_\phi,\\
    \vec{e}_z & = \cos\theta\,\vec{e}_r - \frac{1}{r}\sin\theta\,\vec{e}_\theta.
\end{align}
\end{subequations}

Using the coordinate basis relation (\ref{eq:BLto}), one can express the observer's tetrad (\ref{eq:ZAMOtetrad}) in the  coordinate basis.

\section{Photon four-momentum measured by different observers}\label{app:C}

We are interested in determining the photon four-momentum measured by an observer at rest at infinity. To do so, we have to express it in terms of the four-momentum measured by an observer comoving with the emitter (i.e. the radiating electron), which sees an isotropic radiation field. 

First, we relate the comoving and the LNR frames via a Lorentz boost: 
\begin{equation}\label{eq:comzamo}
    \vec{e}_{(a)} = \Lambda^{\hat{b}}_{\hphantom{\hat{b}}{(a)}}\,\vec{e}_{\hat{b}},
\end{equation}
where the components of the Lorentz boost are
\begin{equation}\label{eq:boost}
    \Lambda^{\hat{0}}_{\hphantom{\hat{0}}{(0)}} = \hat{\gamma},\quad \Lambda^{\hat{0}}_{\hphantom{\hat{0}}{(i)}} = \hat{\gamma} v_{\hat{i}},\quad \Lambda^{\hat{i}}_{\hphantom{\hat{i}}{(j)}} = \delta^i_j + \frac{\hat{\gamma}^2}{\hat{\gamma}+1}v^{\hat{i}}v_{\hat{j}},
\end{equation}
where we denote vector components in the comoving frame with round brackets, and we recall $\hat{\gamma} = (1-v^{\hat{i}} v_{\hat{i}})^{-1/2}$ is the Lorentz factor of the emitter measured by the LNR observer.

The tetrad basis of the LNR frame is related to the one of the coordinate frame by
\begin{equation}
    \vec{e}_{\hat{b}} = e^{\mu}_{\hphantom{\mu}{\hat{b}}} \,\vec{e}_{\mu},
\end{equation}
where the transformation components $e^{\mu}_{\hphantom{\mu}{\hat{b}}}$ are given by Eqs.~(\ref{eq:ZAMOtetrad}).

Therefore, the relations between the comoving tetrad basis and the basis of the observer at rest at infinity are
\begin{equation}
    \vec{e}_{(a)} = \Lambda^{\hat{b}}_{\hphantom{\hat{b}}{(a)}}\,e^{\mu}_{\hphantom{\mu}{\hat{b}}}\,\vec{e}_{\mu},\quad \vec{e}_{\mu} = \Lambda_{\hat{b}}^{\hphantom{\hat{b}}{(a)}}\,e_{\mu}^{\hphantom{\mu}{\hat{b}}}\,\vec{e}_{(a)},
\end{equation}
where $\Lambda_{\hat{b}}^{\hphantom{\hat{b}}{(a)}}$ can be obtained from Eq.~(\ref{eq:boost}) by changing the sign of the velocity components, i.e. $v^{\hat{i}} \to -v^{\hat{i}}$, and the transformation components $e_{\mu}^{\hphantom{\mu}{\hat{b}}}$ are given by Eqs.~(\ref{eq:ZAMOoneforms}).

The photon four-momentum in the coordinate frame, $k^\mu$, is therefore related to the photon four-momentum in the comoving frame, $k^{(a)}$, by 
\begin{equation}
    k^\mu = e^{\mu}_{\hphantom{\mu}{\hat{b}}}\,\Lambda^{\hat{b}}_{\hphantom{\hat{b}}{(a)}}\,k^{(a)},
\end{equation}
and we recall that for photons the above four-momenta satisfy, respectively, $k^\mu k_\mu = 0$ and $k^{(a)} k_{(a)} = 0$. Thus, the four-momenta are related by
\begin{subequations}\label{eq:kmuka}
\begin{align}
    k^0 &= k^{(0)} \hat{\gamma} e^{-\nu} [1 + v_{\hat{i}} \,n^{(i)}],\\
    n^r &= \frac{ k^{(0)}}{k^0}e^{-\mu_1} \left[ \hat{\gamma}v^{\hat{r}} + n^{(1)} + \frac{\hat{\gamma}^2}{\hat{\gamma}+1} v^{\hat{r}} v_{\hat{j}}n^{(j)} \right],\\
    n^\theta &= \frac{ k^{(0)}}{k^0}e^{-\mu_2} \left[ \hat{\gamma}v^{\hat{\theta}} + n^{(2)} + \frac{\hat{\gamma}^2}{\hat{\gamma}+1} v^{\hat{\theta}} v_{\hat{j}}n^{(j)} \right],\\
    n^\phi &= \frac{k^{(0)}}{k^0}e^{-\Psi} \Bigg\{e^{\Psi-\nu}\omega \hat{\gamma}[1+v_{\hat{j}}\,n^{(j)}] +  \hat{\gamma}v^{\hat{\phi}} + n^{(3)} + \frac{\hat{\gamma}^2}{\hat{\gamma}+1} v^{\hat{\phi}}     v_{\hat{j}}n^{(j)} \Bigg\},
\end{align}
\end{subequations}
where we have introduced the normalized four-momenta by $n^\mu = k^\mu/k^0$ and $n^{(a)} = k^{(a)}/k^{(0)}$. We recall that the Lorentz factor measured by the comoving and the observer at rest at infinity are related by:
\begin{equation}\label{eq:gammas}
    \gamma = u^0 = u^{\hat{0}} e^0_{\hphantom{0}{\hat{0}}} = \hat{\gamma} e^{-\nu}.
\end{equation}
For the sake of completeness, we also present the components of the inverse relation
\begin{equation}
    k^{(a)} = e_{\mu}^{\hphantom{\mu}{\hat{b}}}\,\Lambda_{\hat{b}}^{\hphantom{\hat{b}}{(a)}}\,k^{\mu},
\end{equation}
which leads to the explicit components
\begin{subequations}\label{eq:kakmmu}
\begin{align}
    k^{(0)} &= k^{0} \hat{\gamma} e^{\nu} [1 - e^{\mu_1-\nu} v^{\hat{r}} \,n^{r}- e^{\mu_2-\nu} v^{\hat{\theta}} \,n^{\theta} - e^{\Psi-\nu} v^{\hat{\phi}}(n^{\phi}-\omega)],\\
    n^{(1)} &= \frac{k^0}{k^{(0)}} e^{\nu} \Bigg\{-\hat{\gamma} v^{\hat{r}} + e^{\mu_1-\nu} n^{r} + \frac{\hat{\gamma}-1}{\hat{v}^2} v^{\hat{r}} \left[ e^{\mu_1-\nu} v^{\hat{r}} \,n^{r} + e^{\mu_2-\nu} v^{\hat{\theta}} \,n^{\theta} + e^{\Psi-\nu} v^{\hat{\phi}}(n^{\phi}-\omega) \right]\Bigg\},\\
    n^{(2)} &= \frac{k^0}{k^{(0)}} e^{\nu} \Bigg\{-\hat{\gamma} v^{\hat{\theta}} + e^{\mu_2-\nu} n^{\theta} + \frac{\hat{\gamma}-1}{\hat{v}^2} v^{\hat{\theta}} \left[ e^{\mu_1-\nu} v^{\hat{r}} \,n^{r} + e^{\mu_2-\nu} v^{\hat{\theta}} \,n^{\theta} + e^{\Psi-\nu} v^{\hat{\phi}}(n^{\phi}-\omega) \right]\Bigg\},\\
    n^{(3)} &= \frac{k^0}{k^{(0)}} e^{\nu} \Bigg\{-\hat{\gamma} v^{\hat{\phi}} + e^{\Psi-\nu} (n^{\phi}-\omega) + \frac{\hat{\gamma}-1}{\hat{v}^2} v^{\hat{\phi}} \left[ e^{\mu_1-\nu} v^{\hat{r}} \,n^{r} + e^{\mu_2-\nu} v^{\hat{\theta}} \,n^{\theta} + e^{\Psi-\nu} v^{\hat{\phi}}(n^{\phi}-\omega) \right]\Bigg\}.
\end{align}
\end{subequations}


\end{document}